\documentclass[11pt,a4paper]{emulateapj}
\usepackage{natbib}
\usepackage{amssymb, amsmath, amsbsy, slashbox}%, rotating}
\usepackage{color}
\usepackage{epsfig}
\usepackage{graphicx}
\slugcomment{submitted to {\it The Astrophysical Journal}}
\begin{document}

\def\mean#1{\left< #1 \right>}

\title{MApping the Most Massive Overdensities Through Hydrogen (MAMMOTH) I: Methodology}
\author{Zheng Cai\altaffilmark{1,2}, Xiaohui Fan\altaffilmark{1}, Sebastien Peirani\altaffilmark{3}, Fuyan Bian\altaffilmark{4\dag}, Brenda Frye\altaffilmark{1}, Ian McGreer\altaffilmark{1}, J. Xavier Prochaska\altaffilmark{2}, Marie Wingyee Lau \altaffilmark{2}, Nicolas Tejos\altaffilmark{2}, Shirley Ho\altaffilmark{5}, Donald P. Schneider\altaffilmark{6,7} }

\affil{$^1$ Steward Observatory, University of Arizona, 933 North Cherry Avenue, Tucson, AZ, 85721, USA}
\affil{$^2$ UCO/Lick Observatory, University of California, 1156 High Street, Santa Cruz, CA 95064, USA }
\affil{$^3$ Institut d'Astrophysique de Paris, 98 bis, Boulevard Arago - 75014 Paris, France}
\affil{$^4$ Research School of Astronomy \& Astrophysics, Mount Stromlo Observatory, Cotter Road, Weston ACT 2611, Australia}
%\affil{Lawrence Berkeley National Laboratory, 1 Cyclotron Road, Berkeley, CA 94720, USA}
\affil{$^5$McWilliams Center for Cosmology, Department of Physics, Carnegie Mellon University, Pittsburgh, PA 15213, USA}
\affil{$^6$ Department of Astronomy and Astrophysics, The Pennsylvania State University, University Park, PA 16802}
\affil{$^7$ Institute for Gravitation and the Cosmos, The Pennsylvania State University, University Park, PA 16802}
\affil{$^{\dag}$ Stromlo Fellow}
%\date{\today}
%\maketitle

\altaffiltext{1} {Email: zcai@ucolick.org}
% Usually omit these for ApJ or MNRAS style files:
%\tableofcontents
%
%\listoffigures
%
%\listoftables

\begin{abstract}

Modern cosmology predicts that a galaxy overdensity is associated to a large reservoir of the intergalactic gas, which can be traced by the Ly$\alpha$ forest absorption. We have undertaken a systematic study of the relation between Coherently Strong intergalactic Ly$\alpha$ Absorption systems (CoSLAs), which have highest optical depth ($\tau$) in $\tau$ distribution, and mass overdensities on the scales of $\sim$ 10--20 $h^{-1}$ comoving Mpc. On such large scales, our cosmological simulations show a strong correlation between the effective optical depth ($\tau_{\rm{eff}}$) of the CoSLAs and the 3-D mass overdensities. In moderate signal-to-noise spectra, however, the profiles of CoSLAs can be confused with high column density absorbers. For $z>2.6$, where the corresponding Ly$\beta$ is redshifted to the optical, we have developed the technique to differentiate between these two alternatives. We have applied this technique to SDSS-III quasar survey at $z = 2.6$--3.3, and we present a sample of five CoSLA candidates with $\tau_{\rm{eff}}$ on 15 $h^{-1}$ Mpc greater than $4.5\times$ the mean optical depth. At lower redshifts of $z < 2.6$, where the background quasar density is higher, the overdensity can be traced by intergalactic absorption groups using multiple sight lines. Our overdensity searches fully utilize the current and next generation of Ly$\alpha$ forest surveys which cover a survey volume of $> (1\ h^{-1}$ Gpc)$^3$. In addition, systems traced by CoSLAs will build a uniform sample of the most massive overdensities at $z > 2$ to constrain the models of structure formation, and offer a unique laboratory to study the interactions between galaxy overdensities and the intergalactic medium. 

\end{abstract}

%Section heading
\section{Introduction}

The most massive large-scale structures at the peak of cosmic star formation, i.e., $z \sim 2$--3, are unique laboratories for understanding cosmic mass assembly. These structures are excellent sites to study the earliest clusters of galaxies, the formation of highly evolved galaxies at high-redshift, and the complex interactions between galaxies and the intergalactic medium (IGM). Also, the abundance of the extreme tails of the mass overdensity provides a particularly potent constraint on models of structure and galaxy formation. %as it marks the regions where stars and galaxies are likely to be oldest at that (or any) epoch. 
Nevertheless, the task of finding high redshift ($z > 2$) protoclusters is challenging. The majority of protoclusters have been found by targeting rare sources, such as quasars \cite[e.g.,][]{hu96}, radio galaxies \cite[e.g.][]{venemans07}, sub-mm galaxies \cite[e.g.,][]{chapman04} and Ly$\alpha$ ``blobs" \cite[e.g.,][]{yang09}. The utility of these markers is limited by small duty cycles and strong selection biases, and thus the overdensities traced by these rare sources are highly incomplete. An alternative approach to identifying large-scale structures is by targeting blank fields in galaxy redshift surveys  \citep{steidel98,ouchi05,chiang13,lee14}. However, the current deep high-$z$ galaxy redshift surveys are limited by small survey area (up to a few deg$^2$). Due to these difficulties, the overall number of confirmed protoclusters is still too low to allow for robust comparisons to hierarchical structure formation models or for environmental studies of galaxy properties at different redshifts \citep{leeK14, chiang13}. 
A more complete search for galaxy overdensities from larger volume is  highly desirable. We are thus motivated to develop systematic techniques to identify early protoclusters, especially to build a uniform sample of the most biased and evolved examples at $z > 2$. This sample will enable studying how these structures interact with cosmic web filaments, feedback to IGM, and transform into the present-day local clusters, as well as to provide  stringent constraints to models of structure formation.  %Therefore, a more complete technique to search galaxy overdensity from larger volume will be highly desirable.  

Hydrogen in the intergalactic medium (IGM) maintains a high ionization fraction in the post-reionization epoch
under a metagalactic ionizing background originating from star-forming galaxies and active galactic nuclei (AGN) \cite[e.g.][]{mcquinn15}. Lyman-alpha (Ly$\alpha$) forest absorption marks locations where the quasar line-of-sight (LOS) intersects intergalactic neutral hydrogen (HI) at the wavelength of the redshifted Ly$\alpha$ resonance lines \cite[e.g.,][]{gunn65, lynds71, bi93}. The optical depth of HI Ly$\alpha$ smoothly traces the dark matter distribution at scales larger than the Jeans scale of the photoionized IGM \citep{cen94,miralda96,viel12,rauch06,lee14}.
Thus, the Ly$\alpha$ forest has been used as a crucial probe of the IGM and underlying mass distribution of $z>2$ over a sufficiently large scale, of densities ranging from the cosmic mean to highly overdense regions \citep{lee14}. 
Recently, Ly$\alpha$ forest has been used to measure clustering on large-scales of $\sim 100\ h^{-1}$ Mpc, and the Baryonic Acoustic Oscillation (BAO) feature at $z\sim 2.5$ was detected \citep{slosar11,slosar13, busca13}.

A number of  theoretical and observational studies have already probed the correlation between Ly$\alpha$ forest absorptions and galaxies on several comoving Mpc (co-Mpc) scales at $z=2-3$. On the observational side, absorbers with H I column densities $N_{\rm{HI}}\gtrsim 10^{14.5}$ cm$^{-2}$ are correlated with galaxy positions on $\sim$ Mpc scales, and the association is stronger for the higher column density systems \citep{rudie12}. Meanwhile, 
\citet{adelberger03} point out that on still larger scales of 10 $h^{-1}$ comoving Mpc, the mean transmission of the Ly$\alpha$ forest tends to be low in 
volumes that contain an overdensity of Lyman break galaxies (LBGs). 
On the theoretical side, \citet{mcdonald02} use several hydro-particle-mesh simulations with box size 
of 40 $h^{-1}$ comoving Mpc to demonstrate a strong correlation between the 
mass and the Ly$\alpha$ forest transmitted flux on 10 $h^{-1}$ comoving Mpc scale. %This 
%strong correlation does not suffer from uncertainties due to resolution of the simulations. 
At higher redshift, \citet{frye08} discover a strong intergalactic Ly$\alpha$ absorption with the optical depth  close to Gunn-Peterson absorption on 30 $h^{-1}$ Mpc in the spectrum of a galaxy at $z=4.9$. The Ly$\alpha$ optical depth is two times higher the mean optical depth at $z=4.9$ on a large scale of 80 $h^{-1}$ comoving Mpc. \citet{matsuda10} conducted deep narrow-band imaging centered on this unusual absorption, which indeed revealed a high concentration of Ly$\alpha$ galaxies at $z = 4.87$ on large-scale. The structure 
expands over a region of $\sim$ 20 comoving Mpc $\times$ 60 comoving Mpc on the sky with a galaxy overdensity of $\delta \sim$ 4. 
All of the above are evidence that support the idea that on sufficiently large scales of $\gtrsim$ 10 $h^{-1}$ comoving Mpc, the IGM HI gas is a fair tracer of the underlying mass. % {\bf I am not sure if you can yet claim that this is a less biased tracer. How do you measure the bias of this vs. other rare objects?}

The correlation between Ly$\alpha$ optical depth and galaxies on large-scales can be naturally interpreted. 
At galaxy and cluster scales ($\lesssim 1-2$ $h^{-1}$ comoving Mpc), hydrodynamical processes, such as 
AGN feedback, supernova (SN) winds, and shock-heating around cosmic web should significantly 
enhance the strength of the local ionizing radiation. Each of these mechanisms can complicate the relation between Ly$\alpha$  forest absorption and 3-D mass fluctuation. 
At the same time, on larger scales of $\gtrsim 5$ $h^{-1}$ comoving Mpc, AGN feedback could only have small effects on the mean ionizing background \citep{kollimeier03}, and other mechanisms, such as SN winds, shock heating, and metal-line cooling, generally have similar or smaller impacts on Ly$\alpha$ optical 
depth compared with AGN feedback \citep{viel13}. Moreover, 
even at small scales, a few simulations suggest that galactic winds tend to escape into the voids, 
leaving the filaments responsible for the Ly$\alpha$ absorption largely intact \citep{theuns02,tepper-garc12}. Therefore, it is not a surprise that 
at scales larger than $5$ h$^{-1}$ comoving Mpc, the strong Ly$\alpha$ absorption is highly correlated with the underlying mass overdensity, which can too be traced by a galaxy overdensity. %{\bf not sure what this means}

Guided by the theoretical and observational arguments above, we have developed a new approach for identifying the extreme tail of the matter density 
distribution at the typical proto-cluster scale of $10-30$ $h^{-1}$ co-Mpc. This approach utilizes the 
largest library of quasar spectra currently available from the  Baryon Oscillations Spectroscopic Survey (BOSS) \cite[e.g.,][]{dawson13}. 
The BOSS project is part of the Sloan Digital Sky 
Survey III (SDSS-III) \citep{eisenstein11}. This largest spectroscopic quasar dataset enables one to locate extremely rare, high optical depth HI (Ly$\alpha$) absorption due to IGM overdensities at scales of $\sim 10-30 $ $h^{-1}$ co-Mpc. These IGM HI overdensities, in turn, are expected to trace the most massive early proto-clusters and overdense regions. 
This technique allows coverage of significantly larger survey volume and may be less biased, since HI density is correlated with the underlying dark matter density field over large scales.

This is the first of a series of papers presenting the selection technique of the high effective optical-depth, large-scale intergalactic HI (Ly$\alpha$) absorption from the SDSS quasar spectral survey. These systems have coherently strong Ly$\alpha$ absorption on scales of $\sim 10-20\ h^{-1}$ Mpc. Over these scales, we present the strong correlation between the transmitted flux of 1-D intergalactic Ly$\alpha$ absorption and 3-D mass overdensities. We give detailed procedures for selecting these absorption systems. They trace regions that are excellent candidates of the most massive large-scale structures at $z = 2$--3. In the following, we 
refer to these systems as the CoSLAs for ``Coherently Strong Ly$\alpha$ Absorption systems". We refer to the whole project as MAMMOTH for MApping the Most Overdensity Through Hydrogen. 

%{\bf CHANGE!!!}

This paper is structured as follows. In \S 2, we introduce the SDSS-III/BOSS Ly$\alpha$ forest sample and the cosmological simulations that are used for theoretical guidance of SDSS-III/BOSS Ly$\alpha$ forest sample. From simulations, we measure the cross-correlation between the Ly$\alpha$ transmitted 
flux and mass fluctuation. In \S 3, we make realistic mock spectra, by including the high column discrete absorbers (Lyman limit system and damped Ly$\alpha$ absorption systems) to our mock spectra and adding noise as well as considering the continuum fitting errors. In \S 4, we introduce the techniques for selecting coherently strong intergalactic absorption systems (CoSLAs), emphasizing the needs to eliminate high column density discrete absorbers (HCDs), such as damped Ly$\alpha$ systems (DLAs).  In \S5, we provide a discussion on 
Ly$\alpha$ absorption systems associated with previously confirmed overdensities. In \S 6, we present the sample of CoSLA candidates selected from SDSS-III/BOSS and the MMT high S/N spectroscopy. In \S 7, we discuss the implications of our observations and use our observational results to compare with cosmological simulations. 
Throughout this paper when measuring distances, we refer to comoving separations or distances, measured in Mpc, with $H_0= 100\ h$ km s$^{-1}$ Mpc$^{-1}$. We convert redshifts to distances assuming 
a $\rm{\Lambda}$CDM cosmology with $\Omega_m= 0.3$, $\Omega_{\Lambda}=0.7$ and $h=0.70$. %The distances in this paper, 
%if not mentioned, are the comoving distances. 

%\section{Large-scale Ly$\alpha$ Absorption Systems Around Confirmed Galaxy Overdensities}

\section{Correlation between mass fluctuation and Ly$\alpha$ transmitted flux at large scales}

In the following, we explore the general statistical correlation between the transmitted flux ($F= \rm{exp}(-\tau_{\rm{eff}})$) of Ly$\alpha$ forest absorption and the 3-D mass fluctuation, and in particular, the scatter between the two. 
%In this section, using cosmological simulations, we study the relation between the transmitted flux ($F= \rm{exp}(-\tau_{\rm{eff}})$) of Ly$\alpha$ forest absorption and the 3-D mass fluctuation. 

To investigate this cross-correlation,  
we use two sets of large-scale cosmological simulations to model the Ly$\alpha$ forest. 
The large-scale Ly$\alpha$ forest simulations are designed to match the properties and guide the observations of 
large-scale Ly$\alpha$ forest surveys, e.g., the SDSS-III/BOSS Ly$\alpha$ forest sample.

\subsection{SDSS-III/BOSS Ly$\alpha$ Forest Sample}

We use the Ly$\alpha$ absorption spectra observed in SDSS-III Baryon Oscillation Spectroscopic Survey (BOSS) \citep{dawson13, ahn14}. 
BOSS is one of the four spectroscopic
surveys in SDSS-III taken with the 2.5-meter Sloan telescope \citep{gunn06}. 
BOSS measures redshifts of 1.5 million luminous red galaxies and 
Ly$\alpha$ absorption towards 160,000 high redshift quasars \citep{bolton12, ross12, dawson13}. 
The BOSS spectra have a moderate resolution of $R\sim 2,000$ from 3600 \AA\ $-$ 10,400 \AA. 
With a total exposure time of 1-hour for each plate, the BOSS Ly$\alpha$ quasar sample \citep{lee12} has a median signal-to-noise ratio (S/N) of $\sim2$ per pixel (1 pixel corresponds to $\approx 1$\AA) at rest-frame 
wavelength $\lambda = 1041- 1185$ \AA. 

The SDSSIII-BOSS/DR12 quasar catalog \citep{paris14} includes 160,000 quasars over 10,000 deg$^2$, which yields a quasar average density of 1 per (15 arcmin)$^2$, where (15 arcmin)$^2$ $=$ (17 $h^{-1}$ Mpc)$^2$ at $z = 2.5$. Assuming the typical mass overdensity extends a 15 $h^{-1}$ Mpc, and assuming each quasar probes an average of $\Delta z \sim 0.3$ on the sight line, our survey volume for searching overdensities is approximately (1.8 $h^{-1}$ Gpc)$^3$. To measure Ly$\alpha$ optical depth, the mean-flux-regulated principal component analysis (MF-PCA) technique is applied for the continuum fitting \citep{lee13}. In this technique, PCA fitting is performed redward of the quasar Ly$\alpha$ line in order to provide a prediction for the shape of the Ly$\alpha$ forest continuum. The slope and amplitude of this continuum prediction are then corrected using external constraints for the Ly$\alpha$ forest mean flux \citep{lee12, becker13}.

\subsection{Large-scale Cosmological Simulations  on Ly$\alpha$ Forest}

% The cosmological simulations are used to provide the theoretical guidances for the SDSS-III/BOSS Ly$\alpha$ forest sample. 
We used two cosmological simulations to examine the correlation between the Ly$\alpha$ transmitted flux and the mass overdensity. 

The first cosmological simulation has a box size of $1.5\ h^{-1}$ Gpc and a Plummer equivalent smoothing of 36 $h^{-1}$ kpc. It contains 1500$^3$ particles \citep{white11}. The initial particle spacing is 1.0 $h^{-1}$ Mpc. 
This simulation was originally used for predicting the BAO feature at $\approx$ 100 $h^{-1}$ Mpc. 
The deterministic fluctuating Gunn-Peterson approximation (FGPA) was used to generate skewers 
of optical depth with 4,096 pixels for each sight line \cite[e.g.,][]{slosar11}. A temperature at the mean density of 2$\times 10^4$ K 
and an equation of state $T (\Delta) = T_0 \Delta^{\gamma-1}$ with a $\gamma = 1.5$ is assumed, where $\Delta$ is the overdensity \cite[e.g.,][]{lee15}. The optical depth was scaled so that the median transmitted 
flux is $\bar{F} = \mean{\rm{exp}(-\tau)}$ = 0.78, with {$\mean{\tau=0.25}$}, consistent with observations at $z\approx 2.5$ \cite[e.g.,][]{kirkman05, bolton09}. In the following, we refer to this simulation as the deterministic simulation. 

The second simulation was performed with the N-body code GADGET-2 \citep{springel05} 
and used a box length of 1 $h^{-1}$ Gpc and 1024$^3$ dark matter particles. The
Plummer-equivalent force softening adopted is 5\% of the mean inter-particle
distance 48.8 $h^{-1}$ kpc. From this simulation, we derived 
mock Ly$\alpha$ forest at $z\sim 2.5$ using LyMAS  (Ly$\alpha$ Mass Association Scheme). 
 The detailed description of this code is given in \citep{peirani14}. 
Contrary to the FGPA mapping which assumes a deterministic relation between
 dark matter overdensity and continuum normalized Ly$\alpha$ flux,
LyMAS considers a stochastic relation described by a conditional 
probability distribution $P(F_s|\delta_s)$ of the smoothed transmitted flux 
$F_s$, on the smoothed dark matter density 
$\delta_s$. The conditional 
probability distribution has been previously derived from high-resolution hydrodynamic simulations of smaller volumes, 
 including a full treatment of physical
processes such as metal-dependent cooling, star formation, photoionization and heating from a UV background, 
supernova feedback and metal enrichment \citep{peirani14}. \citet{lochhaas15} tested the cross-correlation of Ly$\alpha$ forest predicted by LyMAS simulation 
and found that LyMAS perfectly reproduces the correlation between dark matter and transmitted flux computed from the full hydrodynamic gas distribution. 
In this paper, we use the extended form of LyMAS which produces coherent spectra 
reproducing the 1-dimensional power spectrum and 1-point flux distribution of the hydro simulation spectra (in redshift-space).
 LyMAS is expected to be more accurate than a deterministic density-flux mapping; for instance, the deterministic mapping significantly overpredicts the
flux correlation function relative to the LyMAS scheme \citep{peirani14}.
In the following,
we refer to this simulation as LyMAS. We use both simulations to examine the
 correlation between Ly$\alpha$ absorption and mass overdensities
at scales of ten to a few tens of Mpc.

\subsection{Strong correlation between Mass and Ly$\alpha$ Forest Transmitted Flux over 10 -- 40 $h^{-1}$ Mpc}

Using cosmological simulations, we study the large-scale correlation between mass and Ly$\alpha$ transmitted flux as a function of different smoothing lengths (scales) from 3 $h^{-1}$ Mpc -- 40 $h^{-1}$ Mpc. 

We define the fluctuation of the transmitted flux $\delta_F = 1- F/\bar{F}$, where $F$ is the 
transmitted flux in the Ly$\alpha$ forest after the spectrum smoothed with a top-hat filter, and $\bar{F}$ is the mean transmitted flux. We also define the mass perturbation $\delta_m= m/\bar{m}$. The mass $m$ is defined as the total mass of the particles 
inside the cubes (redshift-space) centered on the pixel of the $\delta_F$ measurement. Each cube has a linear size equal to the smoothing length (scales). 

\begin{figure}[tbp]
\epsscale{1}
\label{fig:corr}
\plotone{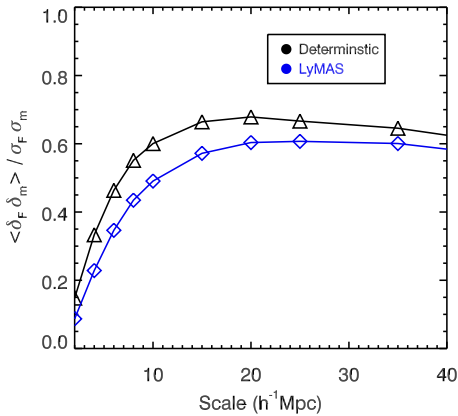}
\caption{The cross-correlation between three-dimensional (3-D) mass and one-dimensional (1-D) transmitted flux as a function of scales. The black curve is calculated from the deterministic simulation (deterministic scheme). Blue represents the LyMAS simulation. The deterministic simulation produces systematically higher correlations than LyMAS. 
 Both cosmological simulations suggest the correlation has a broad peak over the scale of 15-40 $h^{-1}$ Mpc. }
%The scatter is mostly due to the fact that the 3-D large scale structure are elongated and 1-D Ly$\alpha$ forest optical depth is biased towards those seeing along major axis. }
\end{figure}

Figure~\ref{fig:corr} presents the cross-correlation coefficients of transmitted flux $\delta_F$ and mass perturbation $\delta_m$  as a function of different smoothing scales. 
The profile of correlation coefficients predicted by deterministic simulation is generally consistent with those calculated by the LyMAS simulation. Both simulations suggest a strong correlation between the mass overdensity and Ly$\alpha$ absorption on large scales. 
The correlation peaks at scales of 10 $-$ 40 $h^{-1}$ Mpc. 
From the Figure~\ref{fig:corr}, the deterministic scheme gives generally higher 
correlation coefficients between $\delta_F-\delta_m$. %At $10\ h^{-1}$ Mpc, the correlation coefficient given by 
%deterministic scheme is 20\% higher than that provided by LyMAS scheme. 
For example, at $20\ h^{-1}$ Mpc, the deterministic scheme presents a slightly ($\sim$10\%) stronger correlation. %This difference is expected because the FGPA scheme tends to overpredict 
%the correlation and cross-correlation measurements (Peirani et al. 2014).

The correlation coefficient on 10 $h^{-1}$ Mpc derived by the LyMAS scheme 
is in excellent agreement with that calculated by \citet{kollimeier03}, who use a fully hydrodynamical 
simulation with a box size of 40 $h^{-1}$ Mpc. \citet{mcdonald02} also derive the correlation between the Ly$\alpha$ forest absorption and mass fluctuation on $\sim$ 5 $h^{-1}$ Mpc. 
Compared with \citet{mcdonald02} and \citet{kollimeier03}, this paper extends the cross-correlation to scales from $3\ h^{-1}$ Mpc to $40\ h^{-1}$ Mpc. 
For the upcoming analysis, we mainly use the LyMAS simulation.

\figurenum{1}

\figurenum{2}
\begin{figure}[tbp]
\epsscale{1.2}
\label{fig:scatter}
\plotone{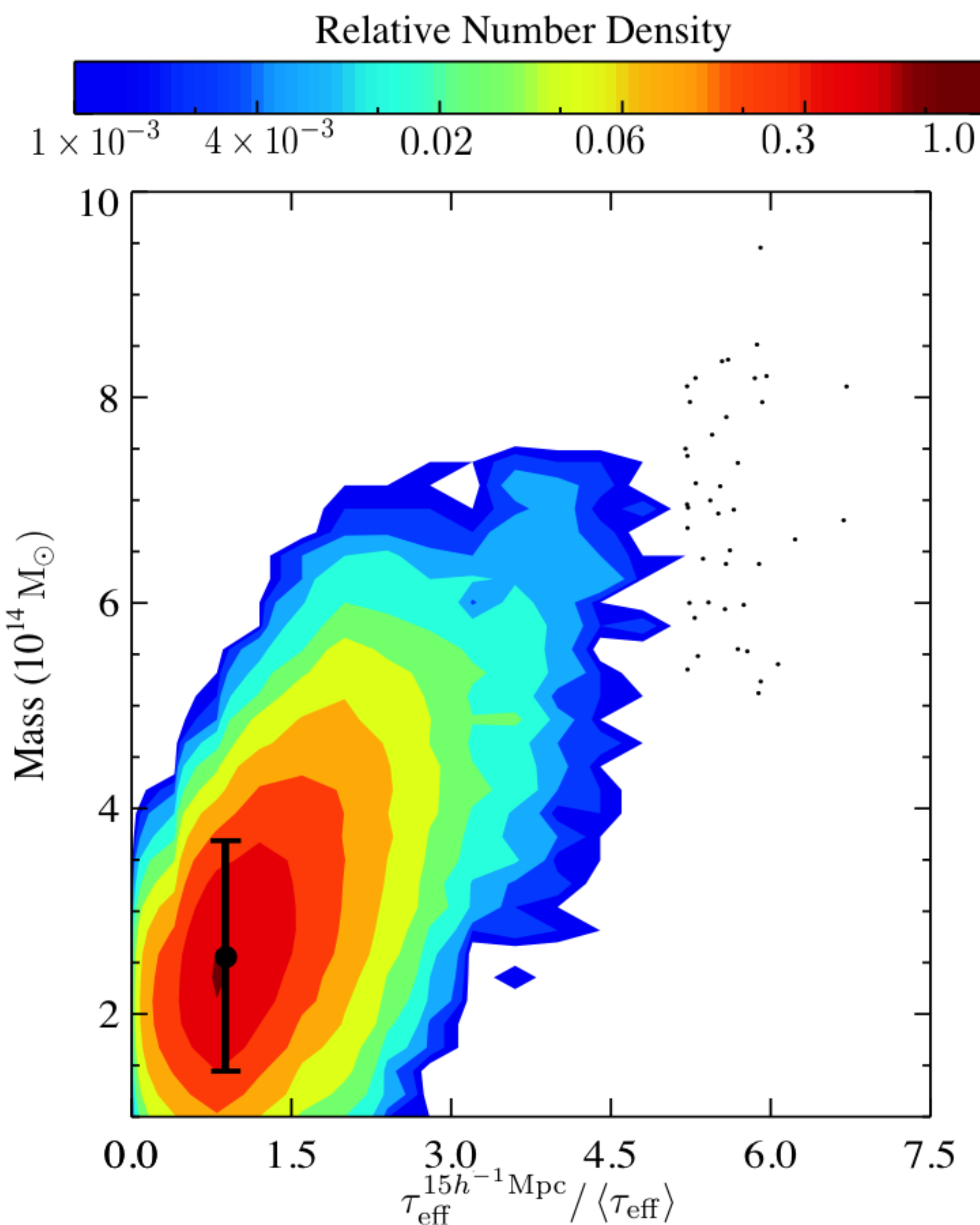}
\caption{The relation between $\tau_{\rm{eff}}$ and mass on $15\ h^{-1}$ Mpc derived using the LyMAS simulation. The filled circle with error bar represent the cosmic median mass within ($15\ h^{-1}$ Mpc)$^3$ and the 1-$\sigma$ error. 
The colors represent the probability density of systems in this diagram.  
We normalised to the systems with the highest number density (brown) to unity. 
Black dots show the mass within $15\ h^{-1}$ Mpc cubes traced by absorption with highest effective optical depth. Most of the black dots reside at the most massive end of the mass distribution. We adopt the mean optical depth of $\mean{\tau_{\rm{eff}}}=0.25$ at $z=2.5$ \cite[e.g.][]{becker13}. }
%This figure shows that most of the black dots locate at the massive-end of the mass distribution. }
\end{figure}

\subsection{The $\delta m$ -- $\delta_F$ correlation on 15 $h^{-1}$ Mpc}
%The correlation between 1-D Ly$\alpha$ transmitted flux and 3-D mass has a broad 
%peak from $10$ to $40$ $h^{-1}$ comoving Mpc (Figure~\ref{fig:corr}). 

To better understand the $\delta m-\delta_F$ correlation, we choose a specific scale, 15 $h^{-1}$ Mpc, to examine the scatter between  $m$ and effective optical depth $\tau_{\rm{eff}}$ ($\tau_{\rm{eff}}= \ln \mean{F}$, where $\mean{F}$ is the average transmitted flux on 15 $h^{-1}$ Mpc). We choose the scale of 15 $h^{-1}$ Mpc, because this scale gives a high correlation between the transmitted transmitted flux and mass overdensity (Fig.~\ref{fig:corr}). Further, the 15 $h^{-1}$ Mpc is the typical extent of the large-scale galaxy overdensities at $z>2$ \cite[e.g.,][]{steidel98, steidel05, ouchi05, matsuda10, chiang13}. We also provide the results on the scales of 10 $h^{-1}$ Mpc and 20 $h^{-1}$ Mpc in the Appendix.  Figure~\ref{fig:scatter} presents a plot of mass vs optical depth on 15 $h^{-1}$ Mpc. 
Different colors represent relative number densities of points in the $\tau_{\rm{eff}}-m$ diagram, normalized to 
systems in regions with the highest density (brown color). The black dots indicate systems with highest effective optical depth.  This figure demonstrates that most of the black points, with high $\tau_{\rm{eff}}$, residing at the most massive end in the mass distribution.  

In this figure, the median mass within ($15\ h^{-1}$ Mpc)$^3$ in the LyMAS simulation is $2.6\times 10^{14}$ M$_\odot$, denoted by the black horizontal line in Figure~\ref{fig:scatter}. %We also use a simple fit to the mass as a function of $\tau_{\rm{eff}}$, and this best-fit (red line in Figure~\ref{fig:scatter}) is described by following equation: 
%the median mass within ($15\ h^{-1}$ Mpc)$^3$ as a function of $\tau_{\rm{eff}}$ can be described by a fit:

%\begin{equation}
%M (15\ h^{-1} \rm{Mpc}) = (4.11\times \tau{_{\rm{eff}}^{15h^{-1} {\rm{Mpc}}}} + 1.58)\ \times 10^{14} M_\odot
%\end{equation}
%where M (15 $h^{-1}$ Mpc) is the mass within ($15h^{-1}$ Mpc)$^3$ and $\tau{_{\rm{eff}}^{15h^{-1}\rm{Mpc}}}$ is the effective optical depth over a scale of $15\ h^{-1}$ Mpc. 
The scatter of the mass-$\tau_{\rm{eff}}$ relation is primarily due to  the 3-D geometrical configurations of large-scale structures, e.g., strong 1-D Ly$\alpha$ forest absorption is biased towards those lines of sight aligned with the major axis of the 3-D structures.

\subsection{Coherently Strong Ly$\alpha$ Absorption Systems}

In the previous subsection, we demonstrated that there exists a correlation between $\delta_F$ $(\tau_{\rm{eff}})$ and mass on large scales of $ 10$ -- $40 \ h^{-1}$ Mpc (\S2.2). 
In this subsection, we focus on the mass distribution traced by extreme systems with largest $\tau_{\rm{eff}}$ over $10 - 20\ h^{-1}$ Mpc (15 -- 30 \AA\  in the spectra at $z=2.5$). We compare the mass overdensities traced by strongest Ly$\alpha$ forest absorption to those traced by other methods (e.g., quasars) in the simulation.

\begin{figure}[tbp]
\epsscale{1.1}
\label{fig:tau_distri}
\plotone{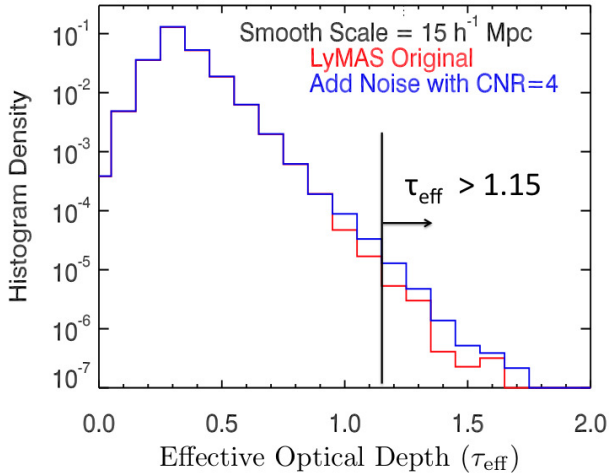}
\caption{The distribution of the effective optical depth ($\tau_{\rm{eff}}$) on the scale of 15 $h^{-1}$ Mpc at $z=2.5$. In MAMMOTH project, we focus on the strongest absorption systems with the highest effective optical depth with 4.5$\times\mean{\tau_{\rm{eff}}}$ ($\tau_{\rm{eff}}>1.15$ at $z=2.5$). In the lognormal distribution of optical depth, these systems have $\tau_{\rm{eff}}$ 4-$\sigma$ higher than the mean optical depth. }
\end{figure}

First, let us define our Ly$\alpha$ absorption sample. We select systems from mock spectra that have intrinsically 
high effective optical depth.  
The red histogram in Figure~\ref{fig:tau_distri} presents the distribution of  $\tau_{\rm{eff}}$ on 15 $h^{-1}$ Mpc scale. 
In the lognormal distribution, the mean optical depth is $\mean{\tau_{\rm{eff}}}= 0.25$, with a standard deviation of 0.20 on the logarithmic scale, which are consistent with observations \cite[e.g.,][]{bolton09,becker13}.   %both in original mock spectra (red) and the noise-added mock spectra (blue). 
We focus on the systems that have $\tau_{\rm{eff}}>1.15$, which resides at the 4-$\sigma$ tail of the lognormal distribution of effective optical depth. 
%and $4.5\times$ the mean optical depth ($\mean{\tau{_\rm{eff}}}$) at this redshift ($z=2.5$). 
This threshold yields 200 systems in the (1 $h^{-1}$ Gpc)$^3$ volume. This number is determined with considerations of observational follow-ups. In the following, we refer to the Ly$\alpha$ absorption systems having $\tau_{\rm{eff}}\ge4\sigma$ higher than $\mean{\tau_{\rm{eff}}}$ 
 as CoSLAs (Coherently Strong Ly$\alpha$ Absorption systems). 
%On 15 $h^{-1}$ Mpc, we also compare the $\tau_{\rm{eff}}$ in the noise-added spectra with continuum-to-noise ratio (CNR) = 4. Most of the absorption systems have intrinsically high optical depth. %In the next section, we will further examine that absorption selected with or without noise trace similar mass distribution. %These systems have coherently strong 
%Ly$\alpha$ absorption due to gas in IGM. 

\begin{figure}[tbp]
\epsscale{1.1}
\label{fig:individual_spectra}
\plotone{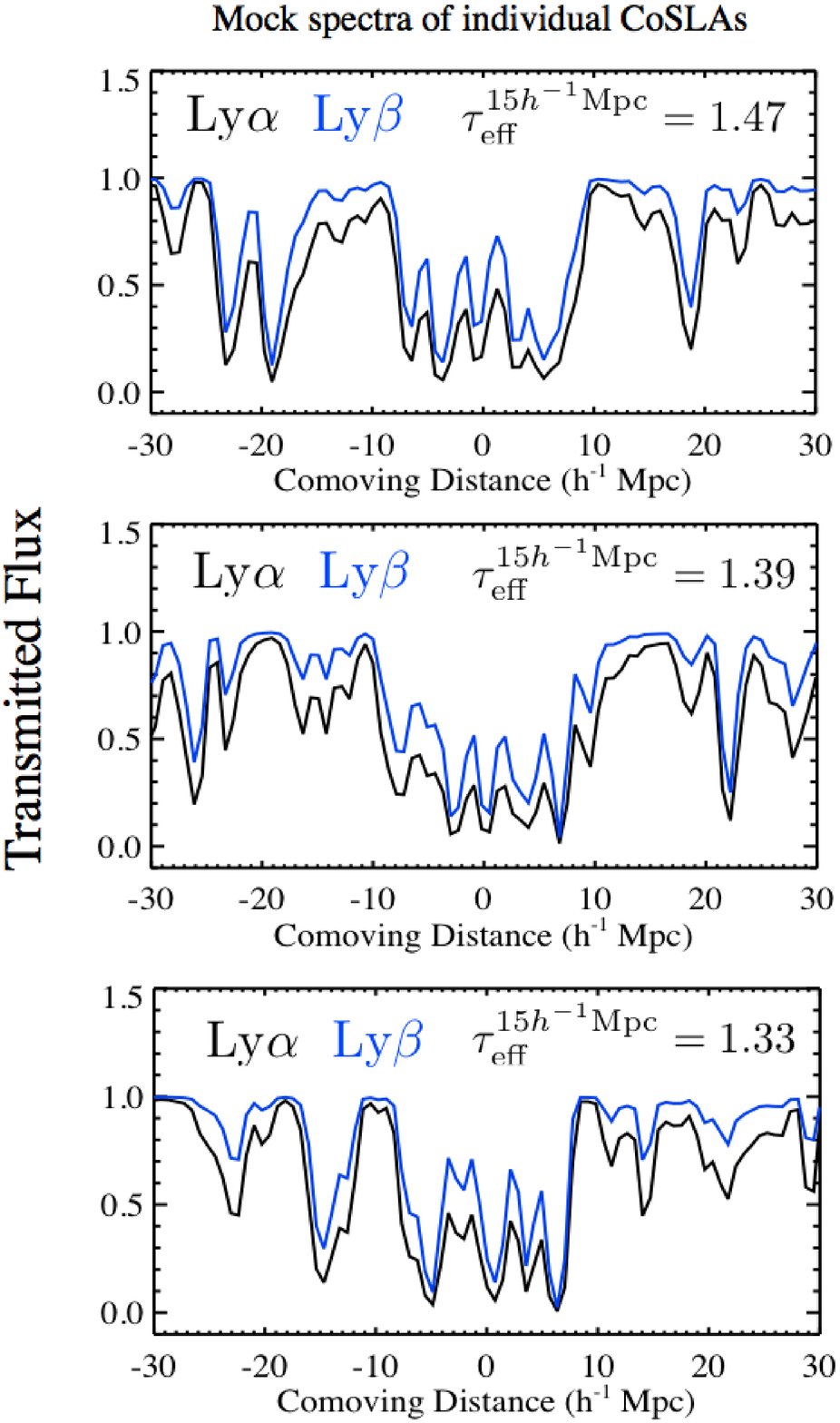}
\caption{Three  examples of the individual mock spectra of CoSLAs selected from the LyMAS simulation at $z=2.5$. The spectra have been convolved to the SDSS-III/BOSS resolution. The black represents Ly$\alpha$ and blue shows the corresponding Ly$\beta$. CoSLAs are  the superposition of Ly$\alpha$ forest lines.}
\end{figure}

Figure~\ref{fig:individual_spectra} presents three examples of CoSLAs in the LyMAS simulation. On 15 $h^{-1}$ Mpc, the CoSLAs have $\tau^{15h^{-1}\rm{Mpc}}_{\rm{eff}}\ge 4.5\times\mean{\tau{_{\rm{eff}}}}$, $4\sigma$ higher than the mean optical depth. The high optical depth is due to the superposition of intergalactic Ly$\alpha$ forest.

Similarly, at the scale of 10 $h^{-1}$ Mpc, CoSLAs have $\tau^{10h^{-1}\rm{Mpc}}_{\rm{eff}}= 1.40$ (see left panel of Appendix Figure~\ref{fig:tau_distri_20}),  beyond 4-$\sigma$ of $\tau_{\rm{eff}}$ distribution. At the scale of 20 $h^{-1}$ Mpc, the $\tau_{\rm{eff}}$ locates at the 4-$\sigma$ tail is  $\tau^{20h^{-1}\rm{Mpc}}_{\rm{eff}}= 1.03$ (see right panel Appendix Figure~\ref{fig:tau_distri_20}).

%%%%%%%%%%HOW TO SAY SNR %%%%%%%%%%%%%%%%%%%%%%%%
%We have also added noise according to continuum-to-noise ratio (CNR) equals 4.  
%%%%%%%%%%%%%%%%%%%%%%%%%%%%%%%%%%%%%%%%%%%%
%SAY MY PAPER ... ANDEAU FONT ...  
%

\figurenum{3}

\figurenum{4}

\subsection{Mass Overdensities traced by CoSLAs}

How effectively do they pinpoint large-scale overdensities compared with other tracers ? In the following, let us examine into these questions in simulations. 

\begin{figure}[tbp]
\epsscale{1.2}
\label{fig:massLyMAS15}
\plotone{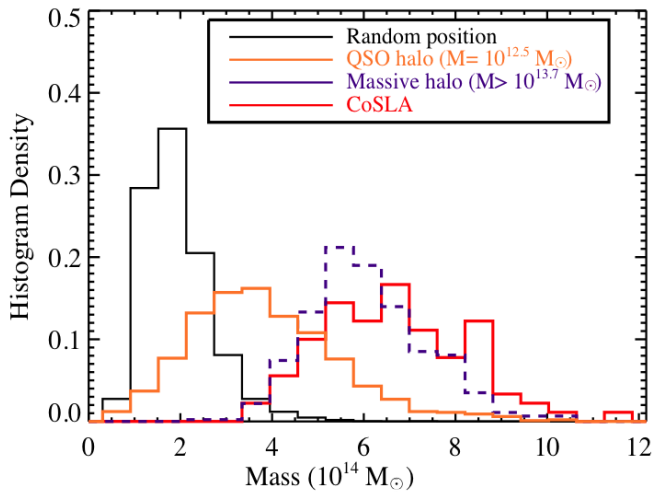}
\caption{The distribution of mass traced by different objects. The x-axis 
is the mass within  (15 $h^{-1}$ Mpc)$^3$ cubes. 
The y-axis is the number of the cubes. 
The black represents mass within (15 $h^{-1}$ Mpc)$^3$ centered at random positions. The yellow histogram shows mass traced by quasar halos. 
Purple dash-dotted histogram represents mass distribution centered at most massive halos in LyMAS simulation 
(M$_{\rm{halo}}> 10^{13.7}\ \rm{M}_\odot$).  
 Red is the mass traced by the CoSLAs on 15 $h^{-1}$ Mpc scale, 
selected from the noise-added mock spectra with a continuum-to-noise ratio of 4. }
\end{figure}

Figure~\ref{fig:massLyMAS15} presents the distribution of mass traced by the CoSLAs on 15 $h^{-1}$ Mpc scale in the
LyMAS simulation.  The x-axis is the mass within (15 $h^{-1}$ Mpc)$^3$ cubes. The y-axis is the number of such cubes. 
The different histograms are explained as follows: 

(1) The first case we consider is the random distribution of the mass within (15 $h^{-1}$ Mpc)$^3$ cubes. 
The black represents the cubes centered at random positions in the simulation box. The mass distribution follows a lognormal distribution, with a median mass of $2.6\times 10^{14}\ M_\odot$, and a standard deviation of $1.2\times10^{14} \ M_\odot$ on the logarithmic scale. 

(2) The second case uses quasars as tracers of the overdensities. 
%{\bf so the question is why use this, rather than just quasar halos?}
The yellow histogram represents the mass within 15 $h^{-1}$ Mpc which is centered on quasar halos with $M_{\rm{halo}}= 2-3\times 10^{12}\ M_\odot$ \cite[e.g.,][]{white12}:  the median mass in 15 $h^{-1}$ Mpc cubes is $3.8\times 10^{14}\ M_\odot$, with a standard deviation of $1.6\times10^{14} \ M_\odot$ on the logarithmic scale. 

(3) The third case uses the most massive halos  in the LyMAS simulation as tracers of the mass overdensity at $15\ h^{-1}$ Mpc. Purple histogram (dotted dash) represents mass within ($15\ h^{-1}$ Mpc)$^3$ cells centered on halos with $M_{\rm{halo}}> 10^{13.7}\ M_\odot$, more than a factor of ten times more massive than typical quasar halos. They are the top 0.01\% most massive halos in the LyMAS simulation, and (1 h$^{-1}$ Gpc)$^3$ volume only contains 256 such halos. Note $10^{13.7}$ $M_\odot$ at $z=2.5$ is about the mass of the progenitors of Coma-like clusters ($10^{15}\ M_\odot$ at $z=0$) \citep{chiang13}. The median mass within (15 $h^{-1}$ Mpc)$^3$ cubes centered on such rarest halos is $6.1\times10^{14}\ M_\odot$, with a standard deviation of $1.0\times10^{14}\ M_\odot$.  

(4) 
The fourth case we consider is the CoSLAs on $15\ h^{-1}$ Mpc. 
The mass distribution traced by CoSLAs selected from the original (noise-free) mock spectra has a median mass of $7.0\times 10^{14}\ M_\odot$, with a standard deviation of $1.6\times 10^{14}\ M_\odot$. If one adds noise according to CNR=4, the mass  traced by CoSLAs selected from the noise-added spectra has a median mass of $6.7\times 10^{14}\ M_\odot$, with a standard deviation of $1.6\times 10^{14}\ M_\odot$ (red histogram). 
Thus, the CoSLAs selected from the noise-added (CNR=4) or noise-free mock spectra trace overdensities with 
similar mass distribution. \footnote{We first set the continuum-to-noise ratio (CNR) = 4, because CNR $\ge 4$ empirically provides a good equivalent width estimate for absorbers (e.g., DLAs) in BOSS dataset (Font et al. 2012, Noterdaeme et al. 2012, Cai et al. 2014). For example, each 15 $h^{-1}$ Mpc contains 20 pixels, with the CNR of 4 on each pixel, the error of mean transmitted flux over $15\ h^{-1}$ Mpc is $<0.05$, making the uncertainty of $\tau_{\rm{eff}}$ 
small. Therefore, we can see noise has a small effect of mass distribution (see last two lines of Table 1). 
Later in this paper, higher CNR $=8$ is also presented according to the needs of target selection efficiency. CoSLAs selected from spectra with {\it CNR $>$ 4} should yield a mass distribution betweeen spectra with CNR=4 and noise-free mock spectra.}  
Similar results have been found at other scales ranging from 10 -- 20 $h^{-1}$ Mpc (see Appendix). 

\begin{figure}[tbp]
\epsscale{1.2}
\label{fig:cumulative}
\plotone{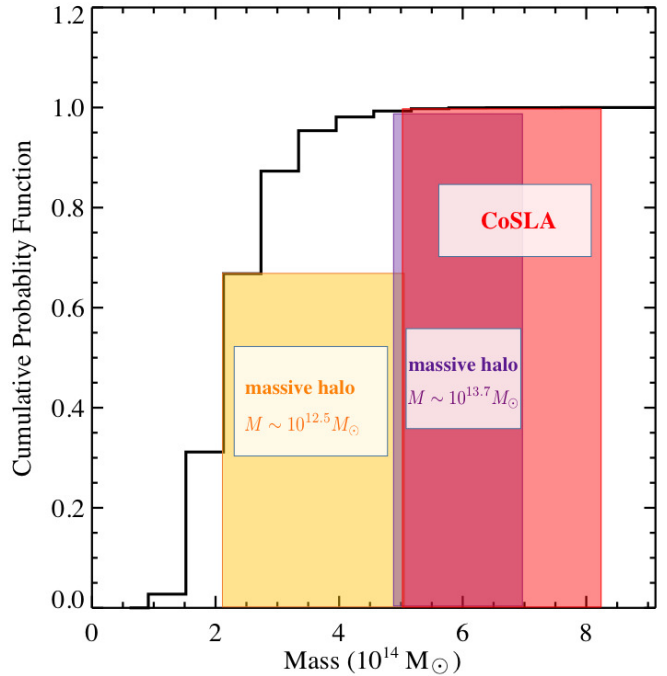}
\caption{The cumulative probability of mass within (15 $h^{-1}$ Mpc)$^3$.   The x-axis 
is the mass within  (15 $h^{-1}$ Mpc)$^3$ cubes. 
The y-axis is the cumulative probability.  Yellow shaded area shows mass distribution ($\pm 1$-$\sigma$) traced by quasar halos. 
Purple shaded area represents mass distribution centered at most massive halos in the LyMAS simulation 
(M$_{\rm{halo}}> 10^{13.7}\ \rm{M}_\odot$).  
 The red shaded area represents the mass traced by the CoSLAs on 15 $h^{-1}$ Mpc scale, selected 
from the noise-added spectra. }
\end{figure}

The figure shows that the CoSLAs effectively trace the most 
 massive overdensities. The overdensities traced by CoSLAs have a comparable mass distribution 
to those centered on the most massive single halos  with $M_{\rm{halo}}>10^{13.7}\ M_\odot$ (purple dot-dashed histogram). More than half of the CoSLAs
represent the top 0.2\%  most massive overdensities ($>$ 3.3-$\sigma$) on 15 $h^{-1}$ Mpc scale. Figure~\ref{fig:cumulative} further illustrates that the systems traced by CoSLAs represent the most massive tail of the mass distribution. 
Table 1 summarizes the distribution of mass within 15 $h^{-1}$ traced by different objects in the LyMAS simulation.

\begin{deluxetable}{lll}  
\tablecolumns{3}
%\longtable
\tablewidth{0pt}
\tablecaption{Mass in 15 $h^{-1}$ Mpc cubes centered on different objects}
\label{table:F1300}	
\tablehead{\colhead{Center} &
	          \colhead{Median mass} &
	          \colhead{$\sigma_{\rm{15 h^{-1}\ Mpc}}$}  \nl
                 \colhead{} & 
                  \colhead{( $10^{14}\ M_\odot$} ) & 
                  \colhead{($10^{14}\ M_\odot$} ) \nl}
\startdata
Random &   2.6  &  1.2 \\
Quasars ($M_{\rm{halo}}= 2$-$3\times10^{12}\ M_\odot$) &3.7  & 1.6 \\
Halos with $M_{\rm{halo}}\ge 5\times10^{13}\ M_\odot$ &  6.1 & 1.0 \\  
CoSLA (no noise)  & 7.0 &  1.6 \\
CoSLA \it{CNR}=4 & 6.7 &  1.6 
\enddata
\tablecomments{Summary of mass within 15 $h^{-1}$ using different tracers in the LyMAS simulation. HI (no noise) represents mass traced by CoSLAs selected from the original mock spectra, no noise being added; CoSLA {\it{CNR}=4} shows the mass traced by strongest absorption selected from noise-added spectra. }
\end{deluxetable}

\figurenum{5}

\figurenum{6}

% We summarize these results in the Appendix. %The left and right panels of Figure~\ref{fig:massLyMAS20} show mass distributions traced by largest Ly$\alpha$ absorption at the scales of 10 $h^{-1}$ Mpc and 20 $h^{-1}$ Mpc in the LyMAS simulation, respectively. On a smoothing length of $20\ h^{-1}$ Mpc, most of the large-scale structures traced by CoSLAs contain mass a factor of $1.6\times$ that of random fields, representing 2.5-$\sigma$ mass overdensities. At $10\ h^{-1}$ Mpc, most of the cubes traced by CoSLAs contain mass a factor of $4\times$ cosmic mean, representing 3.9-$\sigma$ mass overdensities (Figure~\ref{fig:massLyMAS20} in Appendix). 

%Again, we choose systems with lowest transmitted flux defined in the second paragraph of \S3.2 (systems beyond 4.5-$\sigma$ in the $\tau_{\rm{eff}}$ distribution). Similar to Figure~\ref{fig:massLyMAS15}, red represents mass traced by CoSLAs with highest $\tau_{\rm{eff}}$ selected from the original mock spectra (no noise added). Blue shows CoSLAs selected from noise-added mock spectra, with continuum-to-noise ratio (CNR) of 4 per pixel. 

%From Figure~\ref{fig:massLyMAS20}, 

\begin{figure}[tbp]
\epsscale{1.1}
\label{fig:stack_spectra}
\plotone{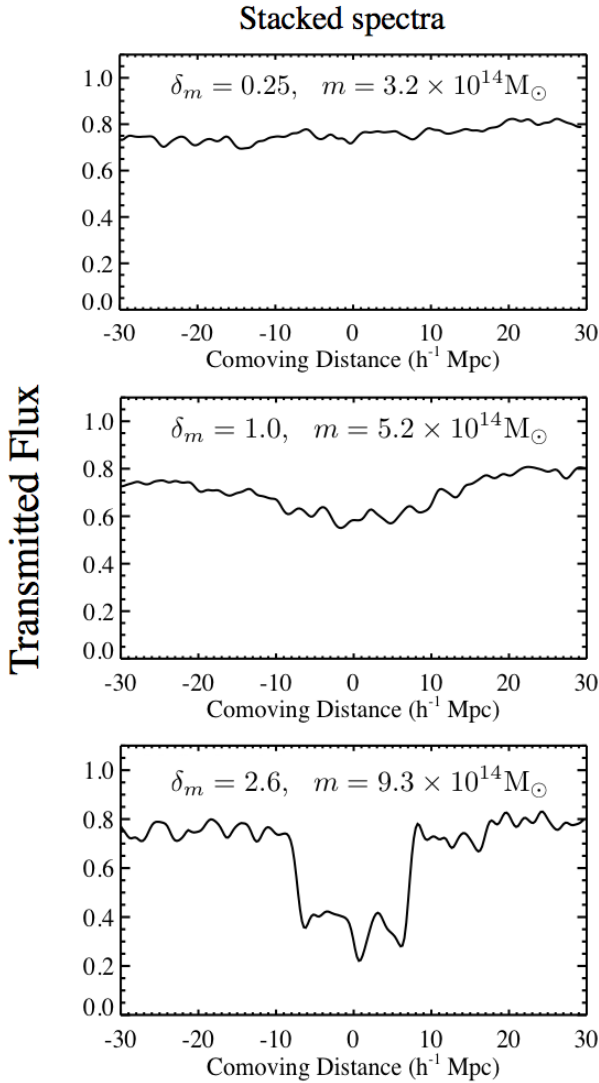}
\caption{The stacked mock Ly$\alpha$ absorption associated with different mass overdensities ($\delta^{15\rm{Mpc}}_m$) at $z=2.5$. 
Each spectrum is the median stacking of 100  simulated spectra associated with similar  $\delta^{15h^{-1}\rm{Mpc}}_m$. 
The effective 
optical depth increases monotonically with mass overdensities.}%The effective 
%optical depth on 15 Mpc ($\tau^{15h^{-1}\rm{Mpc}}_{\rm{eff}}= -\rm{Log}\left<Flux\right>_{\rm{15h^{-1}Mpc}}$) increases with overdensity. }
\end{figure}

\begin{figure}[tbp]
\epsscale{1.2}
\label{fig:example}
\plotone{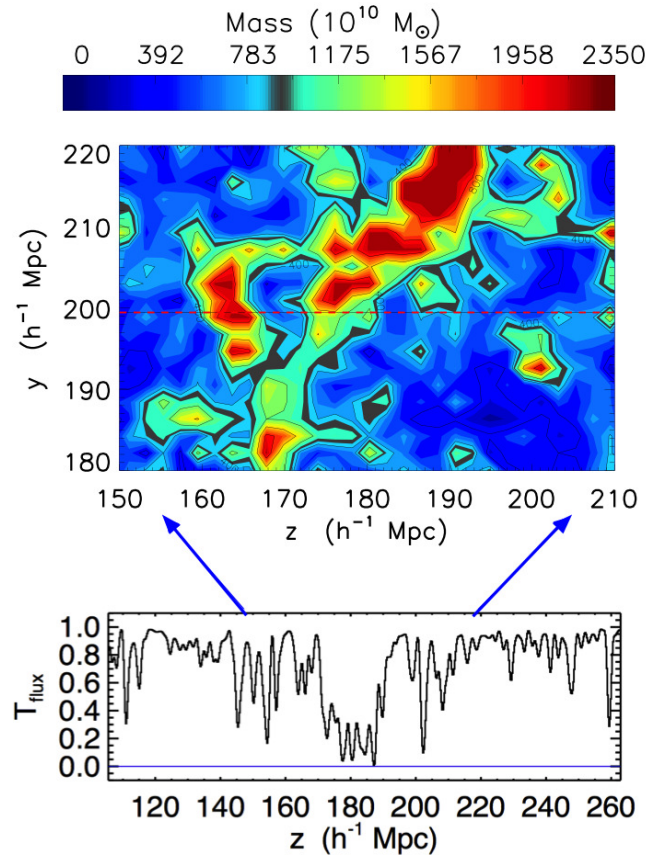}
\caption{A example of CoSLA in the LyMAS simulation. The upper panel presents the (15 $h^{-1}$ Mpc)$^3$ cube 
centered on the Ly$\alpha$ absorption, where the mass is about $4\times$  
the cosmic mean, representing a 4.5-$\sigma$ mass overdensity on $15\ h^{-1}$ Mpc scale. The lower panel presents the simulated 
CoSLA. }
\end{figure}

 Figure~\ref{fig:stack_spectra} further demonstrates the stacked mock Ly$\alpha$ absorption spectra that associated with different mass overdensities ($\delta^{15\rm{Mpc}}_m$) at $z=2.5$. 
Each spectrum is the median stack of 100 individual simulated spectrum associated with similar  $\delta^{15\rm{Mpc}}_m$. The effective 
optical depth on 15 $h^{-1}$ Mpc ($\tau^{15h^{-1}\rm{Mpc}}_{\rm{eff}}= -{\rm{log}\left<Flux\right>}_{\rm{15h^{-1}Mpc}}$) increases monotonically with mass overdensities. Figure~\ref{fig:example} presents an example of an extreme mass overdensity traced by CoSLAs in the LyMAS simulation. This CoSLA effectively trace an extremely massive large-scale structure extends over $\gtrsim$ 40 $h^{-1}$ Mpc.

\figurenum{7}

\figurenum{8}

\section{Mock Spectra $-$ the inclusion of high column density absorbers}

In \S2, we demonstrated that mass overdensities can be effectively traced using the coherently strong Ly$\alpha$ absorption systems (CoSLAs) with  $\tau_{\rm{eff}}\ge4.5\times \mean{\tau_{\rm{eff}}}$ on $15$ $h^{-1}$ Mpc. In real spectra, however, there will be contaminants like DLAs, sub-DLAs, clustered LLSs. In this section, we explore how we can effectively recover genuine CoSLAs from realistic spectra. 
We use the mock spectra to test the selection techniques, and apply these techniques to SDSS-III/BOSS data.

The original mock spectra generated by LyMAS simulations well predict the intergalactic low-density gas, which produces the optically-thin Ly$\alpha$ forest \citep{peirani14}. However, this large-scale cosmological simulations lack sufficient resolution and gas self-shielding schemes to reproduce the number of high column density absorption systems (HCDs) with $N_{\rm{HI}}>10^{17}$ cm$^{-2}$ in fidelity \citep{miralda96,cen03}. These HCDs are self-shielded $-$ the exterior absorbs the ionizing radiation and the interior remain mostly neutral \cite[e.g.,][]{mcdonald05}.
They are  clumps of dense gas in galactic or circumgalactic environments \cite[e.g.,][]{rubin14}, and are observed as damped Ly$\alpha$ systems (DLAs,  $N_{\rm{HI}}>10^{20.3}$ cm$^{-2}$), and Lyman limit systems (LLSs, $N_{\rm{HI}}>10^{17.2}$ cm$^{-2}$). It is important to take into account the realistic impact of HCDs in the mock spectra for our target selection \citep{mcdonald05, font12}.

\begin{figure}[tbp]
\epsscale{1.1}
\label{fig:LyMASprofile}
\plotone{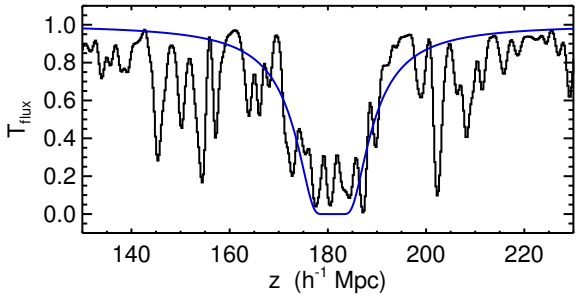}
\caption{A comparison between the simulated absorption due to a CoSLA and a DLA. No noise is added. 
The black spectrum indicates an extreme IGM absorption systems in LyMAS simulation. Blue shows a DLA system with $N_{\rm{HI}}= 10^{20.35}$ cm$^{-2}$. At moderate signal-to-noise ratio, DLAs or overlapping 
sub-DLAs can mimic the large-scale IGM Ly$\alpha$ absorption.}% a. }
\end{figure}

The CoSLAs, which trace massive overdensities, have profiles that resemble of DLAs, especially at moderate S/N. 
This effect is illustrated in  Figure~\ref{fig:LyMASprofile}, 
which compares a CoSLA in the LyMAS simulation to a  DLA with 
$N_{\rm{HI}}\sim 10^{20.3}$ cm$^{-2}$.  
 A significant fraction of DLAs resides in low mass or modest mass  
halos with $M<10^{12}\ M_\odot$ \cite[e.g.,][]{moller13}, and most of these halos do not trace extremely massive large-scale overdensities. Thus, 
 DLAs and strong sub-DLAs (e.g., $N_{\rm{HI}}\gtrsim10^{19.5}$ cm$^{-2}$) are contaminants. 
Also, we need to take into account the clustered LLSs that have the similar equivalent widths to that of CoSLAs, which is another 
population of contaminants.

In the following, 
 we introduce our detailed procedures in generating realistic mock spectra, including inserting HCDs, adding continuum uncertainties, 
convolving spectra to the resolution of SDSS-III/BOSS, and adding noise. Similar procedures were adapted in previous works \citep{font-ribera12, slosar13, delubac15}. 

\figurenum{9}

% we will develop the 
%algorithm to select the large-scale IGM Ly$\alpha$ absorption and rule out HCD contaminants. 
%%differentiate the largest IGM absorption systems from the DLA contaminants. 

\subsection{Procedures of inserting HCDs to the mock spectra}

Below, we give detailed procedures to insert the high column density discrete absorbers (HCDs) into the 
mock spectra. 
%Following the procedures presented in \S4.1, 
%\subsection {Insert LLSs, DLAs and add noise to the mock spectra}
The LyMAS simulation has 65,536 skewers, with the box size of  $1\ h^{-1}$ Gpc on a side. The simulation has a Ly$\alpha$ survey volume about the same as that of BOSS DR9, which contains about 30\% of the total volume of BOSS DR12. 

Assuming the distribution of DLAs ($N_{\rm{HI}} > 10^{20.3}$ cm$^{-2}$) that follows a Gamma function \cite[e.g.,][]{prochaska05}
\begin{equation}
f_{\rm{HI}}^{\rm{DLA}} (N, X)= k_g \Big(\frac{N}{N_g} \Big)^{\alpha_g} \rm{exp}(-N/N_g).
\end{equation}
 Following \citep{noterdaeme09}, at $z\sim 2.5$, we assume  log $k_g$= -22.75, log$N_{g}$=21.26, 
$\alpha_g$= -1.27, where, $dX= \frac{H_0}{H(z)} (1+z)^2 dz.$

The total distance covered by our simulation is $ 9.4\times 10^{7} $ Mpc. At $z=2.5$, this distance corresponds to $X= 9.4\times10^7 / (23.2)\times 0.066= 267,413$. 
The total number of DLAs is: 
\begin{equation}
\begin{split}
\int\int f_{\rm{HI}}^{\rm{DLA}}(N,X) dN dX & = l_{\rm{DLA}, z=2.5}\  X=  \\
           &=0.055\times216,362 = 11,899
\end{split}
\end{equation}
This is the number of DLAs we placed into the simulation, distributed randomly in the spatial direction along the sight lines. 

In reality, DLAs are clustered, but DLA host halos are still under debated.  A number of cross-correlation studies measure the average mass 
of DLA host halos to be widely distributed around 10$^9$ -- 10$^{12}$ M$_\odot$ \cite[e.g.,][]{cooke06, font12,moller13, bird14}.
%and thus DLA clustering is poorly constrained observationally. 
We argue that neglecting DLA clustering has negligible effects on the selection of CoSLAs: 

(1) If the DLA clustering yields an overlapping DLA, the absorption from overlapping DLAs have wider width compared with CoSLAs and we can easily identify it as such (see Figure~\ref{fig:LyMASprofile}). %If two DLAs are not overlapping, the clustering does not affect our selection. %because we just treat the  independent 

%(2) The transverse clustering of DLAs have small effect. The separation of sight lines in our simulation is 6 Mpc. The size of DLA is much smaller than Mpc (e.g., Font-Ribera et al. 2012). It is almost a random distribution of DLA in the transverse direction with the transverse separtions larger than Mpc. 

(2) The probability of having a DLA in the same region as a CoSLA is small.  %We calculate that the probility is very low ($\sim 2\%$) for a DLA absorption superimposed on a CoSLA absorption. 
% Font-Ribera et al. (2012) measure that the typical DLA has proper sizes of 20 kpc in host halos with masses of 
%$\sim$ 10^{12} M$_\odot$. 
From the simulation, we have checked that the 96\% of CoSLAs do not associate with particles ($M>7\times10^{10}\ M_\odot$) within the projected distance of 100 kpc. Therefore, the majority of CoSLAs are not associated to DLAs with host halo with $M>7\times 10^{10}\ M_\odot$. 
However, DLAs with halo mass of $M< 7\times10^{10}\ M_\odot$ cannot be resolved from our simulation. Nevertheless, these low-mass DLAs are expected to have small impact parameters 
$\lesssim10$ kpc \citep{font12}. Even if one assumes the covering fraction of DLA clouds is 100\% within 10 kpc around the halo center, we expect $\sim 0.2\%$ of DLAs to be associated with CoSLAs \footnote{The DLA halo mass is estimated to be 10$^9-10^{12}\ M_\odot$ \cite[e.g.,][]{cooke06}. We conservatively assume that all the DLAs have halo masses between $10^9- 10^{12}\ M_\odot$. Further, we conservatively assume the covering fraction of DLA clouds is 100\% within the impact parameter of 10 kpc. Under these two assumptions, according to \citet{tinker10} the number density of DLA halos is 0.8 $h^3$ Mpc$^{-3}$. Thus, the probability of having a DLA in the cylinder with $10$ kpc $\times$ 10 kpc $\times$ 15 $h^{-1}$ Mpc is 0.002 (0.2\%).}. Therefore, the DLA clustering does not affect our selection of CoSLAs. 

 %from Figure~\ref{fig:LyMASprofile}, the absorption from overlapping DLAs   

Similarly, following the same procedures, we added LLSs to the simulation. From \citet{prochaska05}, assuming that $f_{\rm{HI}}(N,X)$ is a single power law over the $10^{17.2}$ cm$^{-2} < N_{\rm{HI}}< 10^{20.3}$ cm$^{-2}$ interval:  
\begin{equation}
f_{\rm{HI}}^{\rm{LLS}}(N,X)= k_{\rm{LLS}} (\frac{N}{10^{20.3} \rm{cm}^{-2}})^{\alpha_{\rm{LLS}}}
\end{equation}
At $z\sim 2.5$, $\alpha_{\rm{LLS}}=-0.9$, $k_{\rm{LLS}}=10^{-21.43}$ from SDSS DR3 and DR4 samples \citep{prochaska05}, and 
%\begin{equation}
%\int_{10^{17.2}}^{10^{20.3}} f_{\rm{HI}}^{\rm{LLS}} dN dX %= l_{\rm{LLS}} dX- l_{\rm{DLA}} dX
%\end{equation}
%In Peroux et al. (2003). $l_{\rm{LLS}}(z)= 0.07\times(1+z)^{2.45}$, $l_{DLA}=0.053$. 
%Then, 
\begin{equation}
\int_{10^{17.2}}^{10^{20.3}} f_{\rm{HI}}^{\rm{LLS}}(N,X) dN dX= 0.5 \int f_{\rm{HI}}^{\rm{LLS}} (X)dX= 112,007. 
\end{equation}

This number yields $\approx$ 1.7 LLSs per 1 $h^{-1}$ Gpc path length, which is consistent with the measurements of the mean free path of Lyman limit photons \citep{worseck14,prochaska10}. %This LLS number is also consistent with the high redshift LLSs measurements Prochaska et al. (2010). 
We have inserted 112,007 LLSs and 11,899 DLAs to the mock spectra.

In this paper, we randomly inserted LLSs in the mock spectra. But LLSs are not randomly distributed. They prefer to reside in the overdense region. Thus, a fraction of CoSLAs should contain LLSs. This affects the selection of CoSLAs (see target selection in \S4). 
 If we assume a CoSLA  traces an LLS overdenisty of $10$ in a 15 $h^{-1}$ Mpc, and assume in random fields, one can find 1 LLS per 600 $h^{-1}$ path length at $z\sim2.5$ \cite[e.g.][]{prochaska10}, 
we expect 25\% of CoSLAs contain at least 1 LLS within $\pm 7.5\ h^{-1}$ Mpc from the CoSLA center. About 75\% of CoSLAs do not associate with LLSs, and the absorption is only due to superposition of intergalactic Ly$\alpha$ forest. 

{\subsubsection{Effects of LLS clustering}}

The details of  LLS clustering along the quasar sight lines are poorly constrained. For the mock spectrum, we treat LLSs to be randomly 
distributed. Nevertheless, we should note that the LLS clustering could yield the overlapping LLSs that have similar equivalent width and profiles with CoSLAs. Using 50 LLSs from the HD-LLS survey \citep{prochaska15, fumagalli15}, we roughly estimated how clustered LLSs affect our target selection efficiency. The median $\tau^{15h^{-1}\rm{Mpc}}_{\rm{eff}}=1.9\times\mean{\tau_{\rm{eff}}}$, with a standard deviation of 0.7.  Assuming the $\tau^{15h^{-1}\rm{Mpc}}_{\rm{eff}}$ has a lognormal distribution, we estimate that the chance of clustered LLSs that have optical depth higher than CoSLA selection criterion is smaller than 0.3\%. In \S4, we will discuss how this fraction affects our target selection efficiency.

\subsection{Convolve the mock spectra and add noise}

After inserting DLAs and LLSs, we convolved the spectra to BOSS resolution using the Gaussian kernel, with the FWHM equal to the actual dispersion of BOSS spectrum at $\lambda\sim 4250$ \AA\ ($z=2.5$). We then add the noise to the spectra.  
%accroding to  
%Since the continua have been normalized to unity, we put the Gaussian random 
%noise on pixels with standard deviation equal to the noise-to-continuum ratio. 
We produced two sets of mock spectra, one with a continuum-to-noise ratio (CNR) of 4, the other a higher S/N dataset with CNR = 8 for comparison and following discussions. 
%After inserting DLAs, convolving the spectra and adding noise, the realistic mock spectra. 

\subsection{Uncertainties of Continuum-fitting}

The uncertainties of continuum fitting need to be included to the mock spectra, because the observed optical depth of Ly$\alpha$ forest is calculated based on the continuum fitting. In practice, we use 
the mean-flux-regulated principal component analysis (MF-PCA)  to fit the BOSS quasar continua \citep{lee13}. 
Following the discussion in \citet{lee12}, the continuum residual $\delta C(\lambda)$ is defined as $\frac{C_{\rm{fit}}(\lambda_{\rm{rest}})}{C_{\rm{true}}(\lambda_{\rm{rest})}}$. The median r.m.s. continuum fitting error is 4.5\%, for spectra with $6<S/N<10$ at $z=2.5$; a $5.5\%$ for spectra with $4<S/N<6$.  
Therefore, we put a 4.5\% uncertainty around unity for mock spectra with CNR = 8, and a 5.5\% uncertainty for mock spectra with CNR=4, to simulate the fitting errors of continua. 

Also, the presence of CoSLAs biases the quasar continuum to lower amplitude, because the MF-PCA technique fits the amplitude based on the mean optical depth of Ly$\alpha$ forest \citep{lee12}.  Figure~\ref{fig:delta_tau} presents the ($\Delta \tau_{\rm{eff}}$) of 2,000 simulated systems, where the ($\Delta \tau_{\rm{eff}}$) is defined as the optical depth difference using the continua with and without the CoSLAs being masked. 
Statistically, the CoSLAs reduce the amplitude of continua by an average level of $\approx 2.5\%$. In our target selection from a large dataset, we do not know in advance the exact positions of CoSLAs and so we need to take into account the continuum bias introduced by CoSLAs.

\figurenum{10}
\begin{figure}[tbp]
\epsscale{1.1}
\label{fig:delta_tau}
\plotone{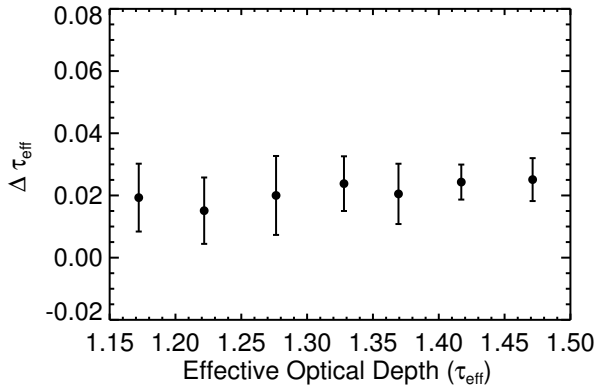}
\caption{The $\tau_{\rm{eff}}$ differences of CoSLAs calculated using two sets of continua fitting. In the first set of continua,  the optical depth of CoSLAs are calculated with CoSLA being masked in the continuum fitting process. In the second set, $\tau_{\rm{eff}}$ are calculated without the CoSLAs being masked in the continuum fitting. The presence of CoSLAs makes the MF-PCA fitted continua bias toward to a lower level by $\approx 2.5\%$. The results are based on the calculation of 2,000 mock spectra.  }% a. }
\end{figure}

After inserting HCD, adding noise, and including the uncertainty of continua fitting, we compare the probability distribution function (PDF) of $\tau^{15h^{-1}\rm{Mpc}}_{\rm{eff}}$ of the LyMAS simulation and the BOSS data. We bootstrap 20,000 systems in both LyMAS simulation 
and BOSS data. The systems in both LyMAS simulation and BOSS data have a median CNR of $8$; and the systems in BOSS data have redshifts of $z=2.5\pm0.2$ to match the redshift of LyMAS simulation. The two PDFs are generally consistent with each other.

\figurenum{11}
\begin{figure}[tbp]
\epsscale{1}
\label{fig:delta_tau}
\plotone{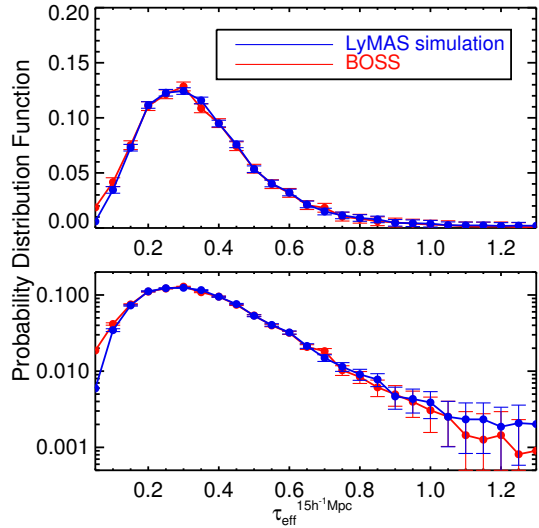}
\caption{The probability distribution function (PDF) of $\tau_{\rm{eff}}$ on 15 $h^{-1}$ in BOSS data (red) and LyMAS simulation (blue), after adding noise, inserting HCDs and including the continnum fitting uncertainties in the LyMAS simulation. We choose 20,000 systems in both LyMAS simulation and BOSS simulation. The noise is included with bootstrap resamping on the data. The upper panel presents the PDF in a linear scale on the y-axis, and the lower panel presents the PDF in a logarithmic scale.  }% a. }
\end{figure}

\section{Selection of CoSLAs}

Using these realistic mock spectra, we now present the detailed technique to select CoSLAs on 15 $h^{-1}$ Mpc. 
The CoSLAs have high effective optical depth ($\tau_{\rm{eff}}$) due to the intergalactic HI overdensities. We need to find 
algorithms to differentiate  
them with the contaminants due to high column density absorbers (HCDs). 
%Most of these absorption systems are expected to trace massive large-scale structures.  
The algorithms developed from these realistic mock spectra can be directly applied to SDSS-III/BOSS data to select CoSLAs (\S6).

\subsection{Number of Targets and Contaminants}

%The procedure is the following: 
First, we define the cutoff for effective optical depth ($\tau_{\rm{eff}}$) of the CoSLAs. After this initial restriction, 
the number of contaminant HCDs is significantly higher than CoSLAs. 
%We calculate the number of targets and the number of contaminants. 
The procedures are as follows: 

%(1) Cut on $\tau_{\rm{eff}}$: the $\tau_{\rm{eff}}$ cutoff is defined for the original mock spectra (no noise being added and no LLSs, DLAs being inserted). On 15 $h^{-1}$ Mpc scale, we define CoSLAs having $\tau_{\rm{eff}}$ on 15 $h^{-1}$ Mpc to be 4.5 $\times\mean{\tau{eff}}$. According to the distribution of the optical depth, these systems have $\tau_{\rm{eff}}$ 4-$\sigma$ beyond the mean optical depth (see \S 2). 

(1) Count the number of targets: smooth the noise-added, continuum-uncertainty included, but no LLSs-, DLAs-inserted mock spectra, and select absorption systems with $\tau^{15h^{-1}\rm{Mpc}}_{\rm{eff}}$ greater than the CoSLA threshold of $4.5\times\mean{\tau_{\rm{eff}}}$. In  LyMAS simulation with (1 $h^{-1}$ Gpc)$^3$, a sample of 303 systems is selected using mock spectra with CNR $=4$. 
For mock spectra with CNR $=8$, 289 systems are selected. %The distribution of mass traced by these systems are summarized 
%in Figure~\ref{fig:massLyMAS15}.  
From Figure~\ref{fig:massLyMAS15}, most of the systems selected from noise-added spectra (CNR=4) trace large-scale structures with mass 
$>$ 2.6$\times$ that in random fields, representing $>$ 3.3-$\sigma$ mass overdensities. This distribution is similar to the CoSLAs selected from original mock 
spectra, without noise being added (Table 1).  We therefore define our targets as CoSLAs selected from the noise-added mock spectra, but no-HCD inserted. %with $\tau_{\rm{eff}}>1.15$ 

%These systems can be regarded as our target. 

(2) Count the number of contaminants: smooth the noise-added, LLSs- and DLAs-inserted spectra. We select the absorption systems with the transmitted flux ($F_t$) we determined in the step (1).  A large number of absorption systems are selected; the vast majority of these systems are due to high column density absorbers (DLAs, sub-DLAs). For spectra with CNR$=4$, 13,635 systems satisfy the $\tau_{\rm{eff}}$ cut proposed in step (1), where 310 of them are CoSLAs. For spectra with CNR$=8$, 12,209 systems are selected (Table 2), where 279 of them are CoSLAs. 
 Because these high column density systems (HCDs) are most likely produced due to individual galaxies \citep{moller13} rather than large-scale IGM overdensities, they are treated as contaminants. 

After the cut in optical depth, the number of contaminants is two orders of magnitude higher than that of targets. We summarize the above results in Table 2. The following of this section describes our attempts at removing the HCD contaminants.

\subsection{Selection Criteria of the Strongest Ly$\alpha$ absorption due to IGM Overdensities}

%From Figure~\ref{fig:LyMASprofile}, at high S/N, we are still able to differentiate the differences between the 
%absorption profiles due to discrete DLAs and IGM overdensities. 

We now describe the algorithm to remove the HCD contaminants and select CoSLAs.

From Figure~\ref{fig:LyMASprofile}, at high S/N, one can identify DLAs by either using the presence of damping wings (Voigt profile) or low-ionization metal lines. However, this is not always possible using data with modest SNR. Our goal is to find selection criteria to eliminate DLAs and sub-DLAs without merely relying on fitting the Voigt profile. 
% PAY ATTENTION !!!!!!!!!!!!!!!!!
%(b) set a limit of metal lines to differentiate DLAs and IGM absorption; (c) further remove high column density systems (HCDs) by either using Ly$\beta$ or the presence of multiple absorbers. 

\begin{deluxetable*}{cccc}  
\tablecolumns{4}
%\longtable
\tablewidth{0pt}
\tablecaption{Strongest 1-D Ly$\alpha$ absorption systems in mock spectra on 15 $h^{-1}$ Mpc}
\label{table:F1300}	
\tablehead{%\colhead{$z_{\rm{abs}}$} &
	          %\colhead{$\mean{\tau_{\rm{eff}}}$} &
	          \colhead{CNR} & 
	          %\colhead{Scale ($h^{-1}$ Mpc) } & 
                  %\colhead{T$_{\rm{flux}}$ } & 
                  \colhead{$\tau_{\rm{eff}}$ } & 
	          \colhead{CoSLAs (noise-added)} &
                  \colhead{Absorption Systems (HCD-inserted)}\nl }
\startdata
  4 &  1.15 -- 1.56 &  303 & 13,635  \\
   8 & 1.15 -- 1.56 &  289 & 12,209 
\enddata
\tablecomments{The last two column data presents the number of the absorption systems selected in noise-added mock spectra and high column density (HCD)-inserted mock spectra. 
For data with CNR $= 4$ and CNR$=8$, a significant number of contaminants are included. In this section, we focus on the CoSLA selection technique from a large number of contaminants.}
\end{deluxetable*}

%{\bf needs to remind reader about 15 Mpc smoothing. }

We explored the following criteria that can effectively 
exclude a significant number of HCD contaminants and efficiently select CoSLAs on 15 $h^{-1}$ Mpc: 

{\bf \it
(a) $w_{0.8}< 70$ \rm{\AA}, where $w_{0.8}$ is denoted as the width at flux/continuum$=0.8$. 
}

{\bf \it
(b) The mean flux of absorption trough ($F_{\rm{trough}})> 0.15$. 
}

{\bf \it
(c) Non-detection of low-ionization metal lines associated with the Ly$\alpha$ absorption systems. 
}

{\bf \it
(d1) For absorbers with $z>2.65$: the Ly$\beta$ transition contains a series of absorbers and the equivalent width (EW) ratio of Ly$\beta$ to Ly$\alpha$ on 15 $h^{-1}$ Mpc is greater than 0.6.}  (at this redshift range,  Ly$\beta$ is covered in 
the optical spectroscopy).

{\bf \it
(d2) For absorbers with $z<2.65$:  the presence of a group of absorption systems at $z<2.65$} (at $z<2.4$, the average BOSS quasar density reaches $\ge$ 15 per deg$^2$). 
We define absorption group as $\ge 4$ absorption systems in a volume of ($25\ h^{-1}$ Mpc)$^3$, with each absorption has a $\tau_{\rm{eff}}$ on 15 $h^{-1}$ Mpc $\ge 3.5\times $ mean optical depth $\mean{\tau(z)}$.

%{\bf how about $2.35 < z < 2.8$, explain.}

In this paper, we mainly consider absorption systems at $z>2.65$. In the next paper of this series, we focus on absorption group and the spectroscopic confirmation of an extremely overdense field. 

In the following, we introduce details for each criterion:

\figurenum{12}
\begin{figure}[tbp]
\epsscale{1.1}
\label{fig:lymasselection}
\plotone{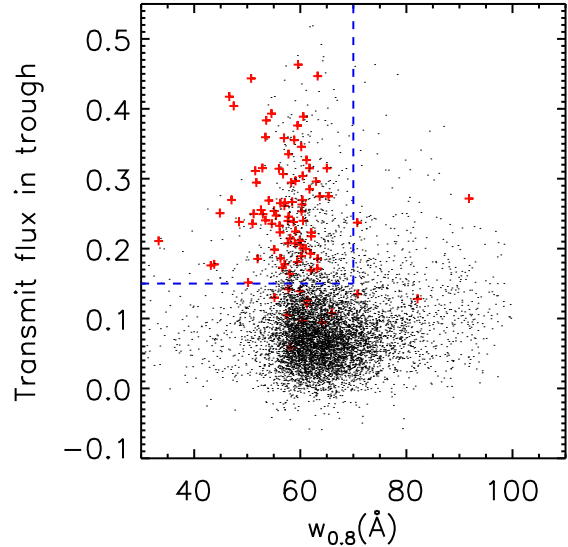}
\caption{$F_{\rm{trough}}$--${\rm{width}}$ diagram for absorption systems with CNR $=8$ per pixel. The red points represent CoSLAs. Black points are  high column density absorbers (HCDs) which have the same optical depth with CoSLAs. Blue dashed box  represents the width selection criterion. This selection criterion selects 70.7\% CoSLAs, and $79.4\%$ of contaminants associated with HCDs are excluded by this selection criterion.  }
\end{figure}

\figurenum{13}
\begin{figure}[tbp]
\epsscale{1.0}
\label{fig:cii}
\plotone{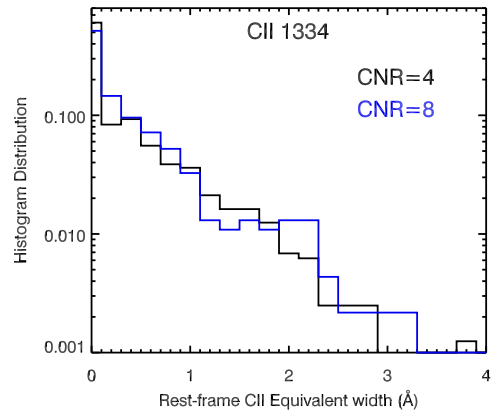}
\caption{The distribution of C II $\lambda$1334 rest-frame equivalent width associated with high column density absorbers (HCDs) with $N_{\rm{HI}}\ge10^{20.0}$ cm$^{-2}$ in SDSS-III/BOSS data \citep{noterdaeme12}. The blue histogram represents HCDs with CNR $>4$, and $\approx 64\%$ HCDs have detectable C II $\lambda1334$ line. Black histogram indicates HCDs with CNR $>8$, and $\approx 75\%$ HCDs have detectable C II $\lambda$1334 absorption line. }
\end{figure}

\figurenum{14}
\begin{figure}[tbp]
\epsscale{1.0}
\label{fig:SiII}
\plotone{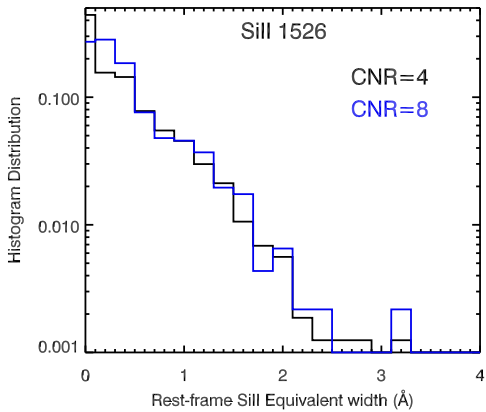}
\caption{The distribution of Si II $\lambda$1526 rest-frame equivalent width associated with high column density absorbers (HCDs) with $N_{\rm{HI}}\ge10^{20.0}$ cm$^{-2}$ in SDSS-III/BOSS data \citep{noterdaeme12}. The blue histogram represents the Si II absorbers associated with HCDs with CNR $>4$, $\sim 74\%$ HCDs have detectable Si II $\lambda1526$ absorption lines.
Black indicates DLAs with CNR $>8$, and $\sim 82\%$ DLAs have detectable Si II $\lambda$1526 absorption line.  }
\end{figure}

{\it \large Criterion(a):} 

Criterion (a) defines the width of the absorption systems. We use (a) to eliminate DLAs with column density $N_{\rm{HI}}>10^{20.6}$ cm$^{-2}$. As shown in the simulation (Figure~\ref{fig:individual_spectra}), even the most extreme IGM Ly$\alpha$ absorbers selected  
from a 1 ($h^{-1}$ Gpc)$^3$ volume have $w_{0.8} < 70$ \AA. Wider absorbers are due to DLAs.

In practice, we define the absorption center as the wavelength giving the smallest transmitted flux within $\pm 10$ \AA\ ($\pm 7.5$ $h^{-1}$ Mpc at $z=2.5$). For a given absorption system, the width at {\it flux/continuum}=0.8 ($w_{0.8}$) is calculated as follows: in the left wing, we calculate the median wavelength at $\frac{\rm{flux}}{\rm{continuum}}=0.80 \pm 0.01$, and denote this wavelength as $w_{l,0.8}$. In the right wing, we also calculate the median wavelength that have flux to continuum $=0.80 \pm 0.01$, and denote this wavelength as $w_{r,0.8}$. 
The width of absorption at $\it{{flux}}/{{continuum}}$=0.8 ($w_{0.8}$) is then defined as $w_{r,0.8}-w_{l,0.8}$, and we require that $w_{0.8} < 60 $ \AA. 

{\it \large Criterion (b): }

Criterion (b) is the constraint for the mean flux in the dark trough. 
For DLAs, the expected level in this trough is zero; however, for CoSLAs, the absorption trough has transmitted flux larger than 0.1. 

In practice, we define the flux of absorption trough ($F_{\rm{trough}})$ as the mean transmitted flux within the region of $\pm 5$ \AA\  from the absorption center. We only select the absorption systems with  $F_{\rm{trough}} > 0.15$. %Note we define the center of specific absorption system as the position where the transmitted flux is the 
%lowest within the $\pm 7.5$ $h^{-1}$ Mpc from this pixel. 

The criteria (a) and (b) are referred to width-trough (w-t) criteria.  
Figure~\ref{fig:lymasselection} shows the separation between HCDs and CoSLAs in $F_{\rm{trough}}$-$w_{0.8}$ diagram. With a CNR of 4 per pixel, using these two criteria, we can exclude 
70.8\% of HCD contaminants. Only 25.3\% of the CoSLAs are eliminated by applying same w-t criteria. 
For spectra with CNR = $8$ per pixel, 79.4\% of contaminants are ruled out, and 80.0\% of the CoSLAs pass the criteria. Therefore, criteria (a) and (b)  effectively eliminate contaminants, and also preserve high completeness for selecting targets.  We summarize this result in the following Table 4. 

{\it \large Criterion (c): }

Low-ionization (low-ion) metal lines trace high column density neutral hydrogen (HI) systems in the interstellar and circumgalactic medium \cite[e.g.,][]{ford13}. However, if the quasar sight line passes through the IGM overdensity, this HI overdensity should not be associated with detectable low-ion metal lines in typical BOSS spectra \cite[e.g.,][]{oppenheimer12}. 

Using low-ion metal absorption lines, we can eliminate a significant fraction of DLAs and sub-DLAs \citep{noterdaeme12}. Specifically, we pay particular attention to OI $\lambda$1302, CII $\lambda$1334, Si II $\lambda$1304, Si II $\lambda$1526, AlIII $\lambda$1670. 
If we detect the low-ion metal lines associated with strong Ly$\alpha$ absorption systems, we designate this system as a contaminant due to discrete DLAs or sub-DLAs. 
In addition, we remove Broad Absorption Lines (BALs) by searching for the presence of strong NV or CIV. 

\citet{noterdaeme12} present the equivalent width (EW) measurements for metal lines associated with HI absorbers with column density $N_{\rm{HI}}>\  10^{20}$ cm$^{-2}$ in SDSS-III/BOSS quasars. 
Using this catalog, we can calculate the fraction of the DLAs and sub-DLAs having detectable low-ion metal lines. 
%i.e., fraction of the DLAs and sub-DLAs that can be ruled out from the low-ionization metal lines. 

For quasar spectra with a median CNR $> 4$, 
Figure~\ref{fig:cii} demonstrates that $ 64\%$ of DLAs with $N_{\rm{HI}}= 10^{20.0}- 10^{20.4}$ cm$^{-2}$
 have detectable CII absorption lines redward of the quasar Ly$\alpha$ emission. Figure~\ref{fig:SiII} shows that $74\%$ of DLAs with 
$N_{\rm{HI}} = 10^{20.0}- 10^{20.4}$ cm$^{-2}$ have Si II absorption lines. 
For quasar spectra with CNR $>8$,  $82\%$ of DLAs with 
$N_{\rm{HI}}= 10^{20.0}- 10^{20.4}$ cm$^{-2}$ can be ruled out using CII absorption or Si II absorption line. 
Therefore, for spectra with CNR $\gtrsim4$, we can remove $75\%$ of DLAs; and for spectra with CNR $\gtrsim 8$, we can 
remove $ 82\%$ of DLAs, simply by utilizing the presence of corresponding C II $\lambda1334$ and Si II $\lambda1526$ lines.
 
Furthermore, for those Ly$\alpha$ absorption systems with non-detected corresponding CII and Si II, we stacked the expected low-ion metal lines redward the quasar Ly$\alpha$ emission.  If we still cannot detect the stacked metal absorption, this system passes the metal-line inspection.  With this technique, we can remove 78\% of DLAs for spectra with CNR= 4, and 84\% DLAs with CNR = 8. % The fraction of the systems associated with metal lines presented in the previous paragraph is only a conservative estimate. 

For sub-DLAs at $z\sim 2.5$ with $N_{\rm{HI}}= 10^{19}- 10^{20}$ cm$^{-2}$,
their metallicity increases as column density decreases \citep{york06, khare07, peroux08}. 
We conservatively  
assume that the average metallicity increases 0.6 dex for $N_{\rm{HI}}= 10^{19} - 10^{20}$ cm$^{-2}$ compared with 
the metallicity for DLAs $N_{\rm{HI}}= 10^{20}- 10^{20.5}$ cm$^{-2}$ \cite[e.g.,][]{khare07}. For spectra 
with CNR=4, $\gtrsim$ 50\% sub-DLAs from $10^{19}- 10^{20}$ cm$^{-2}$ have low-ionization metal lines that can be detected by SDSS-III/BOSS. For spectra with CNR=8, $\gtrsim$ 57\% of sub-DLAs can be ruled out using low-ionization metal lines. 
For LLSs with column density $N_{\rm{HI}}< 10^{19}$ cm$^{-2}$, $\approx$ 20\% of the LLSs have strong CII or Si II absorption with rest-frame 
EW $>$ 0.2 \AA\ and can be detected by BOSS data with CNR $>8$ \citep{prochaska15,fumagalli15}. 
 In our simulation, we conservatively assume that we cannot detect metal lines associated with LLSs with $N_{\rm{HI}}<10^{19}$ cm$^{-2}$ in BOSS data. 

In Table 3 and Table 4, we summarize the results after applying the metal-line selection criteria to our data. 
From Table 4, one can see that the number of unidentified DLAs, sub-DLAs and overlapping LLSs is significantly reduced, but remain more than 
one order of magnitude more than CoSLAs. %Some DLA systems and overlapping 
%LLSs are really hard to differentiate just from their absorption profiles. 
More criteria are needed to improve selection efficiency and 
further eliminate these high column density absorbers. 
%We need the criterion (d1): 

\begin{deluxetable}{lccc}  
\tablecolumns{4}
\tablewidth{0pt}
\tablecaption{DLA and sub-DLA metal line for BOSS } 
\label{table:sources}
\tablehead{\colhead{CNR} & 
		  \colhead{Log[N$_{\rm{col}}$] } & 
                  \colhead{CII or SiII detected }	 & 
                  \colhead{No CII or Si II}}
\startdata
4 & 20.0-20.4  &  75\% &  25\%  \\ 
8 & 20.0-20.4 &   82\% & 18\% 
\enddata
\tablecomments{
Summary of DLAs and sub-DLAs that can be ruled out using low-ionization metal lines. 
Two sets of spectra are presented, one having continuum-to-noise ratio (CNR) = 4, 
the other having CNR = 8. 
}
\end{deluxetable}

{\noindent \it \large Criterion (d1): }

The corresponding Ly$\beta$ absorption can be used to further determine the nature of the systems at $z>2.65$. At $z>2.65$, Ly$\beta$ is covered by BOSS spectra. The comparison between Ly$\alpha$ and Ly$\beta$ can help to distinguish if the absorption system consists of DLAs or the superposition of Ly$\alpha$ forest.

Based on our simulation, the CoSLAs with largest $\tau_{\rm{eff}}$ on 15 $h^{-1}$ Mpc contain the superposition of absorbers with  $N_{\rm{HI}}\sim 10^{15}- 10^{17}$ cm$^{-2}$. The individual Ly$\alpha$ absorbers are in the modestly saturated part of the curve of growth. %Correspondingly, CoSLAs contain the same number of Ly$\beta$ absorbers. 
Using the mock spectra, we have calculated that CoSLAs have EW ratio of Ly$\beta$ to Ly$\alpha$ in the range of $0.70 - 0.85$ (also see Figure~\ref{fig:individual_spectra}). 

\begin{figure}[tbp]
\epsscale{1.1}
\label{fig:cog}
\plotone{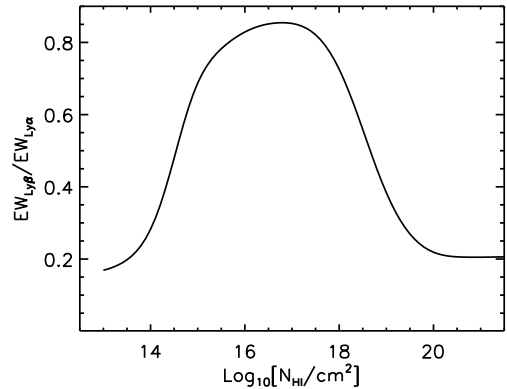}
\caption{The equivalent width (EW) ratio between Ly$\beta$ and Ly$\alpha$ for discrete absorbers as a function of HI column densities. 
When the column density is $N_{\rm{HI}}= 10^{15.5}- 10^{18.0}$ cm$^{-2}$, the EW ratio is greater than 0.75. 
When $N_{\rm{HI}}> 10^{19}$ cm$^{-2}$ or $N_{\rm{HI}}< 14.5$ cm$^{-2}$, the EW ratio drops below 0.25, regardless of the spectral resolution. This EW ratio can be used to significantly rule out absorbers with $N_{\rm{HI}}>10^{19.0}$ cm$^{-2}$.}
\end{figure}

Conversely, without considering LLS clustering, the major contaminant is high column density absorbers with $N_{\rm{HI}}>10^{19.5}$ cm$^{-2}$. The corresponding Ly$\beta$ is a single absorber with a much smaller equivalent width than the Ly$\alpha$ absorption (see Figure~\ref{fig:cog}), since it lies on the damping part of the curve of growth. 
Therefore, 
even at the BOSS SNR and resolution, %we are not able to resolve the individual lines, 
the measured EW ratio of Ly$\beta$/Ly$\alpha$ provides a sensitive test on which part of the curve of growth the absorber lies, and whether it is  a CoSLA or a DLA. 
Figure~\ref{fig:cog} presents that, when the column density $N_{\rm{HI}}= 10^{15.0}- 10^{18.0}$ cm$^{-2}$, the EW ratio of Ly$\beta$ to Ly$\alpha$ is greater than 0.6. 
When $N_{\rm{HI}}> 10^{19.5}$ cm$^{-2}$ or $N_{\rm{HI}}< 10^{14.5}$ cm$^{-2}$, the EW ratio of Ly$\beta$ to Ly$\alpha$ significantly drops below 0.25.
%When $N_{\rm{HI}}< 10^{14.5}$ cm$^{-2}$, the EW ratio of Ly$\beta$ and Ly$\alpha$ also significantly drops, regardless of the spectral resolution.  

Based on the above discussions, we select systems with the ratio of EW$_{\rm{Ly\beta}}$ to EW$_{\rm{Ly\alpha}}$ greater than 
0.6. %and the Ly$\beta$ has the similar width as the Ly$\alpha$ absorption. 

For contaminants with CNR$=4$ that pass the criteria (a), (b), and (c), 72\% can be ruled out using criterion (d1). 
However,  there are still more contaminants than targets after using criteria (a) $-$ (d1). At {\it CNR $=4$}, the remaining contaminants have indistinguishable profiles with CoSLAs using our current criteria. For {\it CNR = 8}, over $ 83\%$ of the systems with $N_{\rm{HI}}>10^{20}$ can be identified. After applying criterion (a) $-$ (d1), the number ratio between targets and contaminants is about 2 (see Table 4), without considering LLS clustering. The remaining 91 systems containing absorbers with $N_{\rm{HI}}\sim10^{17}-10^{18}$ cm$^{-2}$ are hard to rule out from moderate resolution data. 

\begin{figure}[tbp]
\epsscale{1.05}
\label{fig:mass15_afterHCD}
\plotone{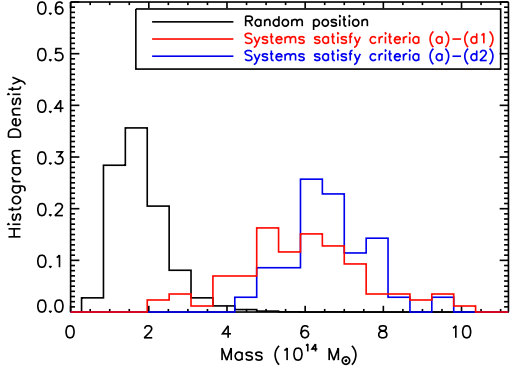}
\caption{The red histogram presents the mass distribution traced by absorption systems that pass 
the criteria (a) -- (d1), without considering the LLS clustering. The absorption systems are selected from the HCD-inserted, noise-added mock spectra in LyMAS simulation. Most of the systems ($\gtrsim$ 50\%) are CoSLAs and trace large-scale structures 
with mass $2.5\times$ that in random fields, representing the $\gtrsim 3.5\sigma$ mass overdensities 
on 15 $h^{-1}$ Mpc. 
The blue histogram shows the mass distribution traced by 17 groups of absorption systems (systems pass criteria (a)-(d2)). 
These 17 groups contain strong IGM Ly$\alpha$ absorption, and effectively trace the mass overdensities. }
\end{figure}

 In Figure~\ref{fig:mass15_afterHCD}, we plot the mass distribution traced by systems that meet criteria (a) -- (d1). Most of the selected 1-D absorption systems 
trace large-scale structures with a median mass of $\ge 6.2\times 10^{14}$ M$_\odot$, a factor of $\gtrsim 2.4\times$ that in random fields, representing $\gtrsim3\sigma$ mass overdensities. We summarize these results in Table 5. 

This is the mass overdensities traced by CoSLAs using single sight lines, i.e., without additional 2-D information. 
Note that these results have not included the LLS clustering. The LLSs with $N_{\rm{HI}}< 10^{18.5}$ cm$^{-2}$ have almost the same Ly$\beta$ to Ly$\alpha$ ratio compared with strong Ly$\alpha$ forests with $N_{\rm{HI}}\sim10^{16}$ cm$^{-2}$ (Figure~\ref{fig:cog}). These clustered LLSs with $N_{\rm{HI}}< 10^{18.5}$ cm$^{-2}$ are hard to eliminate based on our current selection criteria. From \S3.1.1, using high resolution spectra of 50 LLSs with $N_{\rm{HI}}<10^{19}$ cm$^{-2}$ \citep{prochaska15}, we roughly estimate that the LLS clustering yields 0.02\% systems with $\tau_{\rm{eff}}$ higher than our CoSLA selection threshold. Although this is a small fraction, it includes $\approx300$ clustered LLSs with $N_{\rm{HI}}<= 10^{19}$ cm$^{-2}$ that are hard to identify from moderate resolution spectra. These clustered LLSs may reduce our CoSLA selection efficiency to  $\approx40\%$. 
By collecting new data on LLSs, we are preparing a paper to probe the LLS clustering along the line-of-sight in details. 

We apply the criteria (a) -- (d1) to BOSS data to select the strongest absorption systems at $z>2.65$. We summarize the sample selected from SDSS-III/BOSS  in section \S 6.

%{\bf explain the different between these figures and figures 10/11.}

\figurenum{15}

\begin{deluxetable*}{cccccc}  
\tablecolumns{7}
%\longtable
\tablewidth{0pt}
\tablecaption{Selection of Strongest 1-D Ly$\alpha$ absorption systems on 15 $h^{-1}$ Mpc }
\label{table:F1300}	
\tablehead{\colhead{CNR} & 
                 \colhead{ Number} & 
	          %\colhead{Scale ($h^{-1}$ Mpc) } & 
                  \colhead{T$_{\rm{flux}}$ } & 
                  \colhead{$\tau_{\rm{eff}}$ } & 
	          \colhead{Noise-added} &
                  \colhead{HCD-added}\nl }
\startdata
4& before w-t selection  & 0.21- 0.32 & 1.15- 1.56 &  303  &  13,635    \\
4& after w-t selection & 0.21- 0.32 & 1.15- 1.56 &  222 & 4981 \\
4 & after low-ion metal  & 0.21- 0.32 & 1.15- 1.56 & 222 &  1245 \\
4 & after low-ion metal  & 0.21- 0.32 & 1.15- 1.56 & 197 &  336 \\
%CNR=4 & after checking Ly$\beta$ & 15 & 0.21- 0.32 & 1.15- 1.56 & 222 &   \\
8& before w-t selection  & 0.21- 0.32 & 1.15- 1.56 &  289  & 12,210    \\
8& after w-t selection  & 0.21- 0.32 & 1.15- 1.56 &  231 & 4059 \\
8 & after low-ion metal  & 0.21- 0.32 & 1.15- 1.56 & 231 & 730 \\
8 & after checking Ly$\beta$  & 0.21- 0.32 & 1.15- 1.56 & 213 & 91
\enddata
\tablecomments{The last two column present the number of absorption systems in noise-added mock spectra and high column density (HCD)-inserted mock spectra, assuming LLSs are randomly distributed. 
For data with CNR $= 4$ and CNR$=8$, we summarize our results after applying the w-t and metal-line selection criteria to our data. A significant number of DLA contaminants is excluded with applying these criteria. However, one can see that number of unidentified DLAs, sub-DLA remains one order of magnitude larger than that of CoSLAs. More criteria are needed to further rule out high column density absorbers. }
\end{deluxetable*}

\begin{deluxetable*}{lll}  
\tablecolumns{3}
%\longtable
\tablewidth{0pt}
\tablecaption{Mass in 15 $h^{-1}$ Mpc cubes centered on different objects}
\label{table:F1300}	
\tablehead{\colhead{Center} &
	          \colhead{Median mass} &
	          \colhead{$\sigma_{\rm{15 h^{-1}\ Mpc}}$}  \nl
                 \colhead{} & 
                  \colhead{( $10^{14}\ M_\odot$} ) & 
                  \colhead{($10^{14}\ M_\odot$} ) \nl}
\startdata
random &   2.6  &  1.2 \\
quasars ($M_{\rm{halo}}= $2-3$\times10^{12}\ M_\odot$) &3.7  & 1.6 \\
Halos with $M_{\rm{halo}}> 3\times10^{13}\ M_\odot$ &  6.1 & 1.0 \\  
CoSLAs selected from original mock & 7.0 &  1.6 \\
CoSLAs selected from HCD-inserted spectra with \it{CNR=8} & 6.2 & 1.8  
\enddata
\tablecomments{Similar with Table 1, summary of mass within 15 $h^{-1}$ Mpc in different positions in LyMAS simulation. ``CoSLAs from original mock" represents mass traced by CoSLAs selected from original mock spectra, no noise being added or HCDs being inserted; %HI {\it{CNR}=4} shows the mass traced by largest absorption selected from noise-added spectra. 
HCD inserted represents the CoSLAs selected from realistic mock spectra with HCD-inserted and noise added. We apply criteria (a)$-$(d1) to select CoSLAs from HCD-inserted realistic mock spectra, without LLSs clustering considered.}
\end{deluxetable*}

{\noindent \it \large Criterion(d2)}:

For absorbers at $z<2.65$, Ly$\beta$ with good S/N is generally not available from the BOSS data. %The majority of large absorption are still due to HCDs after applying the first three criteria (a) $-$ (c). Therefore, 
We apply criterion (d2): using groups of Ly$\alpha$ absorption systems to pinpoint mass overdensities.

From our simulation, a considerable fraction of CoSLAs is associated with other nearby IGM Ly$\alpha$ forest systems with $\tau_{\rm{eff}}> 0.8\ (3.5\times \mean{\tau}_{\rm{z=2.5}})$. This effect arises because true IGM overdensities trace filamentary structures that could extend a few tens of Mpc (see Figure~\ref{fig:example}). 
However, DLAs or sub-DLAs, which are more likely to trace field galaxies, normally have small HI cross sections ($\sim$ 100 kpc $\times $ 100 kpc, e.g., \citet*{cai14}), and there is a small chance to find a group of HCDs on Mpc scales. For example, with sight-line separations of $\gtrsim$ 15 $h^{-1}$ Mpc, we estimate that the probability of having DLA pair is 0.05\%.  \footnote{We assume that the DLA covering fraction is 100\% within the impact parameter of 10 kpc from the halo center. \citet{cooke06} report that the correlation function $\xi_{\rm{DLA}}(r)$ is $\approx 30\%$ at $r= 15 \ h^{-1}$ Mpc. %Assuming the halomass of DLAs are $10^9- 10^{12}$ M$_\odot$, 
Based on this correlation strength and halo abundance function \citep{tinker10}, the probability of finding two DLAs separated by 15 $h^{-1}$ Mpc is approxmately 0.05\%. } For $\ge 3$ DLAs within a 15 $h^{-1}$ Mpc, the probability is further much smaller. 

At $z<2.35$, the SDSS-III/BOSS has reached an average quasar density of $\gtrsim 10$ deg$^{-2}$. In some sub-regions of BOSS area (e.g., Stripe 82), the background quasar density exceeds 20 deg$^{-2}$. In a significant fraction of SDSS coverage, the background quasar density is sufficiently high to use criterion (d2) to select the most massive HI overdensities over a large scale. 

 In the LyMAS simulation, after reducing the density of sight lines to realistic 2-D BOSS quasar distribution, %at $z\ge2.35$,
17 groups of absorption systems are selected from a $1\ (h^{-1}$ Gpc)$^3$ survey volume,  which regions contain $\ge 4$ absorption systems with 
$3.5\times \mean{\tau}< \tau_{\rm{eff}}< 7\times \mean{\tau}$ in a volume of ($25~h^{-1}$ Mpc)$^3$. 
Each of these selected groups contains at least one CoSLA with $\tau_{\rm{eff}}> 4.5\times \mean{\tau}$, 
and most of these group of absorption systems trace the large-scale structures. 
%We summarize our results in Table 7. 
The 3-D mass distribution associated with these 17 groups of absorption systems are presented in the blue histogram of  Figure~\ref{fig:mass15_afterHCD}. 

This method currently works best for BOSS at $z<2.35$. For $z>2.35$, the overdensity  searches using criterion (d2) could be considerably incomplete, because the 2-D quasar density is significantly lower. 

In our next paper of this series, we will introduce our narrowband imaging results on overdense fields with a group of absorption systems selected from SDSS-III. In this paper, we mainly focus on the spectroscopy results of 1-D Ly$\alpha$ absorption systems in single sight lines, and therefore focus on systems at $z>2.65$. 

%{\bf say something about $2.35<z<2.7$.}

\figurenum{16}

\section{Ly$\alpha$ Absorption Systems around Confirmed Overdensities}

In this section, we examine the Ly$\alpha$ absorption signatures on two well-studied galaxy overdensities. There are a few other overdensities at $z=2-3$ that can be utilized to conduct similar tests (e.g., protoclusters in COSMOS) \citep{chiang14}. The complete results of Ly$\alpha$ absorption around confirmed overdensities will be summarized in the paper that is preparing (Mukae et al. 2016 in prep.).

\subsection{SSA22 Protocluster at $z=3.1$}

SSA22 protocluster is the most intensively studied large-scale galaxy overdensity at high redshift. This overdense field was serendipitously discovered by \citet{steidel98} through deep galaxy redshift survey. Further multiwavelength observations from optical to sub-mm confirmed an overdensity of LBGs, LAEs, LABs and SMGs in this field on $\sim$ 30 $h^{-1}$ Mpc \cite[e.g.,][]{chapman04, matsuda05, tamura09}. 
In the SDSS-III/BOSS quasar library, we found four background quasars within 30 $h^{-1}$ Mpc 
from the center of SSA22. 

\begin{figure}[tbp]
\epsscale{1.2}
\label{fig:SSA}
\plotone{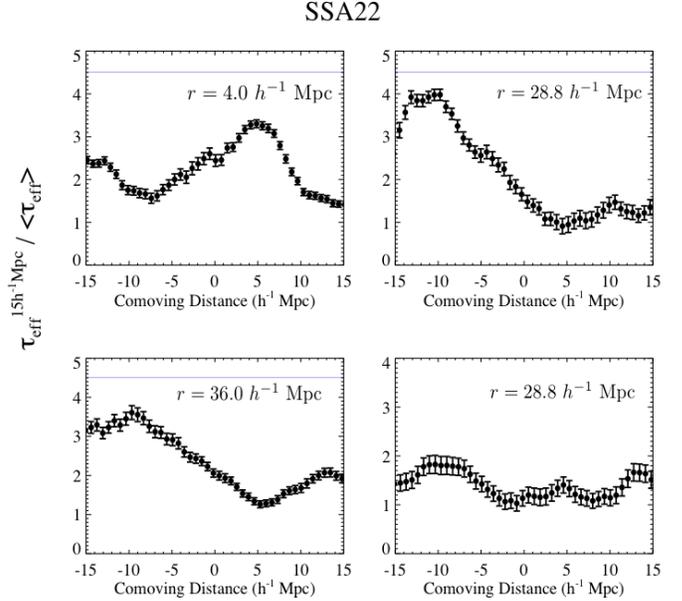}
\caption{A group of four Ly$\alpha$ absorption systems around the SSA22 overdense field at $z=3.1$. These four sight lines have impact parameters $\lesssim 35$ Mpc from the SSA22 field center. Black points with errors represent the optical depth on 15 $h^{-1}$ Mpc  ($\tau^{15h^{-1}\rm{Mpc}}_{\rm{eff}}$).   Blue horizontal lines indicate the CoSLA threshold of 4.5$\times \mean{\tau_{\rm{eff}}}$. We also give the 
impact parameter $r$ of each sight line in every panel.}
\end{figure}

In Figure~\ref{fig:SSA}, we present that the Ly$\alpha$ absorption around the SSA22 overdense field. The upper left panel presents a background quasar 4.0 $h^{-1}$ Mpc away from 
the field center: $\alpha=22$:17:34, $\delta=+00$:17:01 (J2000.0). 
This strong absorption has also been observed using Keck/HIRES \citep{adelberger05}, and confirm that this absorption is consistent with the superposition of the intergalactic Ly$\alpha$ forest rather than LLSs or DLAs. 
The upper right panel present an absorption system at the similar redshift as the galaxy overdensity ($z=3.09$), with an 
effective optical depth on 15 $h^{-1}$ Mpc equals to $\approx4.0\times \mean{\tau_{\rm{eff}}}$. 
 The lower left panel shows a background quasar $\sim$ 36.0 $h^{-1}$ Mpc away from the center of SSA22, with a high $\tau_{\rm{eff}}\approx3.7\times\mean{\tau_{\rm{eff}}}$ on 15 $h^{-1}$ Mpc. The lower right panel presents a modest Ly$\alpha$ absorption at 28.8 $h^{-1}$ Mpc from 
the field center, with an optical depth $\lesssim 2.0\times\mean{\tau_{\rm{eff}}}$.  The SSA22 is traced by a group of Ly$\alpha$ absorption systems.

\figurenum{17}

%{\bf why you removed this?}

\subsection{A Large-scale Ly$\alpha$ Nebula Jackpot Quasar Quartet at $z=2.05$}

Giant Ly$\alpha$ nebulae (also known as “Ly$\alpha$” blobs, LABs) are characterized by a high luminosity of Ly$\alpha$ line emission ($L(\rm{Ly}\alpha)\gtrsim 10^{43}$ erg s$^{-1}$), and a spatially large Ly$\alpha$ emitting region from tens of kiloparsec (kpc) up to intergalactic scales of hundreds of kpc \cite[e.g.,][]{yang09,yang10,yang14, matsuda05, matsuda11}. Previous studies suggest the giant Ly$\alpha$ blobs (LABs) are strongly clustered sources. They are living in massive dark matter halos ($\sim10^{13}$ M$_\odot$), and represent sites of most active galaxy formation and large-scale galaxy overdensities \citep{yang09,yang10}. 

\begin{figure}[tbp]
\epsscale{1.2}
\label{fig:LAB}
\plotone{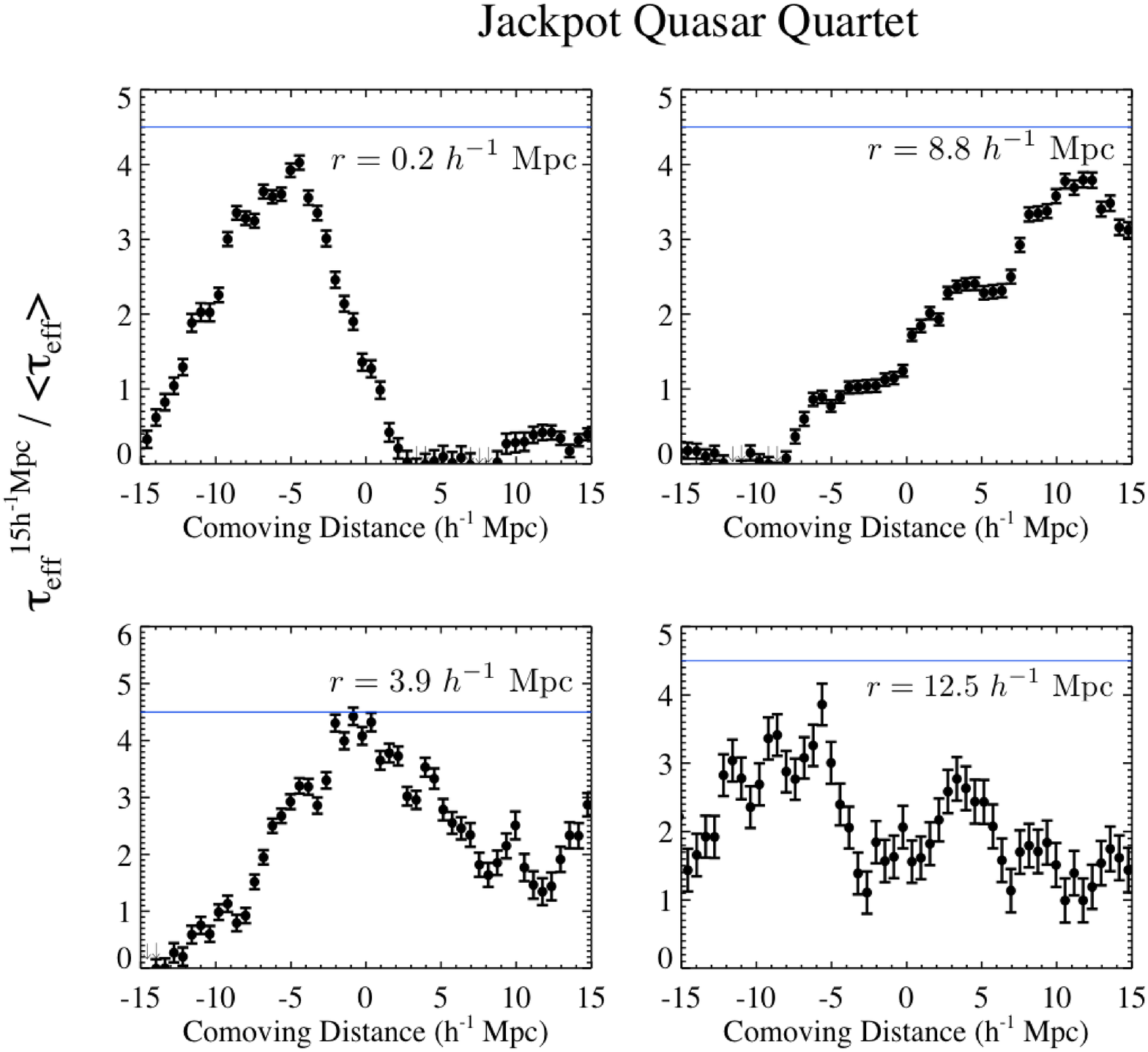}
\caption{The group of Ly$\alpha$ absorption systems around Ly$\alpha$ blob (LAB) Jackpot field at $z=2.055$, with each sight line having transverse separations $\lesssim$ 15 $h^{-1}$ Mpc from Jackpot quasar quartet. These four sight lines have impact parameters $\lesssim 15$ Mpc from the quasar quatet. Black points with errors represent $\tau^{15h^{-1}\rm{Mpc}}_{\rm{eff}}$.   Blue horizontal lines indiciate the CoSLA threshold of 4.5$\times \mean{\tau_{\rm{eff}}}$.}% $2-\sigma$ upper limit of $\tau_{\rm{eff}}$ are presented using the black arrows. }
\end{figure}

\citet{hennawi15} reported a giant and ultraluminous Ly$\alpha$ nebula Jackpot which is associated with a rare quasar quartet at $z=2.05$. 
This structure is embedded in a substantial overdensity of galaxies \citep{hennawi15}. 
At a large scale of 15 $h^{-1}$ Mpc from this LAB, we found fix background quasars in SDSS-III.  Most of these quasar continua have a significant excess of Ly$\alpha$ absorption at $z=2.05$ (Figure~\ref{fig:LAB}). 

In Figure~\ref{fig:LAB}, we present four Ly$\alpha$ absorption surrounding the quasar quartet.
Upper left presents a background quasar 110 physical kpc away from the Ly$\alpha$ nebula, and an absorption at the redshift of the quasar quartet ($z=2.05$, $\lambda = 3710$ \AA). 
Lower left shows another absorption system 3.9 $h^{-1}$ Mpc away from the LAB.  
The upper right panel displays a background quasar 8.8 $h^{-1}$ Mpc away from the Ly$\alpha$ nebula, and lower right 
presents another strong absorption 12.5 $h^{-1}$ Mpc away from the LAB. The $\tau^{15h^{-1}\rm{Mpc}}_{\rm{eff}}$ of these four Ly$\alpha$ absorption range from 3.5 -- 4.2$\times\mean{\tau_{\rm{eff}}}$. 
 The group of absorption systems support the anticipation that a large-scale IGM overdensity is associated with this quasar quartet. %Follow-up narrowband imaging or redshift sruvey at larger scale ($\sim 30'$) will confirm this.  

These results, found from overdensities selected from modest survey volume, encourage us to probe more extreme systems selected from larger survey volume using SDSS-III/BOSS (see \S6). 

\figurenum{18}

\section{CoSLA candidates Selected from SDSS-III/BOSS database}

\subsection{A sample of  candidate IGM absorption systems at $z>2.65$}

In this section, we introduce a sample of CoSLA candidates that are selected from the SDSS-III/BOSS quasar spectral library (see \S2.1). 
For each quasar sight line, we search for CoSLAs over a redshift range 
between $z_{\rm{min}}$, defined as the redshift where the spectral continuum-to-noise ratio (CNR) per pixel reaches 8, and the $z_{\rm{max}}$, defined as $3000$ km s$^{-1}$ blueward of quasar redshift.  The selected absorption systems have highest $\tau_{\rm{eff}}$ over a smoothing distance of 15 $h^{-1}$ Mpc. 
%{\bf again, remind people about scale length.}
We apply our selection criteria (a)$-$ (d1) described in \S 5.2 to select absorbers with $z_{\rm{abs}}>2.65$ where Ly$\beta$ is covered by SDSS-III/BOSS. 
 
\subsubsection{Survey Volume}

Within the RA range (RA $> 20$ h and RA $< 12$ h) that we can reach with our scheduled time in Multiple Mirror Telescope (MMT), we probe the Coherently Strong Ly$\alpha$ Absorption systems (CoSLAs) from  $\approx6,000$ sight lines, a total distance  ($d_{\rm{sight\_line}}$) of $1.53\times 10^6\ h^{-1}$ Mpc along the lines of sight. The significant overdensity of HI gas is expected to extend to at least $\pm 5$ $h^{-1}$ Mpc (e.g., see Figure~\ref{fig:example}), so that each quasar probes a cylinder with a volume of 
$\sim 10\ h^{-1}\times 10\ h^{-1}\times d_{\rm{sight\_line}}$ Mpc$^3$. Thus, overall, we have probed a volume of $\sim 10 \times 10 \times 1.53 \ h^{-3}\ \rm{Mpc^3}= 0.15\ (h^{-1} \rm{Gpc})^3$.  

\begin{figure}[tbp]
\epsscale{1.1}
\label{fig:data_summary}
\plotone{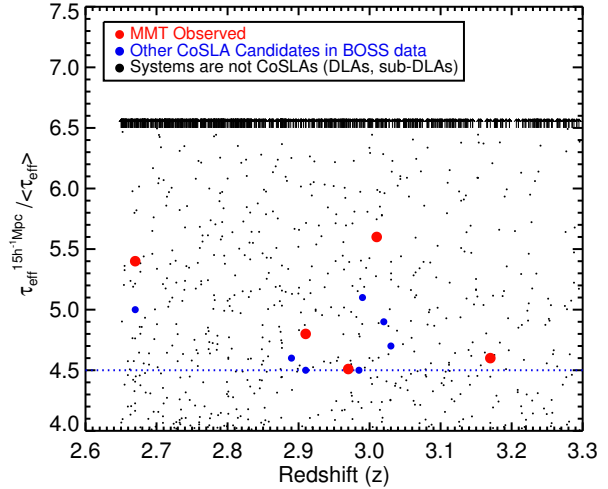}
\caption{The summary of our data. The 5 red points represent the CoSLA candidates selected from SDSS-III/BOSS and have been observed by MMT. 
The 7 blue points represent the other CoSLA candidates selected from SDSS-III/BOSS that passes criteria (a) $-$ (d1). The black dots represent 947 other absorption systems, and the systems with $\tau{_{\rm{eff}}}(z_{\rm{abs}})>$ 6.5$\times\mean{\tau_{\rm{eff}}}$ on 15 $h^{-1}$ Mpc are marked by upper arrows, which are most likely to be HCDs. All of the black symbols do not pass the criteria (a) $-$ (d1). }
\end{figure}

From this volume, applying selection criteria (a) $-$ (d1), we select $12$ CoSLA candidates from 947 absorption systems with $\tau{_{\rm{eff}}}(z_{\rm{abs}})>$ 4.5$\times$ $\mean{\tau_{\rm{eff}}(z_{\rm{abs}})}$ on 15 $h^{-1}$ Mpc. We briefly summarized our data in Figure~\ref{fig:data_summary}. We have obtained high SNR observations for 5 of these 12 absorption systems at Multple Mirror Telescope (MMT) (red points in Figure~\ref{fig:data_summary}, see \S6.2).

%Let us compare the number of CoSLAs we selected from SDSS-III/BOSS to that predicted by simulation. 
%From the Table 4, after using criteria (a) $-$ (d1) to rule out contaminants, 30\% of targets are also eliminated. 
%Therefore,  given an incompleteness of 30\% (see Table 4), the number of CoSLAs from SDSS-III/BOSS is $11/0.7\approx 16$ within the survey volume of 0.15 ($h^{-1}$ Gpc)$^3$.  The 
% LyMAS simulation perdicts $40$ CoSLAs in this volume of 0.15 ($h^{-1}$ Gpc)$^3$, still a factor of $\sim 3\times$ the number of observed value. We should note that the LyMAS simulation we currently used do not include AGN feedback, and could still overpredict the optical depth in overdense regions as well as the number of CoSLAs.  

We do not find any CoSLA with $\tau_{\rm{eff}}>7\times\mean{\tau_{\rm{eff}}(z_{\rm{abs}})}$, where $\mean{\tau_{\rm{eff}}}=0.29- 0.42$ at $z=2.65-3.40$ (Bolton et al. 2009; Faucher-Giguere et al. 2008).  This result is consistent with the LyMAS simulation. Our studies support that there is an upper limit on $\tau_{\rm{eff}}$ of CoSLAs within 0.15 ($h^{-1}$ Gpc)$^3$, which is $6\times \mean{\tau_{\rm{eff}}(z_{\rm{abs}})}$. Any absorbers with $\tau_{\rm{eff}}>6\times \mean{\tau_{\rm{eff}}(z_{\rm{abs}})}$ are contaminated by HCDs.

%\vspace{0.1in}
In the following, we present high SNR spectra of 5 candidates of CoSLAs. 
Table 6 and Table 7 further present the properties of this sample.

\figurenum{19}

%This  sample is selected from part of SDSS-III DR11 and has RA ranging from $00\ \rm{hr}$ to $12\ \rm{hr}$.
%Spectroscopic followup observations are presented in \S6.2. 
%{\bf I am confused, 20 or 15 Mpc, why there are two??}

 \subsection{MMT Spectroscopy of a sample of IGM absorption systems at $z=2.7-3.4$}

%{\bf discuss how this survey volume is defined; did you include incompleteness?} 
%{\bf here present the sample!}

Spectroscopic observations of CoSLA candidates were obtained using the blue channel spectrograph onboard the 6.5 m MMT in 2012 $-$ Jan. 2014. The MMT observations have higher SNR than that of the BOSS data, particularly in the Ly$\beta$ region of CoSLA candidates. 

This section presents a sample of 5 CoSLA candidates at $z=2.7-3.4$ that are strongly suggested by both BOSS and MMT spectra (also see Table 6 and Table 7).

Depending on the central wavelength of the absorption, we either use the
 800 lines mm$^{-1}$ ($R= 2,000$) or the 1200 lines mm$^{-1}$ ($R=3,000$) gratings, with the selection made by the following reasons: (1) to match the central wavelength of the absorption with the most sensitive part of the grating and spectrograph; (2) some absorption systems need a higher resolution to better resolve the absorption features. 
Typically, $90-120$ minutes on-source exposures were taken for
each target, which varied according to the weather conditions and the quasar flux density. 
We divide long exposures into a series of single $20-30$ min individual exposures.
Data reduction was finished with the automatic pipeline during the observations in order to decide 
if more exposures were needed. The wavelength coverage varies according 
to the central wavelength of specific absorption system.  
The airmass of the observations were about
$1.0-1.4$, and we used positions angles close to the parallactic angle. Spectrophotometric standard stars were observed for flux
calibration, and a CuAr arc lamp was used for wavelength
calibration. We have reached a typical CNR of 10 in the absorption region.

\subsubsection{J025252.07+025704.0}

\begin{figure}[tbp]
\epsscale{1.1}
\label{fig:02+02}
\plotone{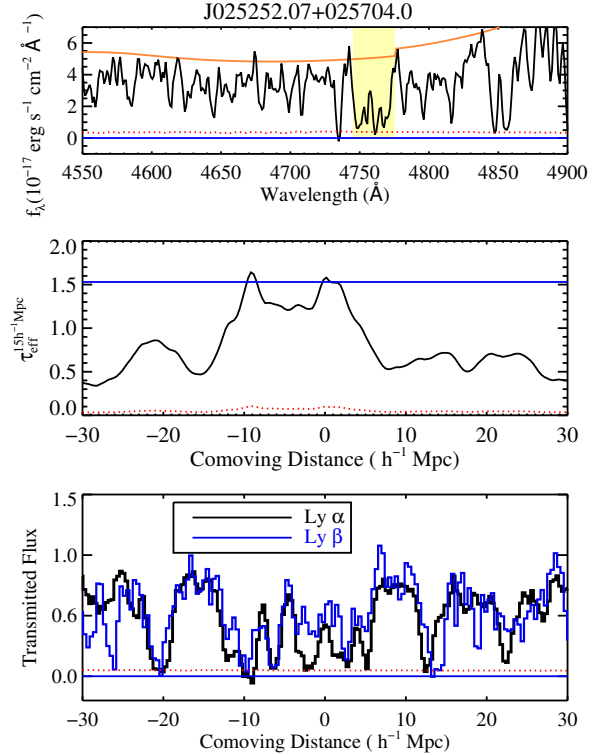}
\caption{The spectra of CoSLA candidate J025252.07+025704.0. The upper panel shows the BOSS spectrum, with absorption marked in yellow shaded area. Orange shows the continuum using mean optical depth regulated PCA fit (Lee et al. 2012). The middel panel presents the $\tau_{\rm{eff}}$ over 15 $h^{-1}$ Mpc centered on the absorption center. The blue horizontal lines indicate the threshold of CoSLAs, which is a 4.5$\times$ the mean optical depth. The lower panel presents MMT spectra ($R=3,000$), expanding the Ly$\alpha$ absorption.  Black presents the Ly$\alpha$ absorption and blue indicates Ly$\beta$ absorption.  
%Black indicates the Ly$\alpha$ and blue indicate Ly$\beta$ absorption. 
From the equivalent width comparison between the 
Ly$\alpha$ (black) and Ly$\beta$ absorption (blue), this system contain multiple absorbers, with 
the column density of each absorber $N_{\rm{HI}}\sim 10^{15}- 10^{18}$ cm$^{-2}$.  }
\end{figure}

Figure~\ref{fig:02+02} presents the spectrum of this target. The upper panel presents the SDSS-III/BOSS spectrum. From the 
BOSS spectrum, this absorption 
system (yellow shaded area) well satisfies the selection criteria of CoSLAs on 15 $h^{-1}$ Mpc 
centered at $4755$ \AA\ ($z=2.91$). From the BOSS data, the effective optical depth of this 
system is $\tau_{\rm{eff}}= 1.61^{+0.10}_{-0.10}$, a factor of $4.8 \times$ the mean optical depth at $z=2.91$.

The middle panel presents the effective optical depth on 15 $h^{-1}$ Mpc  ($\tau^{15h^{-1}\rm{Mpc}}_{\rm{eff}}$), greater 
than our selection threshold (blue horizontal line) of $4.5\times \mean{\tau_{\rm{eff}}}$. 

\begin{figure}[tbp]
\epsscale{1.1}
\label{fig:compare}
\plotone{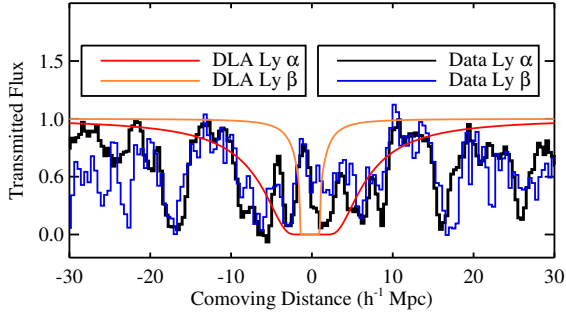}
\caption{A comparison between the CoSLA candidate J025252.07+025704.0 and a sub-DLA (super LLS) with $N_{\rm{HI}}= 10^{20.0}$ cm$^{-2}$. The red spectrum is the DLA Ly$\alpha$; and orange is DLA Ly$\beta$, overplotted with the Ly$\alpha$ absorption (black) and Ly$\beta$ (blue) of    J025252.07+025704.0.   }
\end{figure}

The lower panel presents the MMT follow-up observations of this 
target with $3\times30$ min exposures using 1200 lines mm$^{-1}$ grating. 
With MMT spectra,  we can resolve any Ly$\alpha$ and Ly$\beta$ absorbers with rest-frame 
Doppler parameter $b> 100/(1+z)= 25$ km s$^{-1}$. From the lower left panel, the EW ratio between Ly$\beta$ (blue) and Ly$\alpha$ (black)
is $\frac{EW_{Ly\beta}}{EW_{Ly\alpha}}=0.88^{+0.03}_{-0.03}$, which suggests that this absorption consists of  the superposition of a series of individual absorbers with  $N_{\rm{HI}}\sim 10^{15-18}$ cm$^{-2}$ (Figure~\ref{fig:cog}). %Our fits suggest this absorption system is %From the zoom-in figure, this system 
This absorption system is similar to the 
strongest intergalactic Ly$\alpha$ absorption predicted in LyMAS simulation.  
Figure~\ref{fig:compare} further suggests that the absorption J025252.07+025704.0 strongly deviates from DLA or sub-DLA absorption. 

\begin{figure}[tbp]
\epsscale{1.1}
\label{fig:minflux}
\plotone{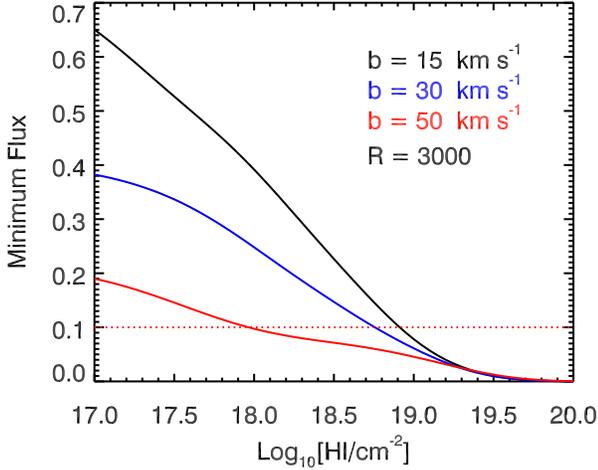}
\caption{The minimum flux as a function of column densities at resolution of $R=3,000$. The black curve represents single Ly$\alpha$ absorber with Doppler parameter $b=15$ km s$^{-1}$. Blue indicates single absorber with $b=30$ km s$^{-1}$, and red shows absorber with $b=50$ km s$^{-1}$. }
\end{figure}

Other evidence also supports the interpretation that the J025252.07+025704.0 arises from blending of intergalactic Ly$\alpha$ forest lines.
 The distribution of Doppler parameter $b$ in the Ly$\alpha$ forest is characterized by Gaussian function with a median $b \approx 30$ km s$^{-1}$ and $\sigma$ = 10 km s$^{-1}$, and cropped below $b \approx 20$ km s$^{-1}$ (e.g. Rudie et al. 2012, Pieri et al. 2014).  Figure~\ref{fig:minflux} presents that, at MMT resolution of 100 km s$^{-1}$ bins (or wider bins, such as BOSS resolution), a Ly$\alpha$ forest  line of typical Doppler parameters does not reach the zero level, regardless of its column density. Only single lines with log$_{10}[\frac{\rm{NHI}}{\rm{cm}^{-2}}] \gtrsim19$ or unusually high Doppler parameters ($\gtrsim50$ km s$^{-1}$) reach a minimum flux of $F_{\rm{min}}\le 0.15$ (also see Pieri et al. 2014). %However, the  $N_{\rm{HI}}\gtrsim10^{19}$ cm$^{-2}$ significantly deviate the profiles, and the EW ratio between Ly$\beta$ to Ly$\alpha$ of J025252.07+025704.0. 
The absorption in J025252.07+025704.0 reaches a $F_{\rm{min}}\le$ 0.1. Combined with the evidence of high EW ratio between Ly$\beta$ to Ly$\alpha$, we conclude that the nature of Ly$\alpha$ absorption in J025252.07+025704.0 is the superposition of Ly$\alpha$ lines. %and even at higher fluxes blends of weak lines may dominate by number over isolated strong lines.

 What is the underlying mass such strong absorption ? 
We define the mass overdensity within 15 $h^{-1}$ comoving Mpc as 

\begin{equation}
M_{15h^{-1}\rm{Mpc}}=(1+\delta_m)\times \mean{M_{15 h^{-1}\ \rm{Mpc}}},
\end{equation}
 where $\mean{M_{\rm{15h^{-1}Mpc}}}$ is $2.6\times 10^{14}\ M_\odot$, the average mass within a  (15 $h^{-1}$ Mpc)$^3$ box. The quantity $\delta_m$ is the mass overdensity. 
On $15 \ h^{-1}$ Mpc scale, the effective optical depth of this absorption is about $4.8\times \mean{\tau_{\rm{eff}}(z=3.1)}$ \citep{bolton09, faucher-giguere08}.  From Figure~\ref{fig:scatter}, the median mass traced by CoSLAs with $\tau_{\rm{eff}}=4.8\times \mean{\tau_{\rm{eff}}}$ 
is around $6.5\times10^{14}\ M_\odot$, equivalent to an overdensity of 1.5 on the large scale of 15 $h^{-1}$ Mpc.  Such large-scale overdensities contain sub-region density peaks that have high mass concentration. The high density peaks represent most overdense environments at high redshifts \cite[e.g.,][]{chiang14, steidel05}. Figure~\ref{fig:example} presents the projected matter distribution in the $x$--$z$ plane, where the $z$-axis is along the line of sight direction. The density peaks have mass overdensities reaching $\delta_m\approx$  10 on ($5\ h^{-1}$ Mpc)$^3$ volume. 

Note the overdensity is estimated with the assumption that J025252.07+025704.0 does not contain LLS. Our current observations cannot rule out the possibility that  J025252.07+025704.0 may contain LLS absorbers with $N_{\rm{HI}}\sim10^{17}-10^{18}$ cm$^{-2}$ (see \S4.2). The presence of LLSs should lower our estimates on the mass overdensity. 

  %Note the mass overdensity in central region is $\delta_m > 10$ (e.g., Figure xx). 

%Figure xx presents an example with large absorption system with $\tau_{\rm{eff}}=4.8\times \mean{\tau_{\rm{eff}}}$ at $z=2.5$, which effectively trace a structure with overdensity of 1.5 over $15\ h^{-1}$ Mpc. 

%  the most overdense environment that have been confirmed at high redshift (e.g., Steidel et al. 2005). 

\figurenum{20}

\figurenum{21}

\figurenum{22}

%\figurenum{19}
%\begin{figure}[tbp]
%\epsscale{1}
%\label{fig:02+02xz}
%\plotone{skewer_43399_xz_thick_30R25.eps}
%\caption{Upper panel: Projected mass distribution within each slice of 0.6 $h^{-1} \times 0.6\ h^{-1}\times 15$  $h^{-1}$ Mpc$^3$. Different color represent different mass in each cell. Lower panel: a simulated strong IGM absorption system from LyMAS simulation that traces the mass overdensity in upper panel.}
%\end{figure}

\subsubsection{J084259.37+365704.3}

\begin{figure}[tbp]
\epsscale{1.1}
\label{fig:08+36}
\plotone{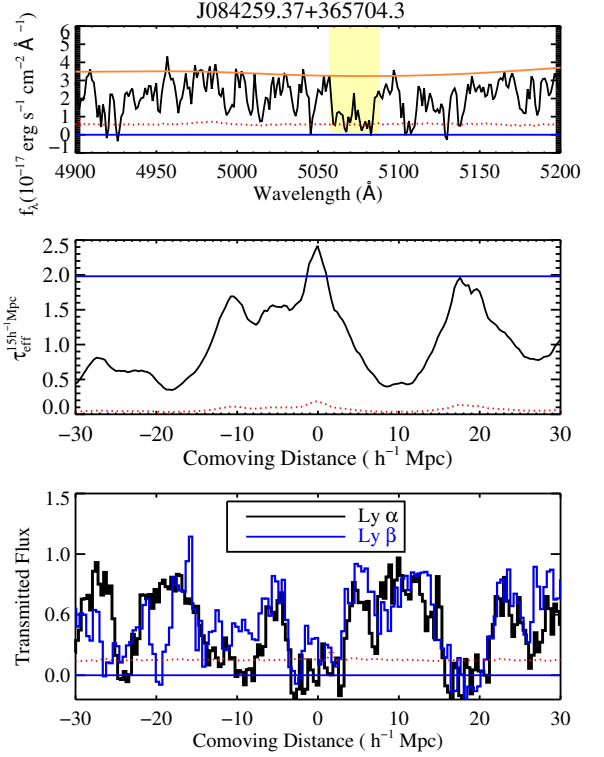}
\caption{Similar to format of Figure~\ref{fig:02+02}, we show the CoSLA candidate J084259.37+365704.3.  The spectra suggest this strong Ly$\alpha$ absorption system is associated with the superposition of Ly$\alpha$ forest in the range $10^{15}\ \rm{cm}^{-2}< N_{\rm{HI}} < 10^{18}\ \rm{cm}^{-2}$.  }
\end{figure}

Figure~\ref{fig:08+36} presents the spectra of J084259.37+365704.3. The upper panel presents the SDSS-III/BOSS spectrum. From the 
BOSS spectrum, J084259.37+365704.3 is a CoSLA candidate centered at $4830$ \AA, and has the effective optical depth of  $\tau_{\rm{eff}}= 1.71^{+0.10}_{-0.10}$, a factor of $4.5 \times$ the mean optical depth at $z=3.17$ (Bolton et al. 2009 \& Dall’Aglio et al. 2008, Faucher-Giguere et al. 2008). 
The mid-panel presents the effective optical depth on 15 $h^{-1}$ Mpc ($\tau^{15h^{-1}\rm{Mpc}}_{\rm{eff}}$). 
The lower panel displays the MMT observations on this 
target with $3\times 30$ min exposures using 1200 lines mm$^{-1}$ grating. 
 On a 15 $h^{-1}$ scale, 
this absorption has rest-frame EW ratio between the Ly$\beta$ to Ly$\alpha$ ($\frac{EW_{Ly\beta}}{EW_{Ly\alpha}})=0.83^{+0.03}_{-0.03}$. 
From the EW ratio, this strong Ly$\alpha$ absorption is not due to DLAs, but is likely to arise from
the superposition of Ly$\alpha$ forest with EW ratio between Ly$\beta$ to Ly$\alpha$ $\gtrsim0.8$. %From with column densities $N_{\rm{HI}}= 10^{16}- 10^{18}$ cm$^{-2}$. 
From Figure~\ref{fig:cog}, our observations suggest this system consists of multiple HI absorbers with column densities $N_{\rm{HI}}= 10^{15}- 10^{18}$ cm$^{-2}$.  

%Higher resolution spectroscopy is required to fully resolve single absorbers associated with this large absorption system. 

 J084259.37+365704.3 is associated with two sight lines with projected separation of 20 $h^{-1}$ Mpc on the sky. 
Strong IGM absorption is present in both sight lines:  J084328.73+364107.4, with a transverse 10 $h^{-1}$ Mpc from the target J084259.37+365704.3, and J084233.26+365129.9, with a transverse 20 $h^{-1}$ Mpc from the target. 
Both sight lines contain Ly$\alpha$ absorption with $\tau_{\rm{eff}}\gtrsim3\times$ the mean optical depth at $z=3$. 

%All these evidences strongly support that this absorption associated 
%with a large mass overdensity of a 15 $h^{-1}$ Mpc. 

%This is a high-priority target to conduct the imaging follow-ups. =

 J084259.37+365704.3, together with this rare absorption group at $z=3.1$, is likely to trace a massive overdensity. The median mass overdensity traced by CoSLAs with $\tau_{\rm{eff}}=4.5\times\mean{\tau_{\rm{eff}}}$ is 1.4 (Figure~\ref{fig:scatter}).

\figurenum{23}

\subsubsection{J081103.27+281621.0}

\begin{figure}[tbp]
\epsscale{1.1}
\label{fig:08+28}
\plotone{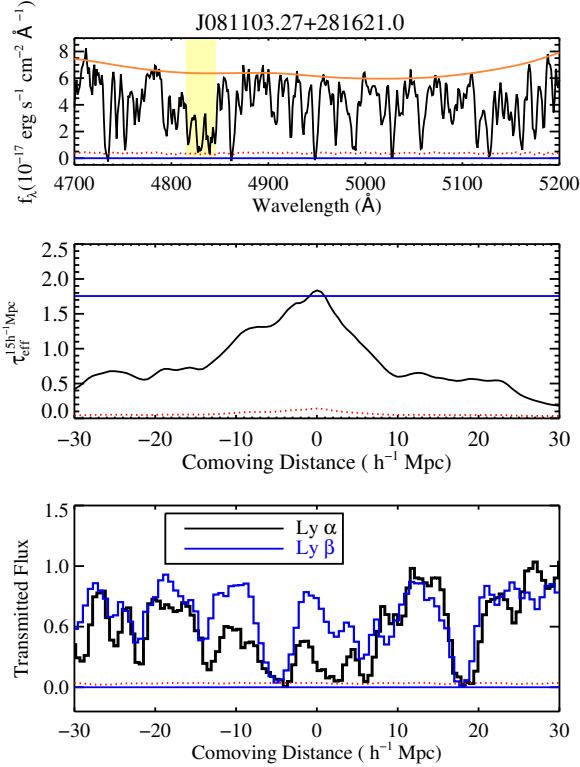}
\caption{Same format as the previous Figure~\ref{fig:02+02}, presenting the CoSLA candidate J081103.27+281621.0.  }
\end{figure}

Figure~\ref{fig:08+28} presents the spectra of CoSLA candidate J081103.27+281621.0 at $z=2.97$. The yellow shaded area of the upper panel presents the BOSS spectrum of this CoSLA candidate centered at $4830$ \AA. From the BOSS data, the effective optical depth of this 
system $\tau_{\rm{eff}}= 1.60^{+0.05}_{-0.05}$, a factor of $4.6 \times$ higher than the mean optical depth at $z=2.97$. %This effective optical depth 
%is consistent with highest effective optical depth in LyMAS scheme.
The mid-panel presents the $\tau^{15h^{-1}\rm{Mpc}}_{\rm{eff}}$ of the absorption system. Red dotted line represents the noise of the $\tau^{15h^{-1}\rm{Mpc}}_{\rm{eff}}$.
The lower panel presents the follow-up MMT observations on this 
target with a $3\times 30$ min exposure using 1200 lines mm$^{-1}$ grating.
The EW ratio between Ly$\beta$ to Ly$\alpha$ is 0.61$^{+0.02}_{-0.02}$. % FIT LYBETA, CHECK PIERI
 From the EW comparison between the 
Ly$\alpha$ (black) and Ly$\beta$ absorption (blue line), this system consists of
the blending of Ly$\alpha$ absorbers with column density $N_{\rm{HI}}= 10^{15}- 10^{18.5}$ cm$^{-2}$.

%In the zoom-in of Ly$\alpha$ absorption, black represents the Ly$\alpha$ transition, and blue overplots the corresponding Ly$\beta$ absorption. The red line is a fit of a strong Ly$\alpha$ absorber in this system. The fit has a column density  $N_{\rm{HI}}= 10^{18.6}$ cm$^{-2}$ with $b=30$ km s$^{-1}$, and this best-fit has similar equivalent width with the observed strong absorber. The brown thinner line is an overplotted Ly$\beta$ fit ($N_{\rm{HI}}= 10^{18.6}$ cm$^{-2}$). Bigger column density would yield larger Ly$\alpha$ EW and fail the Ly$\alpha$ fit, especially on wing part. 
%This fit using single Ly$\alpha$ absorber clearly deviates from the observed results. 
%These fits, together with the EW ratio between Ly$\beta$ to Ly$\alpha$, strongly suggest that this absorber consists of the blending of HI absorbers with $N_{\rm{HI}}= 10^{15}- 10^{18}$ cm$^{-2}$ (also see \citet*{pieri14}). 

The minimum transmitted flux of Ly$\alpha$ and Ly$\beta$ both reach close to zero. Figure~\ref{fig:minflux} suggests that, at a resolution of 100 km s$^{-1}$, an LLS with $N_{\rm{HI}}<10^{19}$ cm$^{-2}$ and typical Doppler parameter does not reach the zero level \citep{pieri14}. Combined with the EW ratio of Ly$\beta$ to Ly$\alpha$, the most likely scenario is that this CoSLA candidate consists of a blending of Ly$\alpha$ forest absorbers with $N_{\rm{HI}}= 10^{15}- 10^{18}$ cm$^{-2}$. 

%Figure~\ref{fig:scatter} suggests the overdensities traced by systems with $\tau_{\rm{eff}}$ have a median mass overdensity of 1.4. 
% Note the overdensity is estimated with the assumption that no LLSs are present in  J025252.07+025704.0. The presence of LLSs should bias the overdensity to lower level. 

% Again, Figure~\ref{fig:02+02xz} present the density peak is in a highly overdense sub-region, and in this density peak of ($5\ h^{-1}$ Mpc)$^3$ the overdensity $\delta_m$ could be greater than 10. 

\figurenum{24} 

\subsubsection{J113647.76+192633.9}

\begin{figure}[tbp]
\epsscale{1.1}
\label{fig:11+19}
\plotone{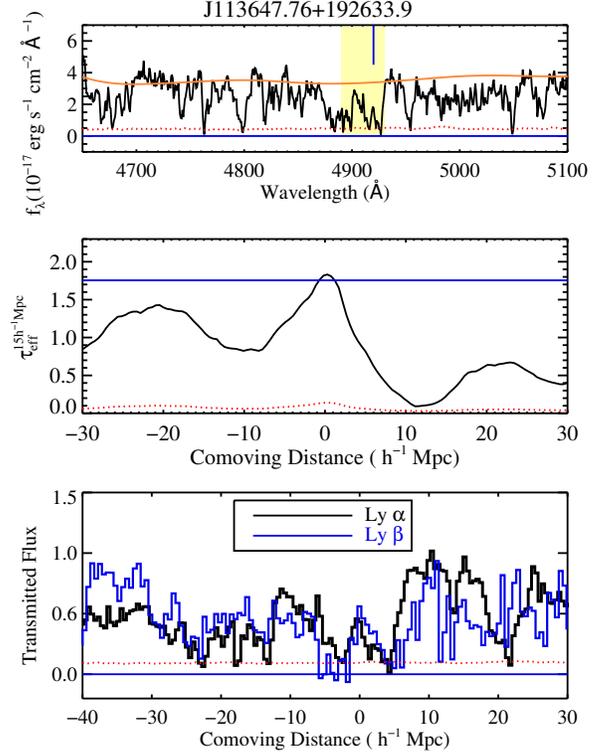}
\caption{Same format as the previous Figure~\ref{fig:02+02}, presenting the CoSLA candidate  J113647.76+192633.9. }
\end{figure}

Figure~\ref{fig:11+19} presents the CoSLA candidate J113647.76+192633.9 at $z=3.03$, which is associated with a rare quasar group at similar redshifts of $z=3.02\pm0.02$. The upper panel presents the BOSS spectrum, which contains a CoSLA candidate on 15 $h^{-1}$ Mpc 
centered at $4912$ \AA\ (yellow shaded area). Also, there is another strong Ly$\alpha$ absorption system centered at $4885$ \AA. At $4912$ \AA, from the BOSS data, the effective optical depth of this 
system is $\tau_{\rm{eff}}= 1.80^{+0.10}_{-0.10}$, a factor of $4.6 \times$ the mean optical depth at $z=3.0$. %This $\tau_{\rm{eff}}$
%is consistent with one of the highest effective optical depth systems in LyMAS scheme.
The lower panel presents the follow-up MMT observations on this 
target with a $2\times 30$ min exposure using 800 lines mm$^{-1}$ grating ($R=2,000$). 

The observed equivalent width ratio between the Ly$\beta$ to Ly$\alpha$ is $0.96^{+0.06}_{-0.06}$. Also, the Ly$\alpha$ and Ly$\beta$ both reach transmitted flux below 0.1. As with the previous 
discussions,  this system is highly likely to be consisting of 
the superposition of Ly$\alpha$ forest lines with HI column density $N_{\rm{HI}}= 10^{15}- 10^{18}$ cm$^{-2}$.

This absorption system is associated with a rare quasar group at same redshifts:  J113630.91+194337.6 at $z=3.04$ with an transverse separation of 2 $h^{-1}$ Mpc; J113653.23+192346.3 at $z=2.99$ with a transverse distance $6\ h^{-1}$ Mpc from the absorption system, and J113602.86+193557.8 at $z=3.01$, with a transverse separation of 16 $h^{-1}$ Mpc. CoSLA candidate J113647.76+192633.9, together with this rare quasar group at $z=3$, trace a massive overdensity. CoSLAs with $4.6 \times$ the mean optical depth trace structures with a median overdensity of 1.5 on 15 $h^{-1}$ Mpc scale (Figure~\ref{fig:scatter}).

\figurenum{25} 

\subsubsection{J122615.09+110543.4}

\begin{figure}[tbp]
\epsscale{1.1}
\label{fig:1226}
\plotone{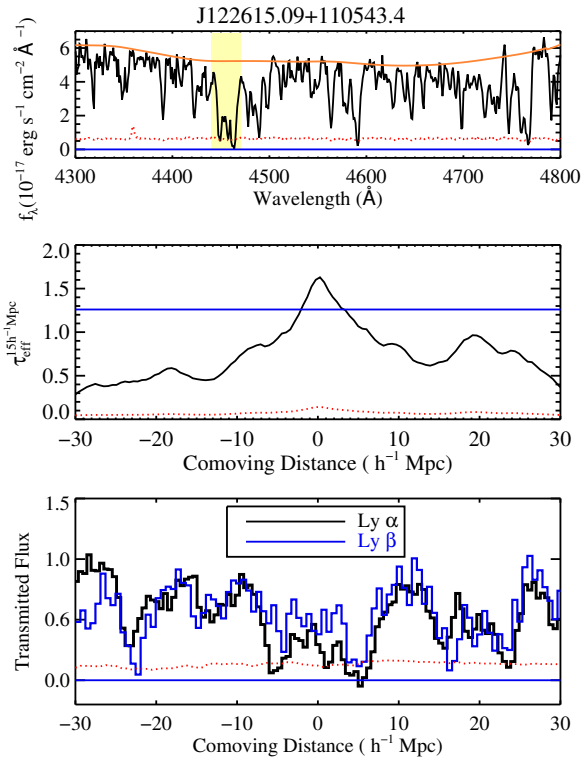}
\caption{Same format as Figure 20, showing strong Ly$\alpha$ absorption system J122615.09+110543.4. }
\end{figure}

Figure~\ref{fig:1226} presents the spectra of the strong IGM absorption system J122615.09+110543.4 at $z=2.67$. This absorption system is associated with two quasars at the same redshift with transverse separation of 30 $h^{-1}$ Mpc. 
Upper panel presents the BOSS spectrum. From this figure, the system satisfies the criteria of IGM absorption over 15 $h^{-1}$ Mpc 
centered at $4454$ \AA. %This effective optical depth 
%is consistent with highest effective optical depth in LyMAS scheme. 

The middle panel presents $\tau^{15h^{-1}\rm{Mpc}}_{\rm{eff}}$ and the lower panel shows the MMT observations on this 
absorption with $2\times 30$ min exposures using 800 lines mm$^{-1}$ grating.  This absorption has an effective optical depth $\tau_{\rm{eff}}= 1.40^{+0.15}_{-0.15}$, a factor of $5.4 \times$ higher than the mean optical depth at $z=2.66$. 
The EW ratio between Ly$\beta$ to Ly$\alpha$ is 0.69$^{+0.04}_{-0.04}$. %The width of Ly$\beta$ absorption is similar with the width of Ly$\alpha$ absorption. 
From Figure~\ref{fig:cog}, this system consists of 
the superposition of Ly$\alpha$ forest with column density $N_{\rm{HI}}= 10^{15}- 10^{18}$ cm$^{-2}$. 

This Ly$\alpha$ absorption is associated with two quasars at the same redshifts with the CoSLA candidate. Two quasars are J122535.57+110423.9 at $z=2.68$ and J122521.12+112248.4  at $z=2.67$, which have the transverse separations of 11 $h^{-1}$ Mpc and 23 $h^{-1}$ Mpc respectively from the CoSLA candidate J122615.09+110543.4. 
The CoSLA candidate J122615.09+110543.4, together with the two quasars at similar redshift, could trace a large-scale structure at $z\approx2.67$. %(Figure~\ref{fig:scatter}). 
%

%J113647.76+192633.9 corresponds to similar system at $z=2.5$ with $\tau_{\rm{eff}}= 5.4\times \mean{\tau_{\rm{eff}}(z=2.5)}= 1.35$. 
%The Absorption systems with this optical depth would trace structures over a large scale of 15 $h^{-1}$ Mpc (Figure 4). 

\figurenum{26} 

\subsection{Contaminant Absorption Systems}

Our selection criteria impose an upper limit to the effective optical depth (see Table 4), it is mainly used for eliminating DLA contaminants. Also, from the simulation with (1 $h^{-1}$ Gpc)$^3$, 
the largest overdensity is traced by an IGM absorption with effective optical depth $\tau_{\rm{eff}}\sim 1.6$ on 15 $h^{-1}$ Mpc scale, 
a factor of 7 $\times$ the mean optical depth at $z=2.5$. %Suggested by simulation, our observed candidates of IGM absorption should have effective optical depth smaller than 1.56 at 15 $h^{-1}$ Mpc, because the observed survey volume is smaller than that of the simulation. 
Systems with $\tau_{\rm{eff}}>7\times \mean{\tau}$ are most likely associated with HCDs rather than CoSLAs. 
Table 8 presents some of the contaminant absorption systems we have observed. From SDSS-III/BOSS spectra, these systems do not show evident DLA damping wings or corresponding low-ionization metal lines, but most of the systems  have $\tau_{\rm{eff}}$ greater than that of the strongest IGM absorption system suggested by our cosmological simulation.  Our follow-up MMT observations have confirmed  that all of these absorbers contain DLA contaminants. We present an example in the appendix. Our observations on these contaminants are consistent with the predictions in the LyMAS simulation. 

\subsection{Requirements for Imaging Follow-up CoSLA candidates}

After identifying the CoSLA candidates, the next step is to confirm them with follow-up imaging observations. We use the star-forming galaxies to quantify the overdensities associated with the 
extreme IGM Ly$\alpha$ absorption systems.  In this paper, we list a few observation requirements for the imaging follow-up. We introduce our imaging follow-up observations in the next paper of this series. 

Through the multi-color broadband imaging with $U$, $G$, and $R$, we can select Lyman break galaxies (LBGs) at $z=2-3.5$ (Steidel et al. 2004).  Another follow-up strategy to select the Ly$\alpha$ emitting galaxies (LAEs) to map and quantify the overdensity. Compared with the broadband selection, the narrowband selected galaxy candidates normally have higher selection efficiency on mapping the structure. But narrowband selection usually produces smaller survey volume. 
%But the proper narrowband filters are not as easily accessible, and .  

To obtain sufficient galaxies to map the overdensities, 
the depth of the broadband imaging should at least reach $L({\rm{UV}})\sim L^*(\rm{UV})$) at $z\sim 3$ (Reddy et al. 2008, Bian et al. 2013). 
Also, the depth of the narrowband (NB) imaging should reach {\it NB} $\sim$ 25 ($L({\rm{Ly\alpha}})\sim L^*(Ly\alpha)$) at $z\sim3$ (Ciardullo et al. 2012). %A wide field camera for probe galaxy overdensities over $> 20\ h^{-1}$ Mpc is ideal. % The projected separation of 
%20 $h^{-1}$ co-Mpc corresponds to a field of view (FoV) of $\sim 30'\times 30'$ at $z\sim2.5-3$. 
Ideally, sufficiently wide field cameras with field of views of $\gtrsim 30'$ are preferred for conducting the imaging observations to quantify the galaxy overdensity on $\gtrsim 20 \ h^{-1}$ Mpc. 
%Therefore, multi-color imaging with cameras, such as LBT/LBC, Subaru/(Hyper-)Suprime-Cam, KPNO-4m/MOSAIC, CTIO-4m/DeCAM, etc., are preferred. 
After deep imaging, multi-slit spectroscopy (MOS) observations are needed to fully map and quantify the massive overdensities. %The bright LBGs to provide other sight lines to conduct IGM tomography (Lee et al. 2013, 2014).  

We have carried out narrowband (NB403)+ broadband (Bw) imaging on multiple fields traced by CoSLAs using the 
KPNO-4m Mayall and LBT/LBC. We have used LBT/MODS to spectroscopically confirm one of these overdense fields. We will present these 
observational results in the next paper of this series (Cai et al. 2016 in prep.).

\section{Summary and Discussions}

Local galaxy clusters are identified by overdensities of galaxies, dark matter and hot ICM \cite[e.g.,][]{fabian06}. 
Galaxy kinematics and gravitational lensing studies show that clusters of galaxies are embedded in a massive 
halo of dark matter with mass $\gtrsim 10^{14}$ M$_\odot$ \cite[e.g.,][]{carlberg97}. 
The progenitors of such galaxy clusters can be identified at high redshifts at $z>2$, 
given large-scale ($\sim10-40\ h^{-1}$  Mpc) density contrasts 
compared to random fields \cite[e.g.,][]{hu96, steidel98, ouchi05, matsuda05, venemans07, matsuda10}. 

At high-redshift, galaxies are believed to interact with surrounding IGM, and galaxies assemble their gas from the intergalactic HI gas \citep{adelberger05, frye08, matsuda10, rudie12, tejos14}. 
A large-scale mass overdensity is associated with an HI reservoir in the IGM, which in turn traces a large galaxy overdensity. 

 In this paper, we systematically study the correlation between mass overdensities and Ly$\alpha$ absorption 
on scales of $\sim 10$ -- $20\ h^{-1}$ Mpc, which scales correspond to typical extents of the large-scale galaxy overdensities at $z>2$ \cite[e.g.,][]{steidel98, steidel05, matsuda05, ouchi08}. 
%(e.g., Steidel et al. 1999, 2005; Ouchi et al. 2007; Matsuda et al. 2010, also see  Figure~\ref{fig:example}). The main conclusion for this paper are as follows: 

(a) Our cosmological simulations suggest that a strong correlation exists between the mass and Ly$\alpha$ transmitted 
flux over large scales, and this correlation peaks at scales of 10- 30 $h^{-1}$ Mpc (see \S3). Using the SDSS-III/BOSS quasar dataset, 
we have confirmed that the group of Ly$\alpha$ absorption systems exist in two well-studied overdense fields: SSA22 and Jackpot nebula fields. 

(b) We focus on the study of Coherently Strong Ly$\alpha$ Absorption systems (CoSLAs). These CoSLAs have highest $\tau_{\rm{eff}}$ on $\sim 15\ h^{-1}$ Mpc. These absorptions are due to intergalactic HI overdensity rather than high column density absorbers (e.g., DLAs and sub-DLAs) that arise from ISM/CGM of  galaxies (\S3). %They are tracing most massive mass overdensities at high-redshift (\S4, Figure~\ref{fig:massLyMAS15}, Figure~\ref{fig:massLyMAS20}, Figure~\ref{fig:mass15_afterHCD}). 

\begin{figure}[tbp]
\epsscale{1}
\label{fig:massLyMAS20}
\plotone{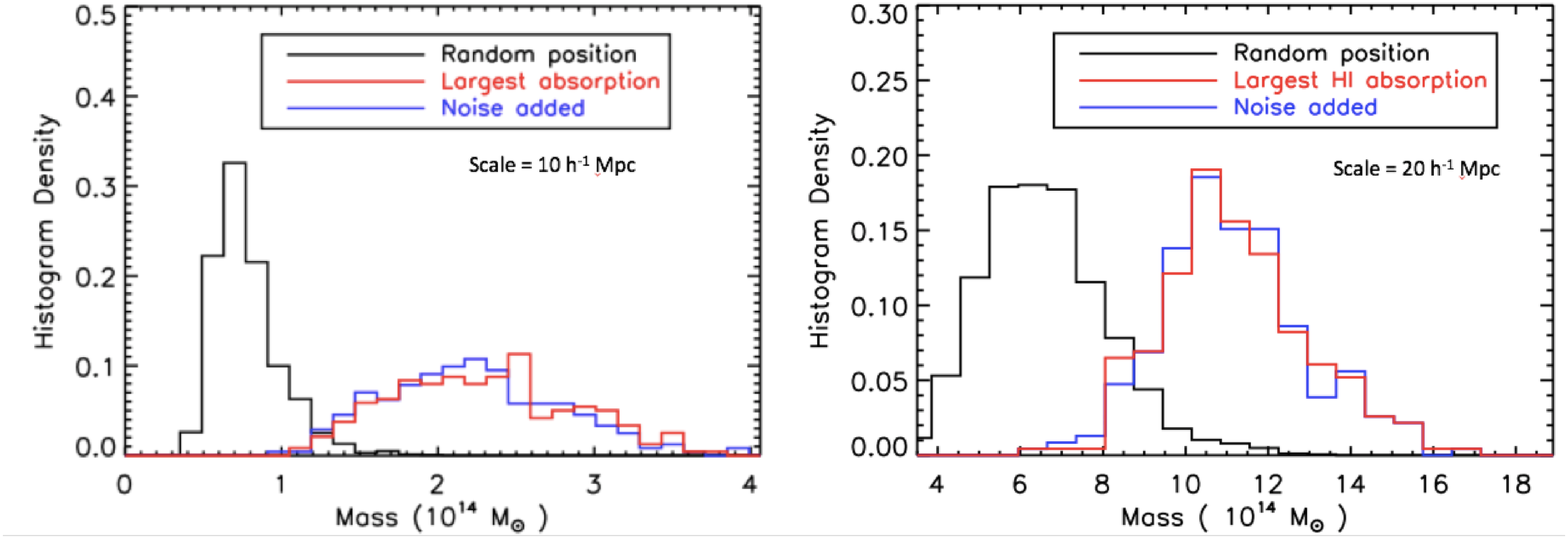}
\caption{The distribution of mass traced by CoSLAs on 10 $h^{-1}$ Mpc (Left) and 20 $h^{-1}$ Mpc (Right)
scale in LyMAS simulation. In both panels, black histogram represents mass centered on random positions. Red represents CoSLAs selected from the original mock spectra (no noise added). Blue shows mass traced by CoSLAs selected from noise-added mock spectra, with continuum-to-noise ratio (CNR) of 4 per pixel. 
The figure shows that the CoSLAs effectively trace a 3-D large-scale structures. On 10 $h^{-1}$ Mpc (Left),  most of the overdensities  traced by CoSLAs contain mass a factor of $3.6\times$ cosmic mean, 
representing 4-$\sigma$ mass overdensities. On 20 $h^{-1}$ Mpc, CoSLAs trace mass a factor of $1.7\times$ cosmic mean, 
representing 3-$\sigma$ mass overdensities. }
\end{figure}

(c) Simulations over a survey volume of (1 $h^{-1}$ Gpc)$^3$ show that the CoSLAs that have effective optical depths $\ge 4.5\times$ the mean optical depth on 15 $h^{-1}$ Mpc, corresponding to systems beyond $4\sigma$ in the optical depth distribution. Most of these absorption systems trace structures with mass overdensities of $\delta_m>1.6$, $>$ 3.3-$\sigma$ beyond the density fluctuation in random fields (Figure~\ref{fig:massLyMAS15}, Figure~\ref{fig:massLyMAS20}, Figure~\ref{fig:mass15_afterHCD}). From our simulations, the density peak regions in these systems have high mass overdensities of 10 over $\sim (5-6\ h^{-1})$ Mpc$^3$ (Figure~\ref{fig:example}). 
In particular, suggested by simulation, the CoSLAs have an upper limit of $\tau_{\rm{eff}}\approx 1.6$, $7-8\times \mean{\tau_{\rm{eff}}}$ over the scale of $15 \ h^{-1}$ Mpc. In a (1~$h^{-1}$ Gpc)$^3$ volume, any absorber with $\tau_{\rm{eff}}$ higher than this upper limit should be associated with high column density absorption systems (HCDs). 

 %(CoSLAs) largest 1-D Ly$\alpha$ absorption systems from currently SDSS-III/BOSS largest Ly$\alpha$ survey (see \S 4). 

(d) Guided by our simulation, we have developed techniques to select these CoSLAs from BOSS spectra. With the absorption trough, absorption wing, corresponding metal lines and Ly$\beta$ absorption, we can effectively select the CoSLAs, and effectively rule out contaminant DLAs and sub-DLAs. These selection criteria work best for absorbers with $z\ge2.65$, where corresponding Ly$\beta$ absorption is covered with BOSS spectra (\S4).  However, the clustered LLSs could also contribute a significant amount of contaminants. We estimate that the 
clustered LLSs could lower our CoSLA selection efficiency to $\approx 40\%$, and better technique to differentiate CoSLAs from clustered LLSs are needed in the future. 

We also propose that with the selection of the Ly$\alpha$ absorption groups, one can effectively 
pinpoint massive galaxy protoclusters (Figure~\ref{fig:mass15_afterHCD}), without the examining the nature of Ly$\beta$ absorption. 
At $z\lesssim2.35$, the average quasar density at 2-D is high enough for us to search for overdensities using such technique (see \S4).

(e)  Based on the selection criteria we proposed in \S4, we have selected a sample of CoSLA candidates from SDSS-III/BOSS, by examining the absorption spectra of $\sim 6,000$ sight lines provided by SDSS-III quasar survey at $z=2.6$ -- $3.3$ with CNR $\gtrsim 8$ (see \S6). In this paper, we present a sample of 5 CoSLA candidates. 5 of them were observed using MMT observations with higher SNR and resolution.  % The high optical depth these systems are suggested to arise from the IGM overdensities.  
These absorption systems are consistent with the predictions in LyMAS simulation, and are expected to pinpoint massive overdensities over $\sim 15\ h^{-1}$ Mpc. %The survey volume is about 0.15 $h^{-1}$ Gpc$^3$. 

{\bf Acknowledgement: } ZC, XF, BF and IM thank the support from the US NSF grant AST 11-07682. ZC and JXP acknowledge support from NSF AST-1412981. 
ZC acknowlendges the insightful comments from Nobunari Kashikawa, Martin White, Linhua Jiang,  Ann Zabludoff, 
Richard Green, Ran Wang, Brant Robertson, Masami Ouchi, Daniel Stark, Zhenya Zheng and Yun-Hsin Huang. 
Funding for SDSS-III has been provided
by the Alfred P. Sloan Foundation, the Participating Institutions,
the National Science Foundation, and the U.S. Department
of Energy Office of Science. The SDSS-III web site is
http://www.sdss3.org/. SDSS-III is managed by the Astrophysical
Research Consortium for the Participating Institutions of
the SDSS-III Collaboration including the University of Arizona,
the Brazilian Participation Group, Brookhaven National
Laboratory, University of Cambridge, Carnegie Mellon University,
University of Florida, the French Participation Group,
the German Participation Group, Harvard University, the Instituto
de Astrofisica de Canarias, the Michigan State/Notre
Dame/JINA Participation Group, Johns Hopkins University,
Lawrence Berkeley National Laboratory, Max Planck Institute
for Astrophysics, Max Planck Institute for Extraterrestrial
Physics, New Mexico State University, New York University, Ohio State University, Pennsylvania State University, University
of Portsmouth, Princeton University, the Spanish Participation
Group, University of Tokyo, University of Utah, Vanderbilt
University, University of Virginia, University of Washington,
and Yale University.  Observations reported here were obtained at the MMT Observatory, a joint facility of the University of Arizona and the Smithsonian Institution.

%\newpage
\paragraph{}

\clearpage

\begin{deluxetable*}{ccccccccccc} 
\tablecolumns{12}
%\longtable
\tablewidth{0pt}
\tablecaption{Summary of the targets satisfy criteria (a)- (d1) and have MMT follow-up observations}
\label{table:F130N_S}    
\tablehead{\colhead{Name} &
                  \colhead{$z_{\rm{abs}}$} &
                  %\colhead{$\mean{\tau_{\rm{eff}}}$} &
                  \colhead{Scale} &
                  \colhead{SDSS} &
                  \colhead{SDSS} &
                  \colhead{MMT} &
                  \colhead{MMT} &
                  \colhead{MMT} &
                  \colhead{MMT} &
                  \colhead{MMT}   \\
                  \colhead{} &
                  \colhead{} &
                  %\colhead{$z=z_{\rm{abs}}$} &
                  \colhead{($h^{-1}$ Mpc)} &
                  \colhead{CNR} &
                  \colhead{$\tau_{\rm{eff}}$} & 
                  \colhead{$\tau_{\rm{eff}}$} &
                  \colhead{Resolution} &
                  \colhead{seeing} &
                  \colhead{Exptime (min)} &
                  \colhead{CNR}  }
\startdata
%J023309.60-003100.3 &  3.46 & 15 & 6 & $1.90^{+0.06}_{-0.05}$ & ... & 2,000 & 1.4'' & $3\times 30$ & 10  \nl
J025252.07+025704.0 & 2.91  & 15 & 11 & $1.60^{+0.10}_{-0.10}$ & $1.61^{+0.03}_{-0.03}$ & 3,000 & 1.0''
& $3\times 20$ & 30 \nl
J081103.27+281621.0 &  2.97 & 15 & 13 & $1.80^{+0.07}_{-0.07}$ & $1.78^{+0.02}_{-0.02}$ & 3,000 & 1.2'' & $3\times 20$ & 20 \nl
J084259.37+365704.3 & 3.17  & 15 &  8 & $2.43^{+0.10}_{-0.10}$ & $2.47^{+0.05}_{-0.05}$ & 3,000 & 1.0'' & $3\times 30 $ & 30 \nl 
J113647.76+192633.9  &   3.01  & 15 & 10 & $1.82^{+0.05}_{-0.05}$ & $1.80^{+0.06}_{-0.06}$ & 3,000 & 1.5'' & $3\times 20$ & 15\nl
J122615.09+110543.4 & 2.67 & 15 & 5 & $1.60^{+0.15}_{-0.15} $ &  $1.67^{+0.15}_{-0.15}$ & 3,000 & 1.5'' & $2\times 30$ & 8 
\enddata
\end{deluxetable*}

\begin{deluxetable*}{lccc} 
\tablecolumns{4}
%\longtable
\tablewidth{0pt}
\tablecaption{Properties of the CoSLA candidates that have MMT observations}
\label{table:F130N_S}    
\tablehead{\colhead{Name} &
                  \colhead{$\tau_{\rm{eff}}$} & 
                  \colhead{median overdensity from simulation}  &
                  \colhead{comments}  \\
                  \colhead{} &
                  \colhead{$\mean{\tau_{\rm{eff}}}$} & 
                  \colhead{over a large scale of $15\ h^{-1}$ Mpc} &
                  \colhead{} }
\startdata
%J023309.60-003100.3 &  4.8 &  1.5 &\nl
J025252.07+025704.0 & 4.6  & 1.5 & \nl
J081103.27+281621.0 &  4.5 & 1.4 & \nl
J084259.37+365704.3 &  4.6 & 1.4  &  associated with two absorption within 20 $h^{-1}$ Mpc\nl 
J113647.76+192633.9  &  5.6 &  1.8 &  associated with two quasars within 20 $h^{-1}$ Mpc\nl
J122615.09+110543.4 & 5.4 & 1.7 & associated with group of three quasars within 20 $h^{-1}$ Mpc 
\enddata
\footnotetext[1]{The overdensity is estimated with assuming the CoSLA candidates are genuine CoSLAs. The presence of LLSs should lower the estimation of mass overdensity. }
\end{deluxetable*}

\begin{deluxetable*}{ccccccccccc} 
\tablecolumns{11}
%\longtable
\tablewidth{0pt}
\tablecaption{Summary of the contaminating DLAs that satisfies criteria (a)-(c)}
\label{table:F130N_S}    
\tablehead{\colhead{Name} &
                  \colhead{$z_{\rm{abs}}$} &
                  \colhead{Scale} &
                  \colhead{SDSS} &
                  \colhead{SDSS} &
                  \colhead{MMT} &
                  %\colhead{MMT} &
                  %\colhead{MMT} &
                  \colhead{MMT} &
                  \colhead{MMT} &
                  \colhead{Column density}  \\
                  \colhead{} &
                  \colhead{} &
                  \colhead{(Mpc)} &
                  \colhead{CNR} &
                  \colhead{$\tau_{\rm{eff}}$} & 
                  %\colhead{$\tau_{\rm{eff}}$} &
                  %\colhead{Resolution} &
                  \colhead{seeing} &
                  \colhead{Exptime (min)} &
                  \colhead{CNR} & 
                  \colhead{}  }
\startdata
J010349.82+032856.1   &  2.64  & 40 & 8 & $1.6^{+0.1}_{-0.1}$  & 1.7'' & $3\times 20$ & 10  & $10^{20.2}$\nl
J021222.01+042745.3 &  2.29  &  30 &  5 & $2.0^{+0.2}_{-0.2}$  & 1.5'' & $6\times 20$ & 12 & $10^{20.7}$\nl
J081453.64+392828.6 &  2.21 & 40 & 4 & 2.5$^{+2.4}_{-0.7}$  & 1.0'' & $3\times20$ & 6 & $10^{20.9}$\nl
J091813.67+205623.7 &  2.41  & 40 & 5 & 4.5$\pm$1.0   & 1.0'' & $6\times20$ & 20 & $10^{21.7}$ \nl
J104033.69+355247.9 &  2.22  & 40 & 3 & $2.7^{+1.2}_{-0.6}$ & 1.2''&  $3\times20$ &10  &  $10^{21.0}$ \nl
J131956.21+363624.1 &  2.24  & 30 & 3 & $1.7^{+0.4}_{-0.4}$ & 1.1'' & $3\times30$ & 7  & $10^{20.2}$ \nl
J145337.15+000410.0 &  2.49  & 30 & 10 & $1.1^{+0.06}_{-0.06}$  & 1.0'' &  $1\times30$ & 10 & $10^{19.8}$ \nl
J143003.15+065719.1 &  2.6 & 40 & 3 & $2.2^{+0.8}_{-0.5}$  & 1.5'' & $3\times 20$ & 5 &  $10^{20}$ \nl
J154511.76+165630.5 &  2.44 & 30 & 3.5& $1.6^{+0.5}_{-0.5}$ & 1.2''&  $3\times20$ &10  & $10^{20.5}$  %\nl
%J161052.40+362333.1 &  2.72 & 0.30 & 30 & 5 & $1.8^{+0.2}_{-0.2}$  & 1.4'' & $3\times 30$ & 8 & $10^{20.3}$ \nl
%J161034.51+285926.4 &  2.72 & 0.30 & 30 & 4 & $2.5^{+1.0}_{-0.5}$ & 1.7'' & $3\times 30$ & 5 & $10^{20.7}$ 
\enddata
\end{deluxetable*}

{\section{Appendix}}

{\subsection{Mass Distribution on 10 $h^{-1}$ and 20 $h^{-1}$ Smoothing Scale}}

In this appendix, we present the mass traced by coherently strong Ly$\alpha$ absorption on 10 $h^{-1}$ Mpc and 20 $h^{-1}$ Mpc scales. 

On 10 $h^{-1}$ and 20 $h^{-1}$ Mpc, mass overdensities can be effectively traced by coherently strong Ly$\alpha$ absorption.  The Left and right panels of Figure~\ref{fig:massLyMAS20} (Appendix) show mass distributions traced by strong Ly$\alpha$ absorption at the scales of 10 $h^{-1}$ Mpc and 20 $h^{-1}$ Mpc in the LyMAS simulation, respectively. 
Again, we choose systems with lowest transmitted flux defined in the second paragraph of \S2.2 (systems beyond 4.5-$\sigma$ in the $\tau_{\rm{eff}}$ distribution). Similar to Figure~\ref{fig:massLyMAS15}, red represents mass traced by CoSLAs with highest $\tau_{\rm{eff}}$ selected from the original mock spectra (no noise added). Blue shows CoSLAs selected from noise-added mock spectra, with a CNR of 4 per pixel.

\figurenum{27}
\begin{figure}[tbp]
\epsscale{1}
\label{fig:tau_distri_20}
\plotone{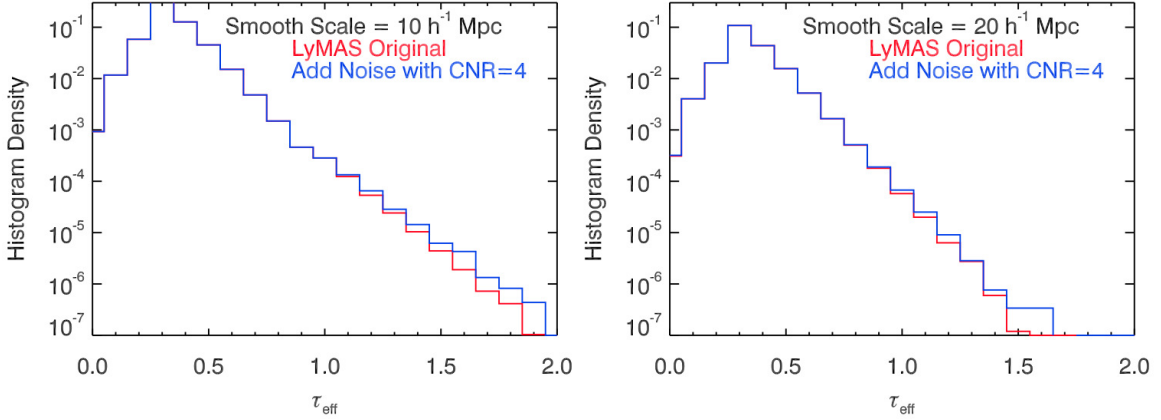}
\caption{Similar plot with Figure 5. Left: the distribution of $\tau_{\rm{eff}}$ on the scale of 10 $h^{-1}$ Mpc. Right: the distribution of $\tau_{\rm{eff}}$ on 20 $h^{-1}$ Mpc. }
\end{figure}

\figurenum{28}

\begin{figure}[tbp]
\epsscale{0.7}
\label{fig:massRunA}
\plotone{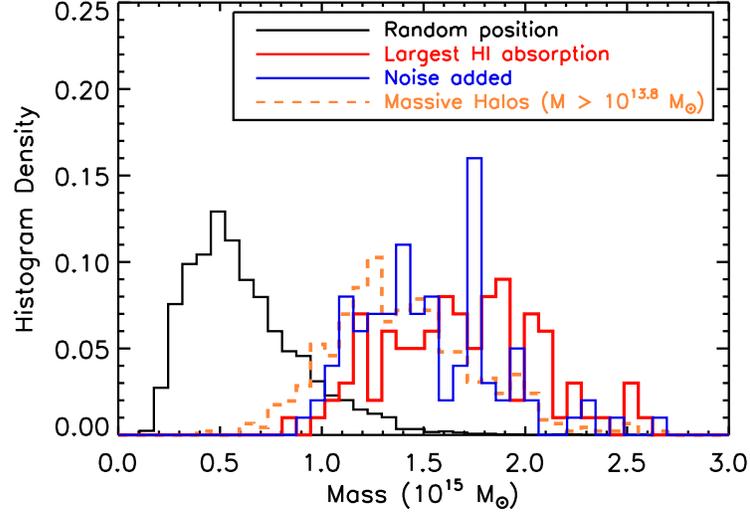}
\caption{Mass distribution within 20 $h^{-1}$ Mpc scale from the deterministic simulation with (1.5 $h^{-1}$Gpc)$^3$. x-axis 
is the mass within the 20 $h^{-1}$ Mpc cube. The y-axis is the number of the cubes. 
Black histogram presents random distribution. 
Yellow shows  mass distribution centered on the most massive halos 
(M$_{\rm{halo}}> 10^{13.8}\ \rm{M}_\odot$).  
 Red is the mass traced by the CoSLAs on 20 $h^{-1}$ Mpc scale, selected 
from the original mock spectra without noise. 
%More than half of the largest Ly$\alpha$ absorption 
%traces the top 0.5\%  most massive overdensities (4-$\sigma$) at 20 $h^{-1}$ Mpc scale. 
Blue is largest Ly$\alpha$ absorption selected from the noise-added mock spectra. The noise is added according to a continuum-to-noise ratio (CNR) of 4.}
\end{figure}

The deterministic simulation has a larger simulation box with 1.5 $h^{-1}$ Gpc.
In the deterministic scheme, Figure~\ref{fig:massRunA} presents similar results that IGM Ly$\alpha$ absorption systems most effectively trace the overdensities over a large scale. Using deterministic simulation, we study the mass overdensities traced by CoSLAs. 

In Figure~\ref{fig:massRunA}, black represents the cubes centered at the random positions in the simulation box: the cosmic mean mass in a 20 $h^{-1}$ Mpc cube is $5\times 10^{14}\ M_\odot$ with a standard deviation of $2.3\times10^{14} \ M_\odot$ on the logarithmic scale.
Yellow represents the mass within 20 $h^{-1}$ Mpc which is centered on the most massive halo in a volume of (1.5 $h^{-1}$ Gpc)$^3$, with M$_{\rm{halo}}> 10^{13.8}\ \rm{M}_\odot$, 0.2 dex larger than the most massive halos in LyMAS simulation because of bigger box volume. %a factor of $\sim$ 30 $\times$ more massive than BOSS quasar).  

Red and blue lines present cases where masses are traced by the CoSLAs selected from (1.5~$h^{-1}$ Gpc)$^3$ box. Red presents mass overdensities traced by CoSLAs at $20\ h^{-1}$ Mpc selected from original mock spectra, without adding noise. 
More than half of the largest Ly$\alpha$ absorption 
traces the top 0.2\%  most massive overdensities ($>$ 3.0-$\sigma$) on 20 $h^{-1}$ Mpc scale. 
Similar to red, blue presents mass distribution traced by CoSLAs selected from noise-added mock spectra with CNR of 4 per pixel. 

Therefore, both simulations 
support that extreme mass overdensities over $\sim 10\ h^{-1}-  20\ h^{-1}$ Mpc can be traced by largest Ly$\alpha$ absorption systems.

\figurenum{29}

{\subsection{Example of  the Contaminant DLAs}}

\begin{figure}[tbp]
\epsscale{0.8}
\label{fig:16+36}
\plotone{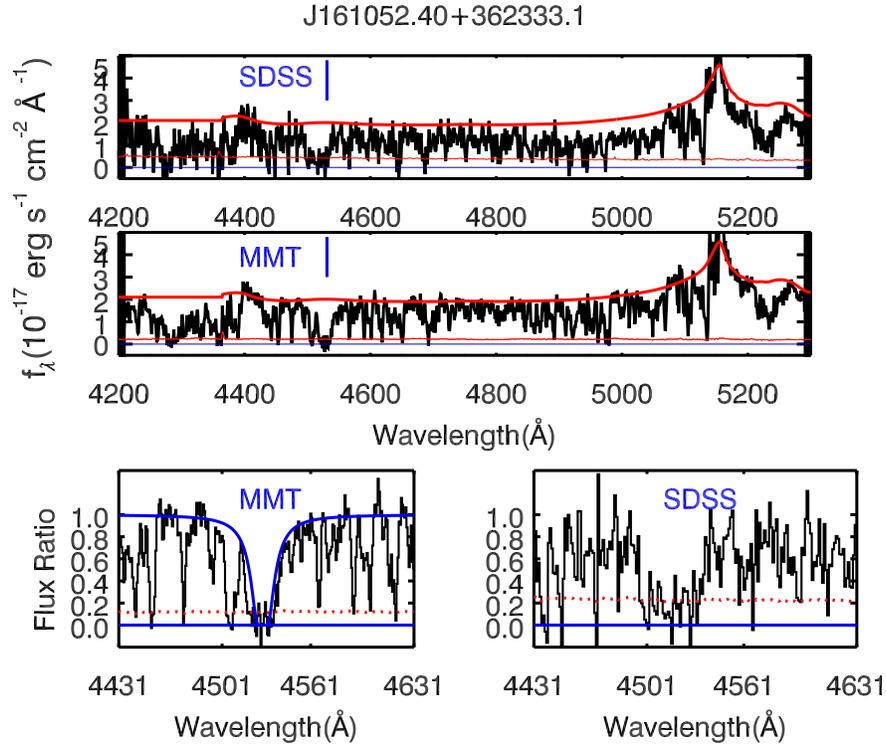}
\caption{The upper panel presents the SDSS-III spectrum.  From the SDSS-III, the effective optical depth of this system $\tau_{\rm{eff}}= 1.8^{+0.2}_{-0.2}$, $6\times$ the mean optical depth at $z=2.7$. The mid-panel presents the spectrum using MMT blue channel with 800 lines/mm ($R=2,000$). Lower left shows the zoom-in of Ly$\alpha$ absorption. 
 The high S/N MMT observation clearly indicate this absorption system is largely contributed by a sub-DLA with column density $N_{\rm{HI}}= 10^{20.0}$ cm$^{-2}$. Also, 
a strong Ly$\alpha$ absorption shows up in the left wing of this sub-DLA. }
\end{figure}

Figure~\ref{fig:16+36} presents an example of contaminants. The complete list of contaminants DLAs that satisfies criteria (a)-(c).  The upper panel presents the SDSS-III spectrum.  This
absorption system in SDSS-III well satisfies the w-t criteria centered at $4520$ \AA: (1) $w_{\rm{0.8}}= 50$ \AA, $<$ the threshold of 70 \AA, and (2) the profile for absorption wing is sharp and $w_{\rm{0.5}}/w_{\rm{0.8}}> 0.9$, well 
meet the requirement of second criterion (\S 4.1). From the SDSS-III spectra, the effective optical depth of this system $\tau_{\rm{eff}}= 1.8^{+0.2}_{-0.2}$, $6\times$ the mean optical depth at $z=2.7$. The mid-panel presents the spectrum using MMT blue channel with 800 lines/mm ($R=2,000$). Lower left shows the zoom-in of Ly$\alpha$ absorption. 
 The high S/N MMT observation clearly indicate this absorption system is largely contributed by a sub-DLA with column density $N_{\rm{col}}= 10^{20.0}$ cm$^{-2}$.  

\figurenum{30}


\begin{thebibliography}{84}
\expandafter\ifx\csname natexlab\endcsname\relax\def\natexlab#1{#1}\fi


\bibitem[Adelberger et al.(2003)]{adelberger03} Adelberger, K.~L., 
Steidel, C.~C., Shapley, A.~E., \& Pettini, M.\ 2003, \apj, 584, 45 


\bibitem[Adelberger et al.(2005)]{adelberger05} Adelberger, K.~L., 
Shapley, A.~E., Steidel, C.~C., et al.\ 2005, \apj, 629, 636 


\bibitem[Ahn et al.(2014)]{ahn14} Ahn, C.~P., Alexandroff, 
R., Allende Prieto, C., et al.\ 2014, \apjs, 211, 17 


\bibitem[Becker et al.(2013)]{becker13} Becker, G.~D., Hewett, 
P.~C., Worseck, G., \& Prochaska, J.~X.\ 2013, \mnras, 430, 2067 



\bibitem[Bi(1993)]{bi93} Bi, H.\ 1993, \apj, 405, 479 

\bibitem[Bian et al.(2013)]{bian13} Bian, F., Fan, X., Jiang, 
L., et al.\ 2013, \apj, 774, 28 


\bibitem[Bird et al.(2014)]{bird14} Bird, S., Vogelsberger, 
M., Haehnelt, M., et al.\ 2014, \mnras, 445, 2313 


\bibitem[Bolton et al.(2009)]{bolton09} Bolton, J.~S., Oh, 
S.~P., \& Furlanetto, S.~R.\ 2009, \mnras, 396, 2405 


\bibitem[Bolton et al.(2012)]{bolton12} Bolton, A.~S., Schlegel, 
D.~J., Aubourg, {\'E}., et al.\ 2012, \aj, 144, 144 



\bibitem[Busca et 
al.(2013)]{busca13} Busca, N.~G., Delubac, T., Rich, J., et al.\ 2013, \aap, 552, A96 


\bibitem[Cai et al.(2014)]{cai14} Cai, Z., Fan, X., 
Noterdaeme, P., et al.\ 2014, \apj, 793, 139 


\bibitem[Carlberg et al.(1997)]{carlberg97} Carlberg, R.~G., Yee, 
H.~K.~C., Ellingson, E., et al.\ 1997, \apjl, 485, L13 


\bibitem[Cantalupo et al.(2014)]{cantalupo14} Cantalupo, S., 
Arrigoni-Battaia, F., Prochaska, J.~X., Hennawi, J.~F., 
\& Madau, P.\ 2014, \nat, 506, 63 


\bibitem[Cen et al.(1994)]{cen94} Cen, R., 
Miralda-Escud{\'e}, J., Ostriker, J.~P., \& Rauch, M.\ 1994, \apjl, 437, L9 

\bibitem[Cen et al.(2003)]{cen03} Cen, R., Ostriker, J.~P., 
Prochaska, J.~X., \& Wolfe, A.~M.\ 2003, \apj, 598, 741 

\bibitem[Chapman et al.(2004)]{chapman04} Chapman, S.~C., Scott, 
D., Windhorst, R.~A., et al.\ 2004, \apj, 606, 85 


\bibitem[Chiang et al.(2013)]{chiang13} Chiang, Y.-K., Overzier, 
R., \& Gebhardt, K.\ 2013, \apj, 779, 127 



\bibitem[Chiang et al.(2014)]{chiang14} Chiang, Y.-K., Overzier, 
R., \& Gebhardt, K.\ 2014, \apjl, 782, L3 


\bibitem[Cooke et al.(2006)]{cooke06} Cooke, J., Wolfe, A.~M., 
Gawiser, E., \& Prochaska, J.~X.\ 2006, \apj, 652, 994 





\bibitem[Dawson et al.(2013)]{dawson13} Dawson, K.~S., Schlegel, 
D.~J., Ahn, C.~P., et al.\ 2013, \aj, 145, 10 

\bibitem[Dall'Aglio et 
al.(2008)]{dallaglio08} Dall'Aglio, A., Wisotzki, L., \& Worseck, G.\ 2008, \aap, 491, 465 

\bibitem[Delubac et 
al.(2015)]{delubac15} Delubac, T., Bautista, J.~E., Busca, N.~G., et al.\ 2015, \aap, 574, A59 



\bibitem[Eisenstein et al.(2011)]{eisenstein11} Eisenstein, D.~J., 
Weinberg, D.~H., Agol, E., et al.\ 2011, \aj, 142, 72 


\bibitem[Fabian et al.(2006)]{fabian06} Fabian, A.~C., Sanders, 
J.~S., Taylor, G.~B., et al.\ 2006, \mnras, 366, 417 


\bibitem[Faucher-Gigu{\`e}re et al.(2008)]{faucher-giguere08} 
Faucher-Gigu{\`e}re, C.-A., Lidz, A., Hernquist, L., 
\& Zaldarriaga, M.\ 2008, \apjl, 682, L9 



\bibitem[Font-Ribera et al.(2012)]{font-ribera12} Font-Ribera, A., 
Miralda-Escud{\'e}, J., Arnau, E., et al.\ 2012, JCAP, 11, 059 


\bibitem[Ford et al.(2013)]{ford13} Ford, A.~B., Oppenheimer, 
B.~D., Dav{\'e}, R., et al.\ 2013, \mnras, 432, 89 


\bibitem[Font-Ribera et al.(2013)]{font13} Font-Ribera, A., 
Arnau, E., Miralda-Escud{\'e}, J., et al.\ 2013, JCAP, 5, 018 


\bibitem[Font-Ribera et al.(2012)]{font12} Font-Ribera, A., 
Miralda-Escud{\'e}, J., Arnau, E., et al.\ 2012, JCAP, 11, 059 



\bibitem[Frye et al.(2008)]{frye08} Frye, B.~L., Bowen, D.~V., 
Hurley, M., et al.\ 2008, \apjl, 685, L5 


\bibitem[Fumagalli et al.(2015)]{fumagalli15} Fumagalli, M., 
O'Meara, J.~M., Prochaska, J.~X., Rafelski, M., 
\& Kanekar, N.\ 2015, \mnras, 446, 3178 


\bibitem[Gunn 
\& Peterson(1965)]{gunn65} Gunn, J.~E., \& Peterson, B.~A.\ 1965, \apj, 142, 1633 



\bibitem[Gunn et al.(2006)]{gunn06} Gunn, J.~E., Siegmund, 
W.~A., Mannery, E.~J., et al.\ 2006, \aj, 131, 2332 


\bibitem[Hennawi et al.(2015)]{hennawi15} Hennawi, J.~F., 
Prochaska, J.~X., Cantalupo, S., 
\& Arrigoni-Battaia, F.\ 2015, Science, 348, 779 


\bibitem[Hu et al.(1996)]{hu96} Hu, E.~M., McMahon, R.~G., 
\& Egami, E.\ 1996, \apjl, 459, L53 

\bibitem[Khare et 
al.(2007)]{khare07} Khare, P., Kulkarni, V.~P., P{\'e}roux, C., et al.\ 2007, \aap, 464, 487 


\bibitem[Kirkman et al.(2005)]{kirkman05} Kirkman, D., Tytler, 
D., Suzuki, N., et al.\ 2005, \mnras, 360, 1373 


\bibitem[Kollmeier et al.(2003)]{kollimeier03} Kollmeier, J.~A., 
Weinberg, D.~H., Dav{\'e}, R., \& Katz, N.\ 2003, \apj, 594, 75 


\bibitem[Le F{\`e}vre et al.(2014)]{fevre14} Le F{\`e}vre, O., 
Adami, C., Arnouts, S., et al.\ 2014, The Messenger, 155, 33 


\bibitem[Lee et al.(2012)]{lee12} Lee, K.-G., Suzuki, N., 
\& Spergel, D.~N.\ 2012, \aj, 143, 51 


\bibitem[Lee et al.(2013)]{lee13} Lee, K.-G., Bailey, S., 
Bartsch, L.~E., et al.\ 2013, \aj, 145, 69 



\bibitem[Lee et al.(2014)]{lee14} Lee, K.-G., Hennawi, J.~F., 
White, M., Croft, R.~A.~C., \& Ozbek, M.\ 2014, \apj, 788, 49 


\bibitem[Lee et al.(2014)]{leeK14} Lee, K.-S., Dey, A., Hong, 
S., et al.\ 2014, \apj, 796, 126 



\bibitem[Lee et al.(2015)]{lee15} Lee, K.-G., Hennawi, J.~F., 
Spergel, D.~N., et al.\ 2015, \apj, 799, 196 



\bibitem[Le Fevre et al.(2014)]{lefevre14} Le Fevre, O., Tasca, 
L.~A.~M., Cassata, P., et al.\ 2014, arXiv:1403.3938 


\bibitem[Lehner et al.(2013)]{lehner13} Lehner, N., Howk, J.~C., 
Tripp, T.~M., et al.\ 2013, \apj, 770, 138 



\bibitem[Liske et al.(2000)]{liske00} Liske, J., Webb, J.~K., 
Williger, G.~M., Fern{\'a}ndez-Soto, A., 
\& Carswell, R.~F.\ 2000, \mnras, 311, 657 


\bibitem[Lochhaas et al.(2015)]{lochhaas15} Lochhaas, C., 
Weinberg, D.~H., Peirani, S., et al.\ 2015, arXiv:1511.04454 



\bibitem[Lynds(1971)]{lynds71} Lynds, R.\ 1971, \apjl, 164, L73 



\bibitem[McDonald et al.(2005)]{mcdonald05} McDonald, P., Seljak, 
U., Cen, R., et al.\ 2005, \apj, 635, 761 



\bibitem[Matsuda et al.(2004)]{matsuda04} Matsuda, Y., Yamada, 
T., Hayashino, T., et al.\ 2004, \aj, 128, 569 

\bibitem[Matsuda et al.(2005)]{matsuda05} Matsuda, Y., Yamada, 
T., Hayashino, T., et al.\ 2005, \apjl, 634, L125 


\bibitem[Matsuda et al.(2010)]{matsuda10} Matsuda, Y., Richard, 
J., Smail, I., et al.\ 2010, \mnras, 403, L54 


\bibitem[Matsuda et al.(2011)]{matsuda11} Matsuda, Y., Yamada, 
T., Hayashino, T., et al.\ 2011, \mnras, 410, L13 


\bibitem[McDonald et al.(2002)]{mcdonald02} McDonald, P., 
Miralda-Escud{\'e}, J., \& Cen, R.\ 2002, \apj, 580, 42 


\bibitem[McQuinn(2015)]{mcquinn15} McQuinn, M.\ 2015, 
arXiv:1512.00086 


\bibitem[Miralda-Escud{\'e} et al.(1996)]{miralda96} 
Miralda-Escud{\'e}, J., Cen, R., Ostriker, J.~P., 
\& Rauch, M.\ 1996, \apj, 471, 582 


\bibitem[M{\o}ller et al.(2013)]{moller13} M{\o}ller, P., Fynbo, 
J.~P.~U., Ledoux, C., \& Nilsson, K.~K.\ 2013, \mnras, 430, 2680 


\bibitem[Noterdaeme et 
al.(2009)]{noterdaeme09} Noterdaeme, P., Petitjean, P., Ledoux, C., \& Srianand, R.\ 2009, \aap, 505, 1087 



\bibitem[Noterdaeme et 
al.(2012)]{noterdaeme12} Noterdaeme, P., Petitjean, P., Carithers, W.~C., et al.\ 2012, \aap, 547, LL1 



\bibitem[Noterdaeme et 
al.(2014)]{noterdaeme14} Noterdaeme, P., Petitjean, P., P{\^a}ris, I., et al.\ 2014, \aap, 566, AA24 


\bibitem[Oppenheimer et al.(2012)]{oppenheimer12} Oppenheimer, B.~D., 
Dav{\'e}, R., Katz, N., Kollmeier, J.~A., 
\& Weinberg, D.~H.\ 2012, \mnras, 420, 829 



\bibitem[Ouchi et al.(2005)]{ouchi05} Ouchi, M., Shimasaku, K., 
Akiyama, M., et al.\ 2005, \apjl, 620, L1 


\bibitem[Ouchi et al.(2008)]{ouchi08} Ouchi, M., Shimasaku, K., 
Akiyama, M., et al.\ 2008, \apjs, 176, 301 



\bibitem[P{\^a}ris et 
al.(2014)]{paris14} P{\^a}ris, I., Petitjean, P., Aubourg, {\'E}., et al.\ 2014, \aap, 563, A54 



\bibitem[Peirani et al.(2014)]{peirani14} Peirani, S., Weinberg, 
D.~H., Colombi, S., et al.\ 2014, \apj, 784, 11 


\bibitem[P{\'e}roux et al.(2008)]{peroux08} P{\'e}roux, C., 
Meiring, J.~D., Kulkarni, V.~P., et al.\ 2008, \mnras, 386, 2209 



\bibitem[Pieri et al.(2014)]{pieri14} Pieri, M.~M., Mortonson, 
M.~J., Frank, S., et al.\ 2014, \mnras, 441, 1718 



\bibitem[Prochaska et al.(2005)]{prochaska05} Prochaska, J.~X., 
Herbert-Fort, S., \& Wolfe, A.~M.\ 2005, \apj, 635, 123 


\bibitem[Prochaska et al.(2010)]{prochaska10} Prochaska, J.~X., 
O'Meara, J.~M., \& Worseck, G.\ 2010, \apj, 718, 392 



\bibitem[Prochaska et al.(2015)]{prochaska15} Prochaska, J.~X., 
O{'}Meara, J.~M., Fumagalli, M., Bernstein, R.~A., 
\& Burles, S.~M.\ 2015, \apjs, 221, 2 




\bibitem[Rauch(2006)]{rauch06} Rauch, M.\ 2006, KITP 
Conference: Applications of Gravitational Lensing: Unique Insights into 
Galaxy Formation and Evolution,  


\bibitem[Rubin et al.(2014)]{rubin14} Rubin, K.~H.~R., Hennawi, 
J.~F., Prochaska, J.~X., et al.\ 2014, arXiv:1411.6016 


\bibitem[Ross et al.(2012)]{ross12} Ross, N.~P., Myers, A.~D., 
Sheldon, E.~S., et al.\ 2012, \apjs, 199, 3 




\bibitem[Rudie et al.(2012)]{rudie12} Rudie, G.~C., Steidel, 
C.~C., Trainor, R.~F., et al.\ 2012, \apj, 750, 67 


\bibitem[Shapley et al.(2003)]{shapley03} Shapley, A.~E., 
Steidel, C.~C., Pettini, M., \& Adelberger, K.~L.\ 2003, \apj, 588, 65 



\bibitem[Steidel et al.(2005)]{steidel05} Steidel, C.~C., 
Adelberger, K.~L., Shapley, A.~E., et al.\ 2005, \apj, 626, 44 


\bibitem[Steidel et al.(2004)]{steidel04} Steidel, C.~C., 
Shapley, A.~E., Pettini, M., et al.\ 2004, \apj, 604, 534 


\bibitem[Steidel et al.(2003)]{steidel03} Steidel, C.~C., 
Adelberger, K.~L., Shapley, A.~E., et al.\ 2003, \apj, 592, 728 


\bibitem[Steidel et al.(1998)]{steidel98} Steidel, C.~C., 
Adelberger, K.~L., Dickinson, M., et al.\ 1998, \apj, 492, 428 



\bibitem[Slosar et al.(2013)]{slosar13} Slosar, A., Ir{\v s}i{\v c}, V., Kirkby, D., et al.\ 2013, JCAP, 4, 26 



\bibitem[Slosar et al.(2011)]{slosar11} Slosar, A., Font-Ribera, 
A., Pieri, M.~M., et al.\ 2011, JCAP, 9, 1 


\bibitem[Springel(2005)]{springel05} Springel, V.\ 2005, \mnras, 
364, 1105 



\bibitem[Steidel et al.(1998)]{steidel98} Steidel, C.~C., 
Adelberger, K.~L., Dickinson, M., et al.\ 1998, \apj, 492, 428 


\bibitem[Tamura et al.(2009)]{tamura09} Tamura, Y., Kohno, K., 
Nakanishi, K., et al.\ 2009, \nat, 459, 61 



\bibitem[Theuns et al.(2002)]{theuns02} Theuns, T., Viel, M., 
Kay, S., et al.\ 2002, \apjl, 578, L5 

\bibitem[Tejos et al.(2014)]{tejos14} Tejos, N., Morris, S.~L., 
Finn, C.~W., et al.\ 2014, \mnras, 437, 2017 



\bibitem[Tepper-Garc{\'{\i}}a et al.(2012)]{tepper-garc12} 
Tepper-Garc{\'{\i}}a, T., Richter, P., Schaye, J., et al.\ 2012, \mnras, 
425, 1640 


\bibitem[Tinker et al.(2010)]{tinker10} Tinker, J.~L., 
Robertson, B.~E., Kravtsov, A.~V., et al.\ 2010, \apj, 724, 878 


\bibitem[White et al.(2012)]{white12} White, M., Myers, A.~D., 
Ross, N.~P., et al.\ 2012, \mnras, 424, 933 



\bibitem[Venemans et 
al.(2007)]{venemans07} Venemans, B.~P., R{\"o}ttgering, H.~J.~A., Miley, G.~K., et al.\ 2007, \aap, 461, 823 


\bibitem[Viel et al.(2012)]{viel12} Viel, M., Markovi{\v c}, 
K., Baldi, M., \& Weller, J.\ 2012, \mnras, 421, 50 


\bibitem[Viel et al.(2013)]{viel13} Viel, M., Schaye, J., 
\& Booth, C.~M.\ 2013, \mnras, 429, 1734 


\bibitem[Worseck et al.(2014)]{worseck14} Worseck, G., Prochaska, 
J.~X., O'Meara, J.~M., et al.\ 2014, \mnras, 445, 1745 



\bibitem[White et al.(2011)]{white11} White, M., Blanton, M., 
Bolton, A., et al.\ 2011, \apj, 728, 126 



\bibitem[Yang et al.(2009)]{yang09} Yang, Y., Zabludoff, A., 
Tremonti, C., Eisenstein, D., \& Dav{\'e}, R.\ 2009, \apj, 693, 1579 





\bibitem[Yang et al.(2009)]{yang09} Yang, Y., Zabludoff, A., 
Tremonti, C., Eisenstein, D., \& Dav{\'e}, R.\ 2009, \apj, 693, 1579 


\bibitem[Yang et al.(2010)]{yang10} Yang, Y., Zabludoff, A., 
Eisenstein, D., \& Dav{\'e}, R.\ 2010, \apj, 719, 1654 


\bibitem[Yang et al.(2014)]{yang14} Yang, Y., Zabludoff, A., 
Jahnke, K., \& Dav{\'e}, R.\ 2014, \apj, 793, 114 


\bibitem[York et al.(2006)]{york06} York, D.~G., Khare, P., 
Vanden Berk, D., et al.\ 2006, \mnras, 367, 945 


\end{thebibliography}
\end{document}